\documentclass[%
 reprint,
 amsmath,amssymb,
 aps,
]{revtex4-2}

\usepackage{graphicx}
\usepackage{dcolumn}
\usepackage{bm}
\usepackage{hyperref}

\usepackage{multirow}

\begin{document}

\preprint{APS/123-QED}

\title{pyRheo: An open-source Python package for complex rheology}

\author{Isaac Y. Miranda-Valdez}
 \email{isaac.mirandavaldez@aalto.fi}
\author{Aaro Niinist{\"o}}%
\author{Tero M{\"a}kinen}%
\author{Juha Lejon}%
\author{Juha Koivisto}%
\author{Mikko J. Alava}%

\affiliation{%
Department of Applied Physics, Aalto University, P.O. Box 15600, 00076 Aalto, Espoo, Finland}%

\date{\today}

\begin{abstract}
Mathematical modeling is a powerful tool in rheology, 
and we present pyRheo, an open-source package for Python designed to streamline the analysis of creep, stress relaxation, oscillation, and rotation tests. pyRheo contains a comprehensive selection of viscoelastic models, including fractional order approaches. It integrates model selection and fitting features and employs machine intelligence to suggest a model to describe a given dataset. The package fits the suggested model or one chosen by the user. An advantage of using pyRheo is that it addresses challenges associated with sensitivity to initial guesses in parameter optimization. It allows the user to iteratively search for the best initial guesses, avoiding convergence to local minima. We discuss the capabilities of pyRheo and compare them to other tools for rheological modeling of biological matter. We demonstrate that pyRheo significantly reduces the computation time required to fit high-performance viscoelastic models. 
\end{abstract}

\maketitle


\section*{\label{sec:intro}Introduction}

Soft matter, 
such as cells and tissues, displays complexity in its composition, structure, and dynamic properties. As such, soft matter exhibits a time-dependent response in its mechanical properties known as viscoelasticity~\citep{bonfanti_kaplan_charras_kabla_2020, song_holten-andersen_mckinley_2023, ricarte_shanbhag_2024}. Quantifying the viscoelastic behavior of soft matter is critical for inferring its dynamics and microstructure. Rheology, as the branch of physics concerned with the deformation of matter, uses mathematical modeling to classify the viscoelastic response of soft matter. Furthermore, rheology abstracts parameters that can enable characterizing and predicting the response of soft matter at short and long timescales. The latter is of pivotal importance in the manufacturing and design of many materials and has widespread applications in fields such as tissue engineering, cell growth, and disease screening~\citep{steinwachs_metzner_skodzek_lang_thievessen_mark_münster_aifantis_fabry_2016, serwane_mongera_rowghanian_kealhofer_lucio_hockenbery_campàs_2017, wu_aroush_asnacios_chen_dokukin_doss_durand-smet_ekpenyong_guck_guz_etal_2018, liegeois_braunreuther_charbit_raymond_tang_woodruff_christenson_castro_erzurum_israel_etal_2024}.

Mathematical modeling in rheology is a gateway to understanding the structure-property relationship for soft matter. However, choosing a model, data processing of the experimental measurements, and curve-fitting routines can represent a steep learning curve in conducting and interpreting viscoelastic experiments, due to the highly non-linear nature of the behavior. For example, rheological models based on fractional order derivatives commonly require a curve-fitting routine that involves computationally expensive operations, such as the Mittag--Leffler function---most commonly represented by an infinite sum of terms containing the gamma function~\citep{haubold_mathai_saxena_2011, jaishankar_mckinley_2014, mainardi_2020}. Another challenge in the mathematical modeling of viscoelasticity is defining a cost function that allows selecting a model and further finding its parameters~\citep{ewoldt_johnston_carreta2015, freund_ewoldt_2015, singh_soulages_ewoldt_2019}. Therefore, choosing a model and inferring its parameters are two critical decisions that are highly uncertain and have non-unique solutions~\citep{singh_soulages_ewoldt_2019}.

Currently, there are limited options for rheological analysis available in the public domain. We need public tools that make rheology research more systematic, transparent, and reproducible. Therefore, this work introduces pyRheo, an open-source Python package that assists in model selection and fitting procedures for several different types of rheology experiments. pyRheo focuses on streamlining the mathematical modeling of rheological data obtained in the linear viscoelastic regime and from flow experiments. First, pyRheo uses machine intelligence to suggest a rheological model likely to describe a provided dataset. Then, it allows users to fit the proposed model or choose a different one. pyRheo enables the user to automatically or manually choose from several rheological models, focusing on fractional order viscoelastic models which have proved to be able to provide valuable insights into soft materials, such as food gels, structured fluids, biological tissues, cells, and polymers~\citep{faber_jaishankar_mckinley_2017b, miranda_cordova_renteria_fliri_2024, caputo_carcione_cavallini_2011, bauland_manna_divoux_gibaud_2024, legrand_baeza_peyla_porcar_fernández-de-alba_manneville_divoux_2024, tiong_crawford_jones_mckinley_2024, miranda-valdez_sourroubille_mäkinen_puente-córdova_puisto_koivisto_alava_2024}. 

pyRheo is highlighted as a tool for analyzing rheological data readily and as an interface that can be coupled with machine learning algorithms widely available for Python~\citep{miranda-valdez_mäkinen_coffeng_päivänsalo_jannuzzi_viitanen_koivisto_alava_2025}. The following sections demonstrate pyRheo's features. We present a set of robust validations against
experimental results and other public toolkits to check the accuracy and computational performance of our code package in characterizing soft materials such as biological tissues, polymers, foams, foodstuff, and gels.  We note that pyRheo is available via its \href{https://github.com/mirandi1/pyRheo.git}{GitHub repository}, where all the Python scripts to compute every single example presented in this paper and its Supplementary Information are available as Jupyter Notebooks. Furthermore, we have created a simple graphical user interface (GUI) for those users whose programming skills may limit their access to pyRheo. The GUI file can be found in the GitHub repository.

\begin{figure*}[htbp]
    \centering
    \includegraphics[width=0.80\linewidth]{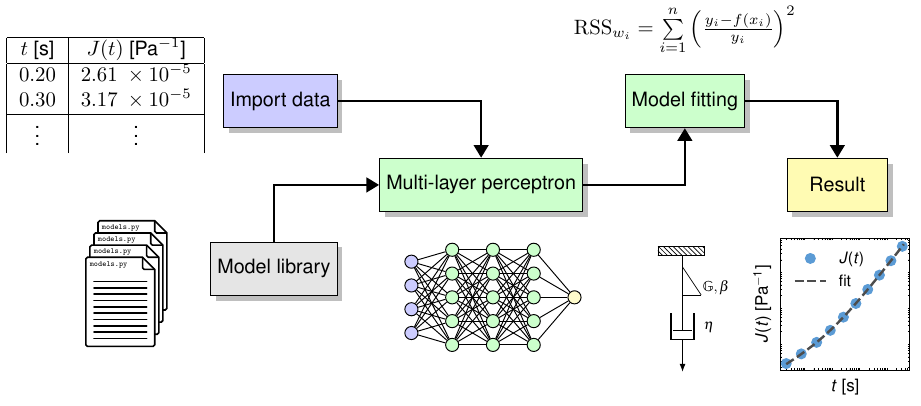}
    \caption{Workflow diagram of the pyRheo package, illustrating the data import process, model selection, and fitting. A time series is imported (left) into pyRheo. A model library provides rheological models depending on the \texttt{class} creep, stress relaxation, oscillation, and rotation. A Multi-Layer Perceptron (center) classifies the imported data, automatically assigning a model to perform the fitting. Model fitting is conducted by minimizing the weighted residual sum of squares ${\rm RSS}_{w_{i}}$ loss function. The final output (right) of the model fitting is stored as an object that can be called for predictions and further visualization.}
    \label{fig:intro_pyrheo_scheme}
\end{figure*}

\section*{\label{sec:results}Results}

Fig.~\ref{fig:intro_pyrheo_scheme} highlights, in green boxes, the two primary features of pyRheo: (i) it utilizes machine learning to determine which model best fits the user's rheological data, and (ii) it fits a rheological model to that data. The workflow of pyRheo is summarized in four steps: (1) importing data, (2) selecting a model, (3) fitting the model, and (4) analyzing the results. Supplementary~Note~1 details these four steps, while Supplementary~Note~2 describes all the models available in pyRheo, including their constitutive equations and representative plots. In step (2), users have the option to call a pre-trained machine learning model, specifically a Multi-Layer Perceptron (MLP), which can help infer the model that most likely describes the provided dataset. Supplementary~Note~3 explains how the MLP model was trained and assesses its performance using synthetic data.

\subsection*{Model prediction and fitting}

To demonstrate that pyRheo's capabilities lead to accurate and computationally efficient performance, we evaluate pyRheo using data from existing literature. In this section, we present the results obtained from fitting creep and stress relaxation data of two biological materials. Furthermore, we show how users of pyRheo can take advantage of available Python packages specifically designed for rheology. For instance, \citet{lennon_mcKinley_swan_2023} has developed a robust algorithm for creating master curves based on Gaussian Process Regression. Here, as well as in some of the demonstrations provided in Supplementary~Note~4, we showcase the results of integrating \citet{lennon_mcKinley_swan_2023}'s package with pyRheo.

\begin{figure}[tbp]
    \centering
    \includegraphics[width=0.99\linewidth]{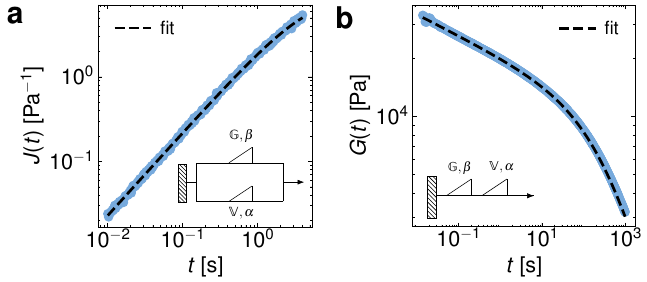}
    \caption{Creep and stress relaxation data classified and fitted with pyRheo. \textbf{a)} Creep compliance $J(t)$ of a perihepatic abscess sample. The curve is fitted using the \texttt{auto} function in pyRheo, which classifies the data as a \texttt{FractionalKelvinVoigt}. \textbf{b)} Relaxation modulus $G(t)$ of a fish muscle classified and fitted with \texttt{FractionalMaxwell}. The raw data of the perihepatic abscess was reproduced from \citet{shih_chung_shende_herwald_vezeridis_fuller_2024} and the data of the fish muscle from \citet{song_holten-andersen_mckinley_2023}.}
    \label{fig:results_pyrheo_fitting_creep_relax}
\end{figure}

First, we test the performance of pyRheo using the creep data measured for a perihepatic abscess reported by \citet{shih_chung_shende_herwald_vezeridis_fuller_2024} and the stress relaxation data of a fish muscle reported by \citet{song_holten-andersen_mckinley_2023}. 
The first step in pyRheo's workflow is to import the rheological data of creep compliance $J(t)$ and relaxation modulus $G(t)$ into the MLP model. In the cases shown in Fig.~\ref{fig:results_pyrheo_fitting_creep_relax}a,b, the MLP model classifies the data from creep as \texttt{FractionalKelvinVoigt} and stress relaxation as \texttt{FractionalMaxwell}. \texttt{FractionalKelvinVoigt} consists of two springpots connected in parallel, whereas \texttt{FractionalMaxwell} is built by two springpots connected in series (see insets in Fig.~\ref{fig:results_pyrheo_fitting_creep_relax})~\citep{song_holten-andersen_mckinley_2023}. The predicted model is automatically fitted to each dataset by pyRheo, and the fitting results are depicted by the dashed lines in Fig.~\ref{fig:results_pyrheo_fitting_creep_relax}a,b.

\begin{figure}[tbp]
    \centering
    \includegraphics[width=0.99\linewidth]{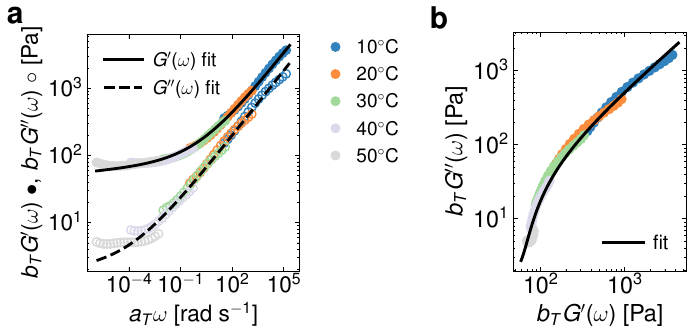}
    \caption{Master curves for the linear viscoelastic behavior of an interpenetrating-network hydrogel made of cellulose nanofibers and methylcellulose. \textbf{a)} Storage modulus $G^{\prime}(\omega)$ and Loss modulus $G^{\prime \prime}(\omega)$ master curves ($T_{\rm ref}=30^{\circ}$C) constructed using the time-temperature superposition (TTS) and fitted with \texttt{FractionalKelvinVoigt}, constituted by two \texttt{SpringPot} models connected in parallel. \textbf{b)} Cole--Cole representation of the master curve.}
    \label{fig:results_pyrheo_gel}
\end{figure}

In Fig.~\ref{fig:results_pyrheo_gel}, we showcase an instance of coupling \citet{lennon_mcKinley_swan_2023}'s package with pyRheo to analyze the oscillation data of an interpenetrating-network hydrogel made of cellulose nanofibers and methylcellulose. We import the master curve data from $G^{\prime}(\omega)$ and $G^{\prime \prime}(\omega)$ to the MLP classifier of pyRheo. 
The MLP classifies the $G^{\prime}(\omega)$ and $G^{\prime \prime}(\omega)$ data as a \texttt{FractionalKelvinVoigt}. The result of fitting this model to the master curve data is depicted by the solid and dashed lines in Fig.~\ref{fig:results_pyrheo_gel}a. Furthermore, we demonstrate in Fig.~\ref{fig:results_pyrheo_gel}b how to utilize the model predictions generated with pyRheo to easily construct other visualizations such as Cole--Cole diagrams. This is feasible because pyRheo stores the model results as an object the user can call to, for example, predict the material response according to a specified $\omega$ range. Consequently, this flexibility allows for estimating model predictions to higher and lower $\omega$ values.

This section is complemented by additional results in Supplementary~Note~4, where we present more fitting routines, which include materials such as mucus, foams, polymer networks, gels, plastics, food colloids, and polysaccharides. The examples included in Supplementary~Note~4 present data analysis from rotation experiments, which are not detailed in the main article due to their lower computational complexity. For transparency, every demonstration with pyRheo is available as a Jupyter Notebook on the pyRheo GitHub page, which users can test and adapt to suit their needs.

\subsection*{Performance of Mittag--Leffler function in pyRheo}

\begin{figure*}[tbp]
    \centering
    \includegraphics[width=0.99\linewidth]{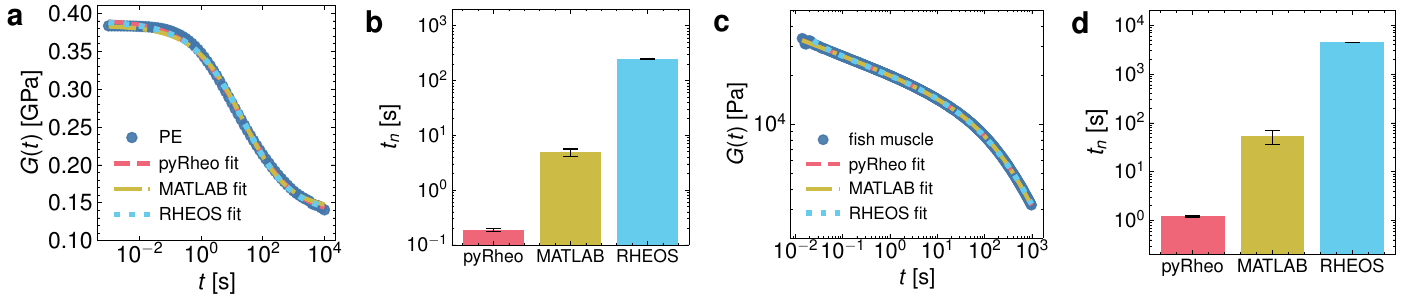}
    \caption{Performance comparison of fitting routines across pyRheo (Python), MATLAB, and RHEOS (Julia) for the relaxation modulus \( G(t) \) of polyethylene (PE) and fish muscle. \textbf{a)} Fitting results for PE from each implementation using \texttt{FractionalZenerSolidS}. \textbf{b)} Normalized computation times \( t_n \) for fitting routines on PE; RHEOS required downsampling to 20\% of the dataset, while pyRheo and MATLAB processed the full dataset. \textbf{c)} Fitting results for fish muscle from each implementation using \texttt{FractionalMaxwell}, based on \citet{song_holten-andersen_mckinley_2023}. \textbf{d)} Normalized computation times \( t_n \) for fish muscle fitting; RHEOS required downsampling to 10\%, while others used the complete dataset. All computations were on a system with an Intel Core i5-12600K CPU at 3.7 GHz, 31 GB RAM, and 1 TB SSD, running Ubuntu 20.04.}
    \label{fig:results_comparison_pyrheo_performance}
\end{figure*}

When using fractional rheological models, one often encounters the Mittag--Leffler (ML) function in the constitutive equation of the rheological model~\citep{mainardi_2020}. The ML function is expensive to compute as it is represented by an infinite sum of terms with gamma functions~$\Gamma$. In its generalized form, the ML function uses the following notation,
\begin{equation}
    E_{a, b}(z) = \sum_{n=0}^{\infty} \frac{z^n}{\Gamma(a n + b)}.
    \label{eq:mittag_leffler_function}
\end{equation}
There are several methods for numerically computing the ML function, either through the numerical inversion of its Laplace transform or by using mixed techniques, including Taylor series, asymptotic series, and integral representations~\citep{zeng_chen_2015}. Notable examples of these methods can be found in the algorithms developed by Garrappa~\citep{garrappa_numerical_2015} and Podlubny~\citep{sierociuk_podlubny_petras_2013}.

In fitting routines, the computational demands of the ML function can become increasingly sensitive to the size of the dataset. This often leads to exponential growth in computation time as the dataset expands. The latter is common in master curve fitting and creep and stress relaxation tests, where the sampling rate is higher than in oscillation and rotation tests. Current methods for reducing the computation time of the ML function typically involve downsampling. However, this process introduces uncertainty and can lead to non-unique outcomes in the fitting. The computational demand of the ML function is also evident when the fitting involves iterative optimization of multiple parameters; poor initial parameter guesses can result in slower convergence or even lead to convergence to a local minima. Consequently, finding the best model parameters may require restarting the optimization with a different initial guess for the model parameters.

To the best of our knowledge, using Padé approximations to compute rheological models has not been extensively researched. Thus, pyRheo exploits the Padé approach to reduce the computation time spent in fitting fractional rheological models by implementing the ML function based on the global Padé approximations proposed by \citet{zeng_chen_2015} and \citet{sarumi_furati_khaliq_2020}. In Fig.~\ref{fig:results_comparison_pyrheo_performance}, we compare the performance of pyRheo against popular methodologies for fitting fractional rheological models, which are based on the MATLAB toolkit published by~\citet{song_holten-andersen_mckinley_2023} and the RHEOS package for Julia programming language~\citep{kaplan_bonfanti_kabla_2019}. The MATLAB toolkit uses Garrappa's algorithm~\citep{garrappa_numerical_2015} to compute the ML function, whereas RHEOS uses~\citet{gorenflo2002mittag}'s approach. 

\begin{table*}[htbp]
    \caption{Comparison of material parameters for polyethylene and fish muscle~\citep{song_holten-andersen_mckinley_2023}. More detailed information about the rheological models are in Supplementary~Note~2.}
    \centering
    \renewcommand{\arraystretch}{1.5}
    \begin{tabular}{c c c c c c c}
    \hline
    Material & Model Scheme & Material Function & Parameter & pyRheo & MATLAB & RHEOS \\
    \hline
    \multirow{5}{*}{Polyethylene} &
    \multirow{5}{*}{\includegraphics[width=0.17\textwidth]{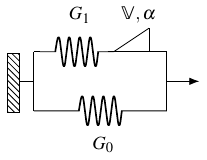}} & 
    \multirow{5}{*}{\(
    \begin{aligned}
    G(t) &= G_0 + G_1 E_{a,b}(z) \\ 
    a &= \alpha \\
    b &= 1 \\
    z &= -(t/\tau_c)^{\alpha} \\
    \end{aligned} \)} &
    $\mathbb{V}$ [Pa~s$^{\alpha}$] & $1.33\times10^{9}$ & $1.38\times10^{9}$ & $1.43\times10^{9}$ \\
    & & & $\alpha$  & $4.57\times10^{-1}$ & $4.20\times10^{-1}$ & $4.85\times10^{-1}$  \\
    & & & $G_0$ [Pa] & $1.32\times10^{8}$ & $1.40\times10^{8}$ & $1.35\times10^{8}$ \\
    & & & $G_1$ [Pa] & $2.60\times10^{8}$ & $2.73\times10^{8}$ & $2.53\times10^{8}$ \\
    & & & ${\rm RSS}_{w_{i}}$ & $6.91\times10^{-3}$  & $1.80\times10^{-2}$  & $9.31\times10^{-3}$ \\
    \hline
    \multirow{5}{*}{Fish muscle} &
    \multirow{5}{*}{\includegraphics[width=0.14\textwidth]{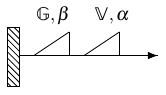}} & 
    \multirow{5}{*}{\(    
    \begin{aligned} 
    G(t) &= G_c(t/\tau_c)^{-\beta} E_{a,b}(z) \\ 
    a &= \alpha - \beta \\
    b &= 1-\beta \\
    z &= -(t/\tau_c)^{\alpha-\beta} \\
    \end{aligned} \)} &
    $\mathbb{V}$ [Pa~s$^{\alpha}$] & $8.70\times10^{5}$ & $8.72\times10^{5}$ & $6.12\times10^{5}$ \\
    & & & $\alpha$  & $6.74\times10^{-1}$ & $6.75\times10^{-1}$ & $6.29\times10^{-1}$  \\
    & & & $\mathbb{G}$ [Pa~s$^{\beta}$] & $2.21\times10^{4}$ & $2.20\times10^{4}$ & $2.25\times10^{4}$ \\
    & & & $\beta$ & $1.11\times10^{-1}$ & $1.10\times10^{-1}$ & $9.97\times10^{-2}$ \\
    & & & ${\rm RSS}_{w_{i}}$ & $1.08\times10^{-2}$  & $1.07\times10^{-2}$  & $9.72\times10^{-2}$ \\
    \hline
    \end{tabular}
\label{tab:material_parameter_comparison}
\end{table*}

Accordingly, in Fig.~\ref{fig:results_comparison_pyrheo_performance}a, we fitted a \texttt{FractionalZenerSolidS} to the relaxation modulus $G(t)$ of a polyethylene (PE) sample to show the applications of pyRheo in other fields of soft matter. For the examples reported here, we chose \texttt{Pade32} in pyRheo; in other words, a second-order global Padé approximation. Fig.~\ref{fig:results_comparison_pyrheo_performance}b displays the computation time spent in fitting the stress relaxation data of PE. The fitting requires computing the one-parameter ML function (i.e., $b=1$).  pyRheo, MATLAB, and RHEOS yield similar parameter values, as seen in Table~\ref{tab:material_parameter_comparison}. However, it is essential to note that the computation times $t_n$ vary significantly among the three implementations. For the stress relaxation data of PE, pyRheo identifies the optimal parameters one to three orders of magnitude faster than MATLAB and RHEOS, respectively. We normalize the computation times by the size of the dataset to enable a fair performance comparison across the different implementations. In the case of RHEOS, we downsampled the PE dataset to contain 20\% ($t_n$ is obtained by multiplying the real time by a factor of five) of the original one since the computation time extended beyond the capabilities of a desktop computer.

In Fig.~\ref{fig:results_comparison_pyrheo_performance}c, we show again the stress relaxation data of the fish muscle from Fig.~\ref{fig:results_pyrheo_fitting_creep_relax}b. Fig.~\ref{fig:results_comparison_pyrheo_performance}d shows that the computation of \texttt{FractionalMaxwell} is more time-consuming than that of the \texttt{FractionalZenerSolidS}. The latter is due to the multi-parametric nature of the ML function used in the \texttt{FractionalMaxwell}. Again, in Table~\ref{tab:material_parameter_comparison}, we observe that the three implementations find similar parameter values for the fish muscle. Nonetheless, the computation times with pyRheo are shorter than those needed by MATLAB and RHEOS. pyRheo leverages the computational efficiency of the ML function thanks to the global Padé approximation. Again, we normalized the computation times for the fish muscle data to enable a fair performance comparison across the different implementations. In the case of RHEOS, we downsampled the datasets to contain 10\% of the original ones (i.e., $t_n$ is the real computation time scaled by a factor of ten).

In addition to the global Padé approximation, we also programmed in pyRheo the option to use Garrappa's algorithm~\citep{garrappa_numerical_2015} for the evaluation of the Mittag--Leffler function. The algorithm was adapted from its MATLAB script~\citep{Garrappa2024}. The inclusion of Garrappa's algorithm allows users to benefit from its robust computation method, especially in cases where the Mittag--Leffler function needs to be evaluated for parameters that pose challenges for the global Padé approximation. This flexibility is crucial as it ensures that accurate and reliable results can be obtained across a broader range of applications and parameter settings, reinforcing pyRheo's position as a trusted tool for researchers and engineers in rheology.

In Supplementary~Note~5 and 6, we detail the global Padé approximations and provide examples demonstrating the differences between the global Padé approximation and Garrappa's algorithm when computing the \texttt{FractionalMaxwellGel}, \texttt{FractionalMaxwellLiquid}, and \texttt{FractionalMaxwell}. These examples showcase the accuracy and reliability of each algorithm in various scenarios. Such detailed comparisons help users make informed decisions about which algorithm to employ for their specific needs, thereby enhancing the overall utility and effectiveness of pyRheo in addressing complex rheological analyses.

\section*{Discussion}

Our Python package, pyRheo, delivers significant computational improvements that enable using fractional order viscoelastic models to describe soft materials effectively. These models have often been overlooked due to their computational complexity, making their implementation non-intuitive and challenging. Our work demonstrates how to reduce the computational cost of fitting routines involving the ML function, which is part of many models constitutive equations for creep compliance $J(t)$ and relaxation modulus $G(t)$. We decreased the computation times by utilizing Padé approximations to compute the ML function. This advancement allows for exploring a wider range of models and methodologies, requiring fewer resources and less time.

Besides faster computation of models with the ML function, the Padé approximation enables pyRheo to offer a solution to a common problem in fitting routines of rheological models, which is sensitivity to the initial guesses. Commonly, a bad choice of initial guess might lead the parameter optimization process to converge to local minima. pyRheo presents two solutions to determine the initial guesses: random search and Bayesian optimization (BO)~\citep{miranda-valdez_mäkinen_coffeng_päivänsalo_jannuzzi_viitanen_koivisto_alava_2025}.  

In the data analyzed in Fig.~\ref{fig:results_pyrheo_gel} and Fig.~\ref{fig:results_comparison_pyrheo_performance}, we have implemented a random search of initial guesses. In all cases, we have fixed a maximum of ten restarts of the optimization algorithm that seeks to minimize the weighted residual sum of squares ${\rm RSS}_{w_{i}}$. As shown in Fig.~\ref{fig:results_comparison_pyrheo_performance}, our brute-force approach combined with the global optimization algorithms from SciPy~\citep{2020SciPy-NMeth} yields faster computations than the MATLAB and RHEOS implementations. On the other hand, in Supplementary~Note~7, our BO approach was also shown to be effective in finding suitable initial guesses for fitting the creep data of a mucus gel. Our work reveals how random search and BO methodologies, techniques that are used in hyperparameter tunning of machine learning, can be adapted to traditional fitting routines.

The fitting tools available in pyRheo, along with its integrated machine intelligence, position it as a valuable resource for developing automated laboratories capable of conducting high-throughput testing and analysis. In the future, the Multi-Layer Perceptron (MLP) classifier could assist high-throughput rheometers in reformulating and optimizing materials. However, it is important to note that the current MLP classifier integrated into pyRheo cannot label rheological data related to Zener models. This limitation arises because the responses of Zener models overlap with those of Maxwell and Kelvin-Voigt models. One potential strategy to address this issue would be to use multi-label classifiers.

\section*{Methods}

pyRheo's methodology has two main features: (i) providing a machine learning decision of what model likely describes the rheological data and (ii) fitting a rheological model to the data. Based on these two features the package can analyze creep, stress relaxation, oscillation, and rotation data. 

\begin{table*}[htbp]
    \caption{Summary of classes in pyRheo, their available models, and fitting methodologies.}
    \centering
    \begin{tabular}{ >{\centering\arraybackslash}m{0.1\linewidth} >{\centering\arraybackslash}m{0.2\linewidth} >{\centering\arraybackslash}m{0.2\linewidth} >{\centering\arraybackslash}m{0.2\linewidth} >{\centering\arraybackslash}m{0.2\linewidth}} 
      \hline
      & \texttt{class} Creep & \texttt{class} Relaxation & \texttt{class} Oscillation & \texttt{class} Rotation \\
      \hline 
      Features \vfill  & $t$, $J(t)$  \vfill   & $t$, $G(t)$  \vfill          & $\omega$, $G^{\prime}(\omega)$, $G^{\prime \prime}(\omega)$  \vfill &  $\dot{\gamma}$, $\eta(\dot{\gamma})$  \vfill\\
      Plot \vfill &  \includegraphics[width=0.85\linewidth]{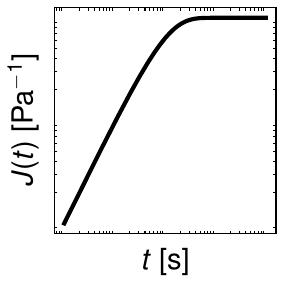} \vfill & \includegraphics[width=0.85\linewidth]{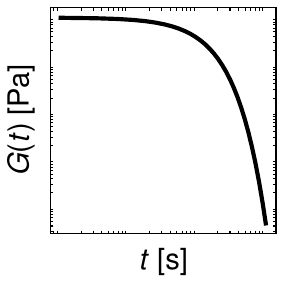} \vfill & \includegraphics[width=0.85\linewidth]{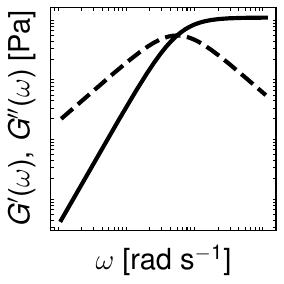} \vfill & \includegraphics[width=0.85\linewidth]{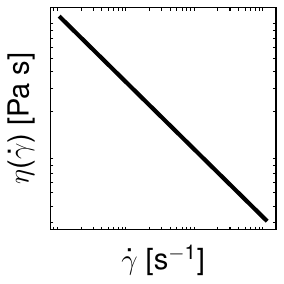} \vfill \\
      Models  \vfill  & \texttt{Maxwell}, \texttt{SpringPot}, \texttt{FractionalMaxwell}, \ldots \vfill & \texttt{Maxwell}, \texttt{SpringPot}, \texttt{FractionalMaxwell}, \ldots \vfill & \texttt{Maxwell}, \texttt{SpringPot}, \texttt{FractionalMaxwell}, \ldots \vfill \vfill & \texttt{HerschelBulkley}, \texttt{Bingham}, \texttt{PowerLaw}, \texttt{CarreauYasuda}, \texttt{Cross}, \texttt{Casson} \vfill\\
      Initial guess \vfill & \texttt{manual}, \texttt{random}, \texttt{bayesian} \vfill & \texttt{manual}, \texttt{random}, \texttt{bayesian} \vfill & \texttt{manual}, \texttt{random}, \texttt{bayesian}  \vfill & \texttt{manual}, \texttt{random}, \texttt{bayesian} \vfill \\
      ML type \vfill & \texttt{Pade32}, \texttt{Pade54}, \texttt{Pade63}, \texttt{Pade72}, \texttt{Garrappa} \vfill & \texttt{Pade32}, \texttt{Pade54}, \texttt{Pade63}, \texttt{Pade72}, \texttt{Garrappa} \vfill & \texttt{Pade32}, \texttt{Pade54}, \texttt{Pade63}, \texttt{Pade72}, \texttt{Garrappa} \vfill & \texttt{Pade32}, \texttt{Pade54}, \texttt{Pade63}, \texttt{Pade72}, \texttt{Garrappa} \vfill \\
    \hline
    \end{tabular}
    \label{tab:features_models_pyrheo}
\end{table*}

\subsubsection*{Step 1: importing data}

First, the user should import the data pertinent to the type of rheological dataset. Depending on the specific nature of the dataset, users should import the relevant features and their corresponding class (creep, stress relaxation, oscillation, or rotation) following Table~\ref{tab:features_models_pyrheo}. pyRheo is designed to work with material functions, so one must provide at least two data vectors. For example, for creep and stress relaxation data, it is expected to import a time $t$ vector together with its corresponding material function $J(t)$ or $G(t)$. Alternatively, for oscillation data, the user must import angular frequency $\omega$ and the materials functions storage modulus $G^{\prime}(\omega)$ and loss modulus $G^{\prime \prime}(\omega)$.

\subsubsection*{Step 2: model selection} 

After importing the data, the user shall select to analyze their data using the \texttt{auto} method or by manually specifying a model according to Table~\ref{tab:features_models_pyrheo}. The \texttt{auto} method uses a pre-trained Neural Network based on a Multi-Layer Perceptron (MLP). Each class has its own MLP classifier, which has been trained using 1 million computations of the corresponding material function (e.g., \(J(t)\)) derived from the constitutive equations of the \texttt{Maxwell}, \texttt{SpringPot}, \texttt{FractionalMaxwellGel}, \texttt{FractionalMaxwellLiquid}, \texttt{FractionalMaxwell}, and \texttt{FractionalKelvinVoigt}. 

The accuracy (with synthetic data) of the MLP classifiers ranges from 70 to 80\%. We suggest using the \texttt{auto} method as a first approximation to identify the type of rheological behavior. More detailed information about the machine learning training process and performance is disclosed in the Supplementary Information.

\subsubsection*{Step 3: model fitting} 

Parameter optimization with pyRheo follows the common practice of minimizing the weighted residual sum of squares~\citep{singh_soulages_ewoldt_2019, song_holten-andersen_mckinley_2023},
\begin{equation}
    {\rm RSS}_{w_{i}} = \sum\limits_{i=1}^{n} \left(\frac{y_i - f(x_i)}{y_i}\right)^2.
    \label{eq:rss_weighed}
\end{equation}
Users may define their own initial guesses and parameter bounds (automatic bounds and random initial guesses by default). Then, to minimize ${\rm RSS}_{w_{i}}$, users can choose from several minimization algorithms implemented on SciPy~\citep{2020SciPy-NMeth}, such as Nelder--Mead, Powell, and \mbox{L-BFGS-B} (Powell by default). 

As shown in Table~\ref{tab:features_models_pyrheo}, an advantage of using pyRheo is that it addresses the challenges associated with sensitivity to initial guesses in parameter optimization. In other words, if an initial guess is close to a local minimum, the minimization algorithm may converge there instead of the global minimum. Therefore, pyRheo allows the user to restart the fitting process multiple times with random initial parameter values and then take as the final result the iteration with the lowest ${\rm RSS}_{w_{i}}$. By generating a diverse set of random starting points, this brute-force approach increases the likelihood of exploring different regions of the parameter space, thus avoiding local minima.

Another method that pyRheo offers for defining initial guesses is Bayesian Optimization (BO)~\citep{head2018scikit, miranda-valdez_viitanen_macintyre_puisto_koivisto_alava_2022, valtteri2024improving}. In this approach, pyRheo creates a mapping from the parameter space $\mathcal{P}$ to the error space $\mathcal{E}$ using Gaussian Process Regression (GPR), represented as  $g: \mathcal{P} \to \mathcal{E}$, where $g$ is the Gaussian Process. The surrogate model $\epsilon = g(p)$ (with $\epsilon \in \mathcal{E}$ and $p \in \mathcal{P}$) is developed by computing the constitutive equation of the target model with fixed parameter values and then recording the difference (residuals) between this computation and the data being analyzed. The goal of BO is to minimize $\epsilon$ by exploring various combinations of parameter values, guided by an acquisition function known as Expected Improvement, which balances exploration and exploitation of the parameter space. Afterward, pyRheo uses the BO solution as the initial guess for the minimization algorithm.

\subsubsection*{Step 4: analysis of results}

After fitting the target model to the rheological data, the results are stored as an object variable that contains all the necessary components for prediction, visualization, and further data analysis. Users can learn more from pyRheo's documentation and the examples available on its GitHub repository.

\section*{Author Contributions}
{\bf Isaac Y. Miranda-Valdez:} Conceptualization, Methodology, Software, Formal analysis, Visualization, Investigation, Data curation, Writing -- Original draft, Writing -- Review \& Editing, Project administration, and Funding acquisition.
{\bf Aaro Niinistö:} Investigation and Software.
{\bf Tero Mäkinen:} Validation and Writing -- Review \& Editing.
{\bf Juha Lejon:} Investigation.
{\bf Juha Koivisto:} Supervision, Validation, Writing -- Review \& Editing, Funding acquisition.
{\bf Mikko J. Alava:} Supervision, Validation, Writing -- Review \& Editing, Funding acquisition, Project administration.

\section*{Conflicts of interest}
There are no conflicts to declare.

\section*{Data availability}
The code for pyRheo, data analysis scripts of this article, and the data for this article are available at pyRheo's Github at \url{https://github.com/mirandi1/pyRheo.git} – format https://doi.org/DOI].

\begin{acknowledgments}
The authors thank Dr. Jesús G. Puente-Córdova for his suggestions and feedback during the conceptualization of pyRheo as well as for sharing experimental data for testing tasks. 
I.M.V. acknowledges the Vilho, Yrjö, and Kalle Väisälä Foundation of the Finnish Academy of Science and Letters for personal funding. 
M.J.A. acknowledges funding from the Finnish Cultural Foundation.
M.J.A., J.K., T.M. and I.M.V. acknowledge funding from Business Finland (211909, 211989).
M.J.A. and J.K. acknowledge funding from FinnCERES flagship (151830423), Business Finland (211835), and Future Makers programs.
Aalto Science-IT project is acknowledged for computational resources.
\end{acknowledgments}

\bibliography{main}

\end{document}


\begin{center}
    \Large
    \textbf{Supplementary Information}
\end{center}
\normalsize

\hspace{2cm}

\noindent {\large \bf pyRheo: An open-source Python package for complex rheology} \\
   
\noindent Isaac Y. Miranda-Valdez,\textit{$^{a,}$}$^{\ast}$ Aaro Niinist\"{o},\textit{$^{a}$} Tero M\"{a}kinen,\textit{$^{a}$}, Juha Lejon,\textit{$^{a}$}, Juha Koivisto,\textit{$^{a}$} and Mikko J. Alava,\textit{$^{a}$}\\

\noindent $^a$Complex Systems and Materials, Department of Applied Physics, Aalto University, P.O. Box 11000, FI-00076 Aalto, Espoo, Finland \\
\noindent $^\ast$ Corresponding author: isaac.mirandavaldez@aalto.fi\\

\noindent {\bf Abstract}  \\
This Supplementary Information accompanies the article ``pyRheo: An open-source Python package for complex rheology''. It provides an overview of the theoretical models implemented in pyRheo. Detailed equations and representative plots for various models are presented to elucidate their functionality within pyRheo. Here we cover the Maxwell, Springpot, Fractional Maxwell, Fractional Kelvin-Voigt, and Fractional Zener models. For each model, the documentation entails the model diagram, constitutive equation, creep compliance, relaxation modulus, and storage and loss moduli. Additionally, representative plots illustrate the practical application of these models. We also discuss how the machine learning models were trained for pyRheo, which is accompanied by confusion matrices showing the machine learning accuracy. The Supplementary Information provides several examples of fitting done with pyRheo. Furthermore, here we describe how the Mittag--Leffler function is implemented in pyRheo, highlighting the fourth-order global Padé approximation and comparing it with Garrappa's algorithm. The Supplementary Information presents an example of using Bayesian optimization to infer the initial guesses for the parameter fitting. This document also presents a tutorial on how to use pyRheo graphical user interface (GUI).
\\

\clearpage



\section{pyRheo workflow}
\label{sec:workflow_methodology}

The general purpose of this package is to analyze creep, stress relaxation, oscillation, and rotation data. pyRheo has two main features: (i) providing a machine learning decision of what model likely describes the rheological data and (ii) fitting a rheological model to the data.  Next, we describe pyRheo's workflow.

\subsubsection*{Step 1: importing data}

First, the user should import the data pertinent to the type of rheological dataset. Depending on the specific nature of the dataset, users should import the relevant features and their corresponding class (creep, stress relaxation, oscillation, or rotation). pyRheo is designed to work with material functions, so one must provide at least two data vectors. For example, for creep and stress relaxation data, it is expected to import a time $t$ vector together with its corresponding material function $J(t)$ or $G(t)$. Alternatively, for oscillation data, the user must import angular frequency $\omega$ and the materials functions storage modulus $G^{\prime}(\omega)$ and loss modulus $G^{\prime \prime}(\omega)$.

\subsubsection*{Step 2: model selection} 

After importing the data, the user shall select to analyze their data using the \texttt{auto} method or by manually specifying a modeL. The \texttt{auto} method uses a pre-trained Neural Network based on a Multi-Layer Perceptron (MLP). Each class has its own MLP classifier, which has been trained using 1 million computations of the corresponding material function (e.g., \(J(t)\)) derived from the constitutive equations of the \texttt{Maxwell}, \texttt{SpringPot}, \texttt{FractionalMaxwell}, and \texttt{FractionalKelvinVoigt}. 

The accuracy (with synthetic data) of the MLP classifiers ranges from 70 to 80\%. We suggest using the \texttt{auto} method as a first approximation to identify the type of rheological behavior. More detailed information about the machine learning training process and performance is disclosed in the Supplementary Note 3.

\subsubsection*{Step 3: model fitting} 

Parameter optimization with pyRheo follows the common practice of minimizing the weighted residual sum of squares~\citep{singh_soulages_ewoldt_2019, song_holten-andersen_mckinley_2023},
\begin{equation}
    {\rm RSS}_{w_{i}} = \sum\limits_{i=1}^{n} \left(\frac{y_i - f(x_i)}{y_i}\right)^2.
    \label{eq:rss_weighed}
\end{equation}
Users may define their own initial guesses and parameter bounds (automatic bounds and random initial guesses by default). Then, to minimize ${\rm RSS}_{w_{i}}$, users can choose from several minimization algorithms implemented on SciPy~\citep{2020SciPy-NMeth}, such as Nelder--Mead, Powell, and \mbox{L-BFGS-B} (Powell by default). 

An advantage of using pyRheo is that it addresses the challenges associated with sensitivity to initial guesses in parameter optimization. In other words, if an initial guess is close to a local minimum, the minimization algorithm may converge there instead of the global minimum. Therefore, pyRheo allows the user to restart the fitting process multiple times with random initial parameter values and then take as the final result the iteration with the lowest ${\rm RSS}_{w_{i}}$. By generating a diverse set of random starting points, this brute-force approach increases the likelihood of exploring different regions of the parameter space, thus avoiding local minima.

In the Supplementary Note 2, we detail all the models implemented in pyRheo. The reader can find the constitutive equations of each model and representative plots showing how material functions change as a function of varying parameters. Furthermore, the reader can find information about how the Mittag--Leffler function is implemented in pyRheo using \citet{zeng_chen_2015}, \citet{sarumi_furati_khaliq_2020}, and \citet{garrappa_numerical_2015} concepts.

\subsubsection*{Step 4: analysis of results}

After fitting the target model to the rheological data, the results are stored as an object variable that contains all the necessary components for prediction, visualization, and further data analysis. Users can learn more from pyRheo's documentation and the examples available on its GitHub repository.

\clearpage

\section{Equations of the models implemented in pyRheo}

\subsection*{Maxwell model (\texttt{Maxwell})}

\subsubsection*{Model diagram}

\begin{figure}[h]
    \centering
    \begin{tikzpicture}
        \begin{circuitikz}
            \ctikzset{bipoles/thickness=1} 

            \node (spring) at (0,0) {};
            \draw (0,0) to[spring] ++(0.85,0);

            \node (dashpot) at (0.85,0) {};
            \draw (0.85,0) to[damper] ++(1.05,0);

            \draw [-] (0.0, 0) -- (-0.25, 0);
            \draw [Latex-] (2.25, 0) -- (1.7, 0);

            \draw[black, pattern=north west lines, opacity=1.0] (-0.45, 0.47) rectangle ++(0.2, -1.0);
        \end{circuitikz}

        \node [] at (-1.75, -0.1) {\small $G$}; 
        \node [] at (-0.85, -0.1) {\small $\eta$}; 
    \end{tikzpicture}
\end{figure}

\subsubsection*{Constitutive equation}
\begin{equation}
    \begin{aligned}
    \sigma(t) + \frac{\eta}{G} \frac{{\rm d}\sigma(t)}{{\rm d}t} &= \eta \frac{{\rm d}\gamma(t)}{{\rm d}t} \\
    \tau_c &= \frac{\eta}{G} \\
    G_c &= G
    \end{aligned}
\end{equation}

\subsubsection*{Creep compliance}
\begin{equation}
    J(t) = \frac{t}{\eta} + \frac{1}{G}
\end{equation}

\subsubsection*{Relaxation modulus}
\begin{equation}
    G(t) = G_c \exp{\left(-\frac{t}{\tau_c} \right)}
\end{equation}

\subsubsection*{Storage modulus and Loss modulus}
\begin{equation}
    \begin{aligned}
    G^{\prime}(\omega) &= G_c \frac{(\omega \tau_c)^{2}}{1 + (\omega \tau_c)^{2}} \\
    G^{\prime \prime}(\omega) &= G_c \frac{\omega \tau_c}{1 + (\omega \tau_c)^{2}} \\
    \end{aligned}
\end{equation}

\subsubsection*{Representative plots}
\begin{figure}[htbp]
    \centering
    \begin{subfigure}{0.3\textwidth}
        \centering
        \begin{picture}(0,0)
            \put(-67,0){\textbf{\Large a}} 
        \end{picture}
        \includegraphics[width=\linewidth]{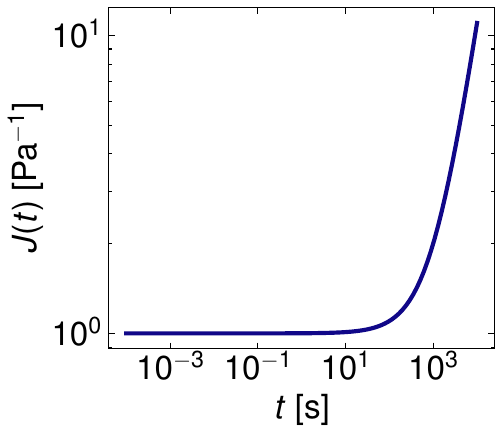} 
    \end{subfigure}
    \hfill
    \begin{subfigure}{0.3\textwidth}
        \centering
        \begin{picture}(0,0)
            \put(-67,0){\textbf{\Large b}} 
        \end{picture}
        \includegraphics[width=\linewidth]{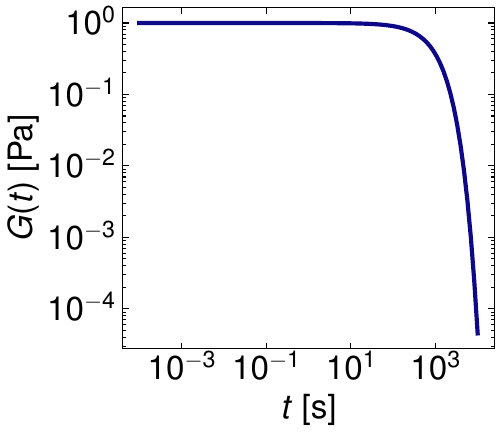} 
    \end{subfigure}
    \hfill
    \begin{subfigure}{0.3\textwidth}
        \centering
        \begin{picture}(0,0)
            \put(-67,0){\textbf{\Large c}} 
        \end{picture}
        \includegraphics[width=\linewidth]{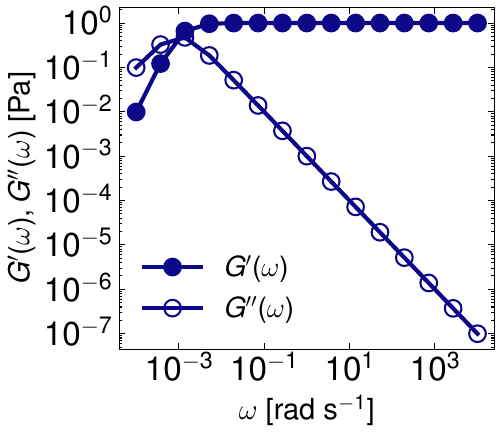} 
    \end{subfigure}
    \caption{Representative plots of the Maxwell model illustrating different viscoelastic behaviors under various types of stimulus. The plots were computed using fixed parameter values of $G = 1$~Pa and $\eta = 1000$~Pa~s. \textbf{a)} Time-dependent creep compliance $J(t)$ showing the variation of strain over time under constant stress. \textbf{b)} Stress relaxation behavior depicting a decrease in the relaxation modulus $G(t)$ under constant strain over time. \textbf{c)} Oscillatory response demonstrating the storage modulus $G^{\prime}(\omega)$ and loss modulus $G^{\prime \prime}(\omega)$ under frequency-changing load.}
    \label{fig:mm_plots}
\end{figure}

\clearpage
\subsection*{Springpot model (\texttt{SpringPot})}

\subsubsection*{Model diagram}

\begin{figure}[h]
    \centering
    \begin{tikzpicture}
        \begin{circuitikz}
            \ctikzset{bipoles/thickness=1} 

            \draw (0.35, -0.0) -- (0, -0.0);
            \draw (1.15, -0.0) -- (1.15, 0.4);
            \draw (1.15, 0.4) -- (0.6, 0.0);

            \draw [-] (-0.25, 0) -- (2.25, 0);
            \draw [Latex-] (2.25, 0) -- (1.7, 0);

            \draw[black, pattern=north west lines, opacity=1.0] (-0.45, 0.47) rectangle ++(0.2, -1.0);
        \end{circuitikz}

        \node [] at (-1.3, -0.0) {\small $\mathbb{V},\alpha$}; 
    \end{tikzpicture}
\end{figure}

\subsubsection*{Constitutive equation}
\begin{equation}
    \begin{aligned}
    \sigma(t)  &= \mathbb{V} \frac{{\rm d}^{\alpha}\gamma(t)}{{\rm d}t^{\alpha}} \\
    \end{aligned}
\end{equation}

\subsubsection*{Creep compliance}
\begin{equation}
    J(t) = \frac{1}{\mathbb{V}} \frac{t^{\alpha}}{\Gamma(1+\alpha)}
\end{equation}

\subsubsection*{Relaxation modulus}
\begin{equation}
    G(t) = \mathbb{V} \frac{t^{-\alpha}}{\Gamma(1-\alpha)}
\end{equation}

\subsubsection*{Storage modulus and Loss modulus}
\begin{equation}
    \begin{aligned}
    G^{\prime}(\omega) &= \mathbb{V} \omega^{\alpha} \cos{(\frac{\pi}{2} \alpha)} \\
    G^{\prime \prime}(\omega) &= \mathbb{V} \omega^{\alpha} \sin{(\frac{\pi}{2} \alpha)} \\
    \end{aligned}
\end{equation}

\subsubsection*{Representative plots}
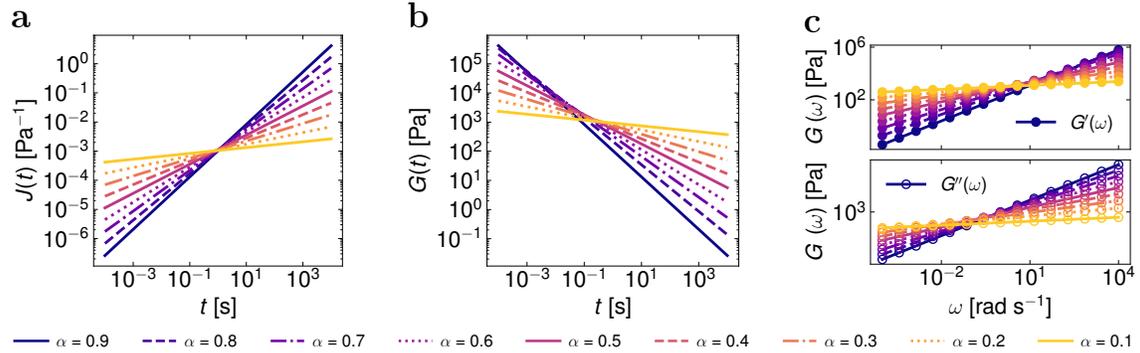
\begin{figure}[htbp]
    \centering
    \begin{subfigure}{0.3\textwidth}
        \centering
        \begin{picture}(0,0)
            \put(-67,0){\textbf{\Large a}} 
        \end{picture}
        \includegraphics[width=\linewidth]{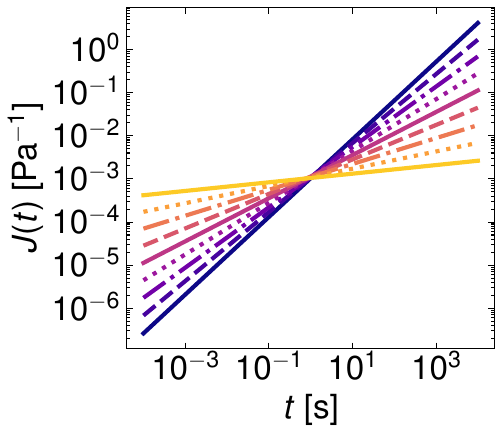} 
    \end{subfigure}
    \hfill
    \begin{subfigure}{0.3\textwidth}
        \centering
        \begin{picture}(0,0)
            \put(-67,0){\textbf{\Large b}} 
        \end{picture}
        \includegraphics[width=\linewidth]{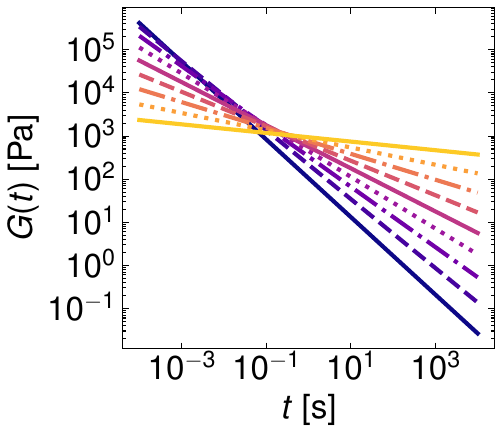} 
    \end{subfigure}
    \hfill
    \begin{subfigure}{0.3\textwidth}
        \centering
        \begin{picture}(0,0)
            \put(-67,0){\textbf{\Large c}} 
        \end{picture}
        \includegraphics[width=\linewidth]{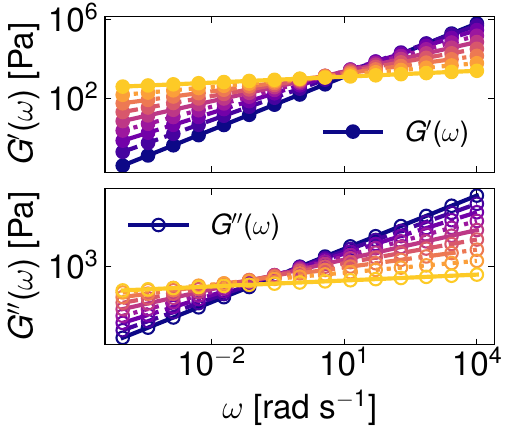} 
    \end{subfigure}
    \begin{subfigure}{1\textwidth}
        \centering
        \includegraphics[width=\linewidth]{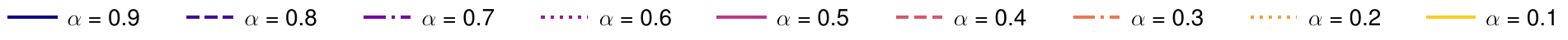} 
    \end{subfigure}
    \caption{Representative plots of the Springpot (critical gel) model illustrating different viscoelastic behaviors under various types of stimulus. The plots were computed using fixed parameter values of $\mathbb{V} = 1000$~Pa~s$^{\alpha}$ and varying $\alpha$. \textbf{a)} Time-dependent creep compliance $J(t)$ showing the variation of strain over time under constant stress. \textbf{b)} Stress relaxation behavior depicting a decrease in the relaxation modulus $G(t)$ under constant strain over time. \textbf{c)} Oscillatory response demonstrating the storage modulus $G^{\prime}(\omega)$ and loss modulus $G^{\prime \prime}(\omega)$ under frequency-changing load.}    
    \label{fig:sp_plots}
\end{figure}

\clearpage
\subsection*{Fractional Maxwell Gel model (\texttt{FractionalMaxwellGel})}

\subsubsection*{Model diagram}

\begin{figure}[H]
    \centering
    \begin{tikzpicture}
        \begin{circuitikz}
            \ctikzset{bipoles/thickness=1} 

            \draw (0,0) to[spring] ++(1.,0)
            to ++(1.,0); 
            \draw (1.75, -0.0) -- (1.75, 0.4);
            \draw (1.75, 0.4) -- (1.2, 0.0);
            \draw (-0.25, -0.0) -- (0, -0.0);
            \draw [-Latex] (2.0, -0.0) -- (2.25, -0.0 );
        
            \draw[black, pattern=north west lines, opacity=1.0] (-0.45,0.47) rectangle ++(0.2,-1.0);
        \end{circuitikz}
    \node [] at (-1.75, -0.1) {\small $G$}; 
    \node [] at (-0.75, -0.1) {\small $\mathbb{V},\alpha$}; 
    \end{tikzpicture}
\end{figure}

\subsubsection*{Constitutive equation}
\begin{equation}
    \begin{aligned}
    \sigma(t) + \frac{\mathbb{V}}{G} \frac{{\rm d}^{\alpha}\sigma(t)}{{\rm d}t^{\alpha}} &= \mathbb{V} \frac{{\rm d}^{\alpha}\gamma(t)}{{\rm d}t^{\alpha}} \\
    \tau_c &= \left(\frac{\mathbb{V}}{G} \right)^{\frac{1}{\alpha}}\\
    G_c &= \mathbb{V} \tau_c^{-\alpha}
    \end{aligned}
    \label{eq:constitutive_fmg}
\end{equation}

\subsubsection*{Creep compliance}
\begin{equation}
    J(t) = \frac{1}{\mathbb{V}} \frac{t^{\alpha}}{\Gamma(1+\alpha)} + \frac{1}{G}
    \label{eq:creep_fmg}
\end{equation}

\subsubsection*{Relaxation modulus}
\begin{equation}
    \begin{aligned}
    G(t) &= G_c E_{a,b}(z)\\
    a &= \alpha \\
    b &= 1 \\
    z &= -\left(\frac{t}{\tau_c}\right)^{\alpha}
    \end{aligned}
    \label{eq:relaxation_fmg}
\end{equation}

\subsubsection*{Storage modulus and Loss modulus}
\begin{equation}
    \begin{aligned}
    G^{\prime}(\omega) &= G_c \frac{(\omega \tau_c)^{2\alpha} + (\omega \tau_c)^{\alpha} \cos{(\frac{\pi}{2}\alpha)}}{1 + (\omega \tau_c)^{2\alpha} + 2(\omega \tau_c)^{\alpha} \cos{(\frac{\pi}{2}\alpha)}} \\
    G^{\prime \prime}(\omega) &= G_c \frac{(\omega \tau_c)^{\alpha} \sin{(\frac{\pi}{2}\alpha)}}{1 + (\omega \tau_c)^{2\alpha} + 2(\omega \tau_c)^{\alpha} \cos{(\frac{\pi}{2}\alpha)}} \\
    \end{aligned}
    \label{eq:oscillation_fmg}
\end{equation}

\subsubsection*{Representative plots}
\begin{figure}[htbp]
    \centering
    \begin{subfigure}{0.3\textwidth}
        \centering
        \begin{picture}(0,0)
            \put(-67,0){\textbf{\Large a}} 
        \end{picture}
        \includegraphics[width=\linewidth]{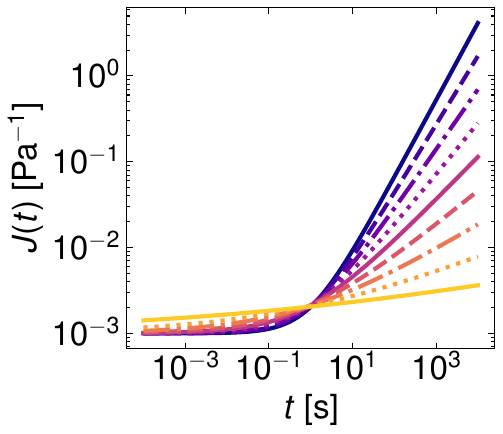} 
    \end{subfigure}
    \hfill
    \begin{subfigure}{0.3\textwidth}
        \centering
        \begin{picture}(0,0)
            \put(-67,0){\textbf{\Large b}} 
        \end{picture}
        \includegraphics[width=\linewidth]{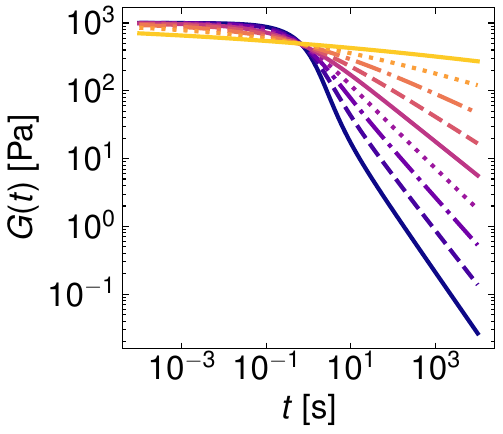} 
    \end{subfigure}
    \hfill
    \begin{subfigure}{0.3\textwidth}
        \centering
        \begin{picture}(0,0)
            \put(-67,0){\textbf{\Large c}} 
        \end{picture}
        \includegraphics[width=\linewidth]{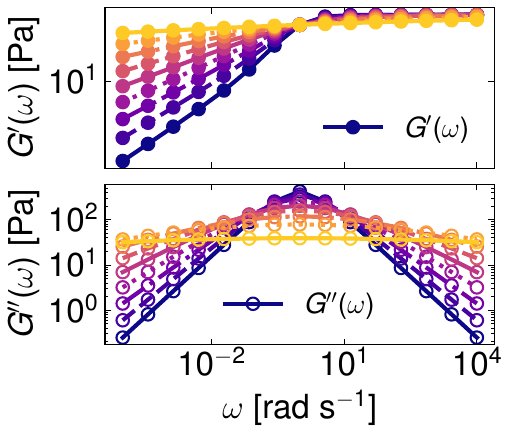} 
    \end{subfigure}
    \begin{subfigure}{1\textwidth}
        \centering
        \includegraphics[width=\linewidth]{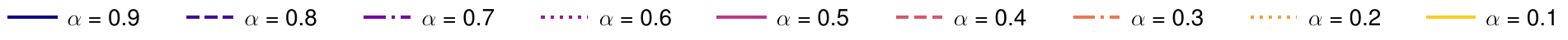} 
    \end{subfigure}
    \caption{Representative plots of the Fractional Maxwell Gel model illustrating different viscoelastic behaviors under various types of stimulus. The plots were computed using fixed parameter values of $\mathbb{V} = 1000$~Pa~s$^{\alpha}$ and $G = 1000$~Pa as well as varying $\alpha$. \textbf{a)} Time-dependent creep compliance $J(t)$ showing the variation of strain over time under constant stress. \textbf{b)} Stress relaxation behavior depicting a decrease in the relaxation modulus $G(t)$ under constant strain over time. \textbf{c)} Oscillatory response demonstrating the storage modulus $G^{\prime}(\omega)$ and loss modulus $G^{\prime \prime}(\omega)$ under frequency-changing load.}      
    \label{fig:fmg_plots}
\end{figure}
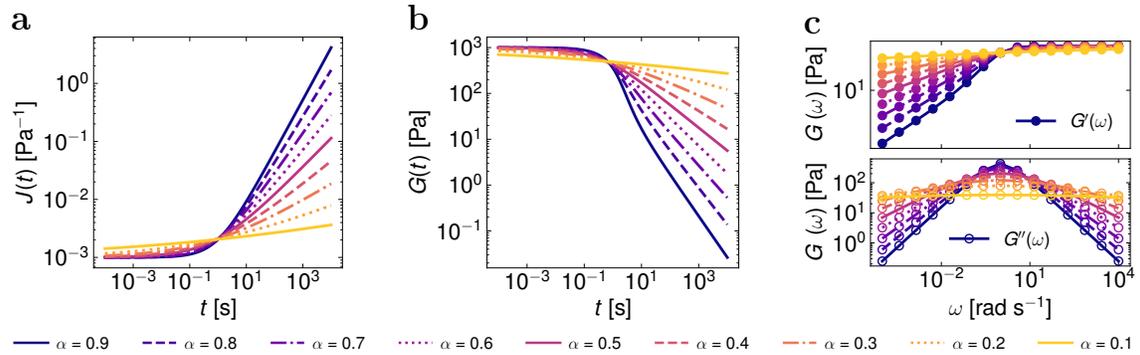

\clearpage
\subsection*{Fractional Maxwell Liquid model (\texttt{FractionalMaxwellLiquid})}

\subsubsection*{Model diagram}

\begin{figure}[H]
    \centering
    \begin{tikzpicture}
        \begin{circuitikz}
            \ctikzset{bipoles/thickness=1} 

            \draw (-0.25, -0.0) -- (0, -0.0);
            \draw (0.6, -0.0) -- (0.6, 0.4);
            \draw (0.6, 0.4) -- (0.0, 0.0);
        
            \draw (0.85,0) to[damper] ++(0.85,0);

            \draw [-] (0.0, -0.0) -- (0.85, -0.0 );
            \draw [Latex-] (2.25, 0.0) -- (1.7, 0.0 );
        
            \draw[black, pattern=north west lines, opacity=1.0] (-0.45,0.47) rectangle ++(0.2,-1.0);
        \end{circuitikz}
    \node [] at (-1.95, -0.1) {\small $\mathbb{G},\beta$}; 
    \node [] at (-0.95, -0.1) {\small $\eta$}; 
    \end{tikzpicture}
\end{figure}

\subsubsection*{Constitutive equation}
\begin{equation}
    \begin{aligned}
    \sigma(t) + \frac{\eta}{\mathbb{G}} \frac{{\rm d}^{1-\beta}\sigma(t)}{{\rm d}t^{1-\beta}} &= \eta \frac{{\rm d}\gamma(t)}{{\rm d}t} \\
    \tau_c &= \left(\frac{\eta}{\mathbb{G}} \right)^{\frac{1}{1-\beta}}\\
    G_c &= \eta \tau_c^{-1}
    \end{aligned}
    \label{eq:constitutive_fml}
\end{equation}

\subsubsection*{Creep compliance}
\begin{equation}
    J(t) = \frac{t}{\eta} + \frac{1}{\mathbb{G}}\frac{t^{\beta}}{\Gamma(1+\beta)}
    \label{eq:creep_fml}
\end{equation}

\subsubsection*{Relaxation modulus}
\begin{equation}
    \begin{aligned}
    G(t) &= G_c \left(\frac{t}{\tau_c}\right)^{-\beta}E_{a,b}(z)\\
    a &= 1 - \beta \\
    b &= 1 - \beta \\
    z &= -\left(\frac{t}{\tau_c}\right)^{1 - \beta}
    \end{aligned}
    \label{eq:relaxation_fml}
\end{equation}

\subsubsection*{Storage modulus and Loss modulus}
\begin{equation}
    \begin{aligned}
    G^{\prime}(\omega) &= G_c \frac{(\omega \tau_c)^{2-\beta} \cos{(\frac{\pi}{2}\beta)}}{1 + (\omega \tau_c)^{2(1-\beta)} + 2(\omega \tau_c)^{1-\beta} \cos{(\frac{\pi}{2}(1-\beta))}} \\
    G^{\prime \prime}(\omega) &= G_c \frac{(\omega \tau_c) + (\omega \tau_c)^{2-\beta} \sin{(\frac{\pi}{2}\beta)}}{1 + (\omega \tau_c)^{2(1-\beta)} + 2(\omega \tau_c)^{1-\beta} \cos{(\frac{\pi}{2}(1-\beta))}}\\
    \end{aligned}
    \label{eq:oscillation_fml}
\end{equation}

\subsubsection*{Representative plots}
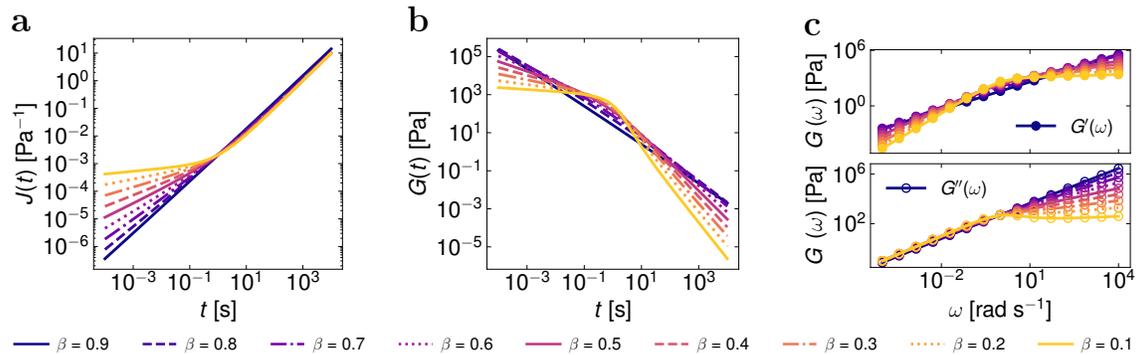
\begin{figure}[htbp]
    \centering
    \begin{subfigure}{0.3\textwidth}
        \centering
        \begin{picture}(0,0)
            \put(-67,0){\textbf{\Large a}} 
        \end{picture}
        \includegraphics[width=\linewidth]{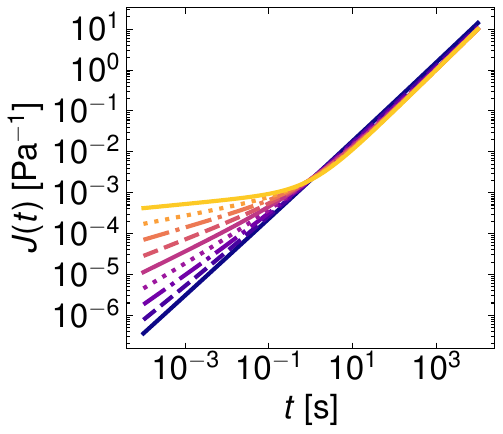} 
    \end{subfigure}
    \hfill
    \begin{subfigure}{0.3\textwidth}
        \centering
        \begin{picture}(0,0)
            \put(-67,0){\textbf{\Large b}} 
        \end{picture}
        \includegraphics[width=\linewidth]{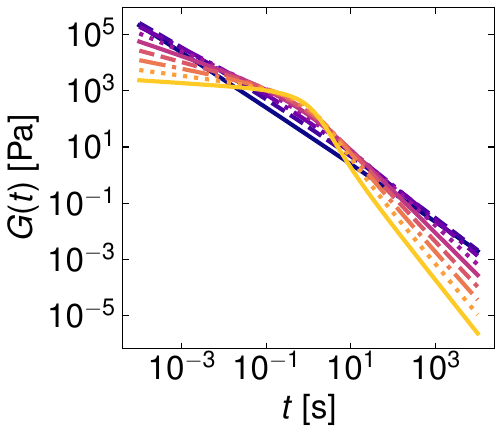} 
    \end{subfigure}
    \hfill
    \begin{subfigure}{0.3\textwidth}
        \centering
        \begin{picture}(0,0)
            \put(-67,0){\textbf{\Large c}} 
        \end{picture}
        \includegraphics[width=\linewidth]{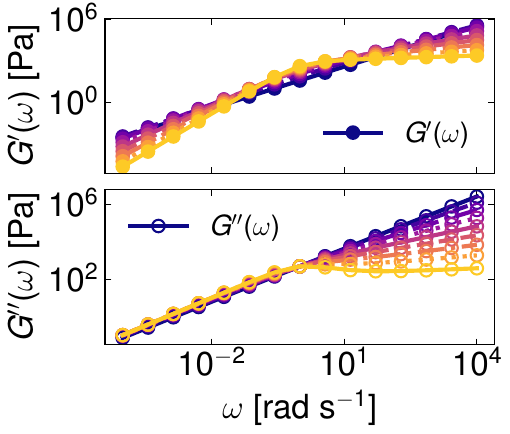} 
    \end{subfigure}
    \begin{subfigure}{1\textwidth}
        \centering
        \includegraphics[width=\linewidth]{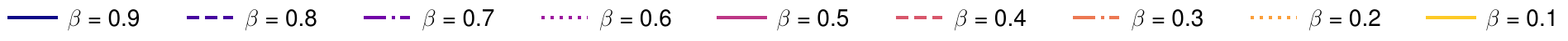} 
    \end{subfigure}
    \caption{Representative plots of the Fractional Maxwell Liquid model illustrating different viscoelastic behaviors under various types of stimulus. The plots were computed using fixed parameter values of $\mathbb{G} = 1000$~Pa~s$^{\beta}$ and $\eta = 1000$~Pa as well as varying $\beta$. \textbf{a)} Time-dependent creep compliance $J(t)$ showing the variation of strain over time under constant stress. \textbf{b)} Stress relaxation behavior depicting a decrease in the relaxation modulus $G(t)$ under constant strain over time. \textbf{c)} Oscillatory response demonstrating the storage modulus $G^{\prime}(\omega)$ and loss modulus $G^{\prime \prime}(\omega)$ under frequency-changing load.}
    \label{fig:fml_plots}
\end{figure}

\clearpage
\subsection*{Fractional Maxwell model (\texttt{FractionalMaxwell})}

\subsubsection*{Model diagram}

\begin{figure}[H]
    \centering
    \begin{tikzpicture}
        \begin{circuitikz}
            \ctikzset{bipoles/thickness=1} 

            \draw (-0.25, -0.0) -- (0, -0.0);
            \draw (0.6, -0.0) -- (0.6, 0.4);
            \draw (0.6, 0.4) -- (0.0, 0.0);
        
            \draw (0.85, 0.0) -- (1.45, 0.4);
            \draw (1.45, 0.4) -- (1.45, 0.0);
            \draw (0.6, 0.0) -- (2.1, 0.0);

            \draw [-] (0.0, -0.0) -- (0.85, -0.0 );
            \draw [Latex-] (2.1, 0.0) -- (1.7, 0.0 );
        
            \draw[black, pattern=north west lines, opacity=1.0] (-0.45,0.47) rectangle ++(0.2,-1.0);
        \end{circuitikz}
    \node [] at (-1.85, -0.1) {\small $\mathbb{G},\beta$}; 
    \node [] at (-0.85, -0.1) {\small $\mathbb{V},\alpha$}; 
    \end{tikzpicture}
\end{figure}

\subsubsection*{Constitutive equation}
\begin{equation}
    \begin{aligned}
    \sigma(t) + \frac{\mathbb{V}}{\mathbb{G}} \frac{{\rm d}^{\alpha-\beta}\sigma(t)}{{\rm d}t^{\alpha-\beta}} &= \mathbb{V} \frac{{\rm d}^{\alpha}\gamma(t)}{{\rm d}t^{\alpha}} \\
    \tau_c &= \left(\frac{\mathbb{V}}{\mathbb{G}} \right)^{\frac{1}{\alpha-\beta}}\\
    G_c &= \mathbb{V} \tau_c^{-\alpha}
    \end{aligned}
    \label{eq:constitutive_fmm}
\end{equation}

\subsubsection*{Creep compliance}
\begin{equation}
    J(t) = \frac{1}{\mathbb{V}}\frac{t^{\alpha}}{\Gamma(1+\alpha)} + \frac{1}{\mathbb{G}}\frac{t^{\beta}}{\Gamma(1+\beta)}
    \label{eq:creep_fmm}
\end{equation}

\subsubsection*{Relaxation modulus}
\begin{equation}
    \begin{aligned}
    G(t) &= G_c \left(\frac{t}{\tau_c}\right)^{-\beta}E_{a,b}(z)\\
    a &= \alpha - \beta \\
    b &= 1 - \beta \\
    z &= -\left(\frac{t}{\tau_c}\right)^{\alpha - \beta}
    \end{aligned}
    \label{eq:relaxation_fmm}
\end{equation}

\subsubsection*{Storage modulus and Loss modulus}
\begin{equation}
    \begin{aligned}
    G^{\prime}(\omega) &= G_c \frac{(\omega \tau_c)^{\alpha} \cos{(\frac{\pi}{2}\alpha)} + (\omega \tau_c)^{2\alpha-\beta} \cos{(\frac{\pi}{2}\beta)}}{1 + (\omega \tau_c)^{2(\alpha-\beta)} + 2(\omega \tau_c)^{\alpha-\beta} \cos{(\frac{\pi}{2}(\alpha-\beta))}} \\
    G^{\prime \prime}(\omega) &= G_c \frac{(\omega \tau_c)^{\alpha} \sin{(\frac{\pi}{2}\alpha)} + (\omega \tau_c)^{2\alpha-\beta} \sin{(\frac{\pi}{2}\beta)}}{1 + (\omega \tau_c)^{2(\alpha-\beta)} + 2(\omega \tau_c)^{\alpha-\beta} \cos{(\frac{\pi}{2}(\alpha-\beta))}}\\
    \end{aligned}
    \label{eq:oscillation_fmm}
\end{equation}

\subsubsection*{Representative plots}
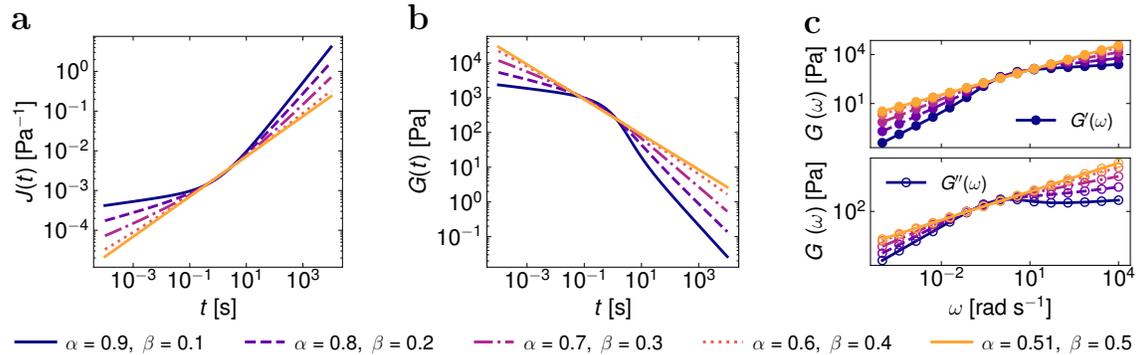
\begin{figure}[htbp]
    \centering
    \begin{subfigure}{0.3\textwidth}
        \centering
        \begin{picture}(0,0)
            \put(-67,0){\textbf{\Large a}} 
        \end{picture}
        \includegraphics[width=\linewidth]{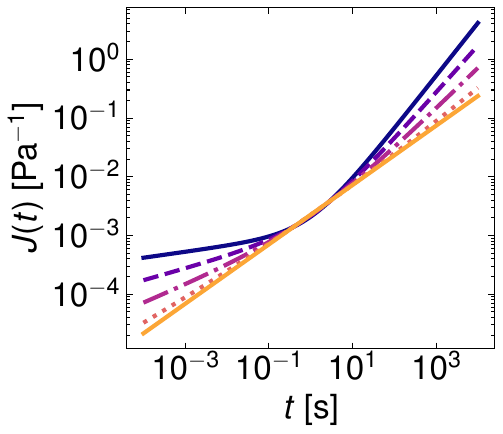} 
    \end{subfigure}
    \hfill
    \begin{subfigure}{0.3\textwidth}
        \centering
        \begin{picture}(0,0)
            \put(-67,0){\textbf{\Large b}} 
        \end{picture}
        \includegraphics[width=\linewidth]{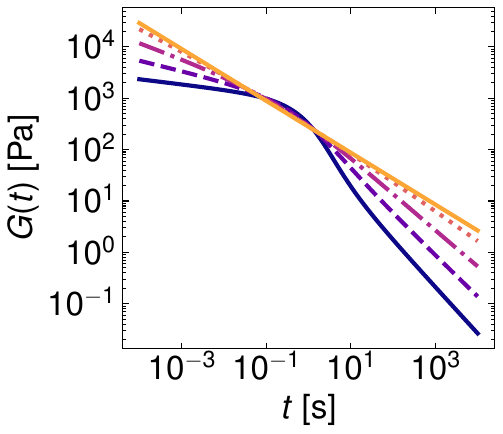} 
    \end{subfigure}
    \hfill
    \begin{subfigure}{0.3\textwidth}
        \centering
        \begin{picture}(0,0)
            \put(-67,0){\textbf{\Large c}} 
        \end{picture}
        \includegraphics[width=\linewidth]{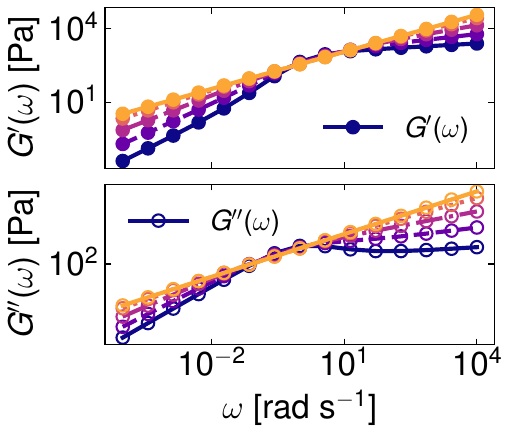} 
    \end{subfigure}
    \begin{subfigure}{1\textwidth}
        \centering
        \includegraphics[width=\linewidth]{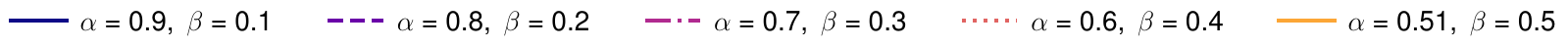} 
    \end{subfigure}
    \caption{Representative plots of the Fractional Maxwell model illustrating different viscoelastic behaviors under various types of stimulus. The plots were computed using fixed parameter values of $\mathbb{V} = 1000$~Pa~s$^{\alpha}$ and $\mathbb{G} = 1000$~Pa~s$^{\beta}$ as well as varying $\alpha$ and $\beta$. \textbf{a)} Time-dependent creep compliance $J(t)$ showing the variation of strain over time under constant stress. \textbf{b)} Stress relaxation $G(t)$ under constant strain over time. \textbf{c)} Oscillatory response demonstrating the storage modulus $G^{\prime}(\omega)$ and loss modulus $G^{\prime \prime}(\omega)$ under frequency-changing load.}
    \label{fig:fmm_plots}
\end{figure}

\clearpage
\subsection*{Fractional Kelvin-Voigt-S model (\texttt{FractionalKelvinVoigtS})}

\subsubsection*{Model diagram}

\begin{figure}[H]
    \centering
    \begin{tikzpicture}
        \begin{circuitikz}
            \ctikzsetstyle{romano}
            \draw (-0.25, -0.0) -- (0, -0.0);
            \draw (0.0, -0.5) -- (0.0, 0.5);
            \draw (2.25, -0.5) -- (2.25, 0.5);
            \draw (0, 0.5) -- (0.65, 0.5);
            \draw (1.55, 0.5) -- (2.25, 0.5);
            \draw (0, -0.5) -- (2.25, -0.5);
        
            \draw (0.65,0.5) to[spring] ++(1.,0.0);

            \draw (1.45, -0.5) -- (1.45, -0.1);
            \draw (1.45, -0.1) -- (0.85, -0.5);

            \draw [Latex-] (2.80, 0.0) -- (2.25, 0.0 );
        
            \draw[black, pattern=north west lines, opacity=1.0] (-0.45,0.47) rectangle ++(0.2,-1.0);
        \end{circuitikz}
    \node [] at (-1.65, 1.7) {\small $G$}; 
    \node [] at (-1.65, -.5) {\small $\mathbb{V},\alpha$}; 
    \end{tikzpicture}
\end{figure}

\subsubsection*{Constitutive equation}
\begin{equation}
    \begin{aligned}
    \sigma(t) &= \mathbb{V}\frac{{\rm d}^{\alpha}\gamma(t)}{{\rm d}t^{\alpha}} + G\gamma(t)\\
    \tau_c &= \left(\frac{\mathbb{V}}{G} \right)^{\frac{1}{\alpha}}\\
    G_c &= \mathbb{V} \tau_c^{-\alpha}
    \end{aligned}
    \label{eq:constitutive_fvks}
\end{equation}

\subsubsection*{Creep compliance}
\begin{equation}
    \begin{aligned}
    J(t) &= \frac{t^\alpha}{\mathbb{V}} E_{a,b}(z) \\
    a &= \alpha \\
    b &= 1 + \alpha \\
    z &= -\left(\frac{t}{\tau_c}\right)^{\alpha}
    \end{aligned}
    \label{eq:creep_fvks}
\end{equation}

\subsubsection*{Relaxation modulus}
\begin{equation}
    G(t) = \mathbb{V}\frac{t^{-\alpha}}{\Gamma(1-\alpha)} + G\\
    \label{eq:relaxation_fvks}
\end{equation}

\subsubsection*{Storage modulus and Loss modulus}
\begin{equation}
    \begin{aligned}
    G^{\prime}(\omega) &= \mathbb{V}\omega^{\alpha}\cos{(\frac{\pi}{2}\alpha)} + G \\
    G^{\prime \prime}(\omega) &= \mathbb{V}\omega^{\alpha}\sin{(\frac{\pi}{2}\alpha)}\\
    \end{aligned}
    \label{eq:oscillation_fvks}
\end{equation}

\subsubsection*{Representative plots}
\begin{figure}[htbp]
    \centering
    \begin{subfigure}{0.26\textwidth}
        \centering
        \includegraphics[width=\linewidth]{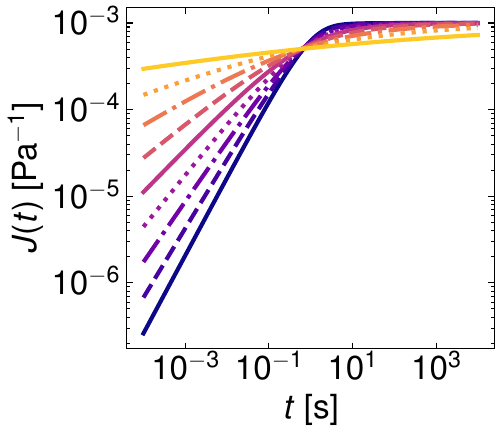} 
        \begin{picture}(0,0)
            \put(-67,120){\textbf{\Large a}} 
        \end{picture}
    \end{subfigure}
    \hfill
    \begin{subfigure}{0.26\textwidth}
        \centering
        \includegraphics[width=\linewidth]{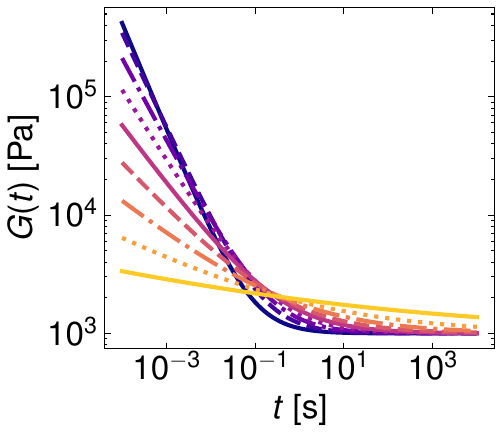} 
        \begin{picture}(0,0)
            \put(-67,120){\textbf{\Large b}} 
        \end{picture}
    \end{subfigure}
    \hfill
    \begin{subfigure}{0.26\textwidth}
        \centering
        \includegraphics[width=\linewidth]{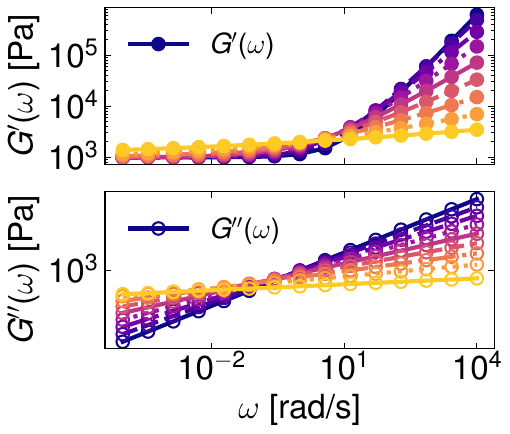} 
        \begin{picture}(0,0)
            \put(-80,120){\textbf{\Large c}} 
        \end{picture}
    \end{subfigure}
    \begin{subfigure}{1\textwidth}
        \centering
        \includegraphics[width=\linewidth]{legend_fmg.pdf} 
    \end{subfigure}
    \caption{Representative plots of the Fractional Kelvin-Voigt-S model. The plots were computed using $\mathbb{V} = 1000$~Pa~s$^{\alpha}$ and $G = 1000$~Pa, varying $\alpha$. \textbf{a)} Time-dependent creep compliance $J(t)$ under constant stress. \textbf{b)} Stress relaxation behavior depicting a decrease in the relaxation modulus $G(t)$ under constant strain over time. \textbf{c)} Oscillatory response demonstrating the storage modulus $G^{\prime}(\omega)$ and loss modulus $G^{\prime \prime}(\omega)$.}
    \label{fig:fkvs_plots}
\end{figure}

\clearpage
\subsection*{Fractional Kelvin-Voigt-D model (\texttt{FractionalKelvinVoigtD})}

\subsubsection*{Model diagram}

\begin{figure}[H]
    \centering
    \begin{tikzpicture}
        \begin{circuitikz}
            \ctikzsetstyle{romano}
            \draw (-0.25, -0.0) -- (0, -0.0);
            \draw (0.0, -0.5) -- (0.0, 0.5);
            \draw (2.25, -0.5) -- (2.25, 0.5);
            \draw (0, 0.5) -- (0.65, 0.5);
            \draw (1.55, 0.5) -- (2.25, 0.5);
            \draw (0, 0.5) -- (2.25, 0.5);
            \draw (0, -0.5) -- (1.25, -0.5);
            \draw (1.55, -0.5) -- (2.25, -0.5);

            \draw (0.65,-0.5) to[damper] ++(1.,0.0);

            \draw (1.45, 0.5) -- (1.45, 0.9);
            \draw (1.45, .9) -- (0.85, 0.5);

            \draw [Latex-] (2.80, 0.0) -- (2.25, 0.0 );
        
            \draw[black, pattern=north west lines, opacity=1.0] (-0.45,0.47) rectangle ++(0.2,-1.0);
        \end{circuitikz}
    \node [] at (-1.65, -0.4) {\small $\eta$}; 
    \node [] at (-1.65, 1.9) {\small $\mathbb{G},\beta$}; 
    \end{tikzpicture}
\end{figure}

\subsubsection*{Constitutive equation}
\begin{equation}
    \begin{aligned}
    \sigma(t) &= \eta\frac{{\rm d}\gamma(t)}{{\rm d}t} + \mathbb{G} \frac{{\rm d}^{\beta}\gamma(t)}{{\rm d}t^{\beta}} \\
    \tau_c &= \left(\frac{\eta}{\mathbb{G}} \right)^{\frac{1}{1-\beta}}\\
    G_c &= \eta \tau_c^{-1}
    \end{aligned}
    \label{eq:constitutive_fvkd}
\end{equation}

\subsubsection*{Creep compliance}
\begin{equation}
    \begin{aligned}
    J(t) &= \frac{t}{\eta}E_{a,b}(z)\\
    a &= 1 - \beta \\
    b &= 1 + 1 \\
    z &= -\left(\frac{t}{\tau_c}\right)^{1 - \beta}
    \end{aligned}
    \label{eq:creep_fvkd}
\end{equation}

\subsubsection*{Relaxation modulus}
\begin{equation}
    G(t) = \eta\Delta(t) + \mathbb{G}\frac{t^{-\beta}}{\Gamma(1-\beta)}\\
    \label{eq:relaxation_fvkd}
\end{equation}

\subsubsection*{Storage modulus and Loss modulus}
\begin{equation}
    \begin{aligned}
    G^{\prime}(\omega) &= \mathbb{G}\omega^{\beta}\cos{(\frac{\pi}{2}\beta)} \\
    G^{\prime \prime}(\omega) &= \eta\omega + \mathbb{G}\omega^{\beta}\sin{(\frac{\pi}{2}\beta)}\\
    \end{aligned}
    \label{eq:oscillation_fvkd}
\end{equation}

\subsubsection*{Representative plots}
\begin{figure}[htbp]
    \centering
    \begin{subfigure}{0.26\textwidth}
        \centering
        \includegraphics[width=\linewidth]{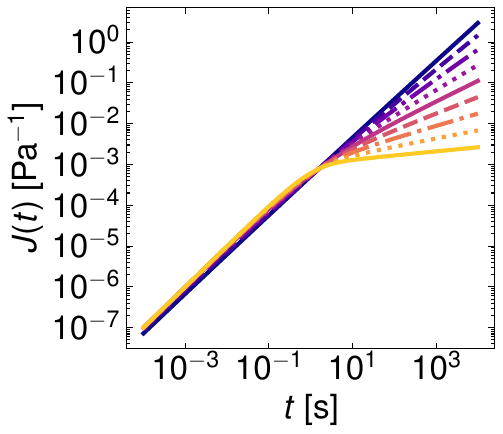} 
        \begin{picture}(0,0)
            \put(-67,120){\textbf{\Large a}} 
        \end{picture}
    \end{subfigure}
    \hfill
    \begin{subfigure}{0.26\textwidth}
        \centering
        \includegraphics[width=\linewidth]{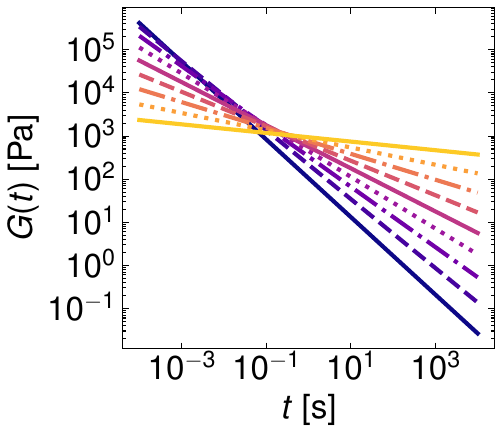} 
        \begin{picture}(0,0)
            \put(-67,120){\textbf{\Large b}} 
        \end{picture}
    \end{subfigure}
    \hfill
    \begin{subfigure}{0.26\textwidth}
        \centering
        \includegraphics[width=\linewidth]{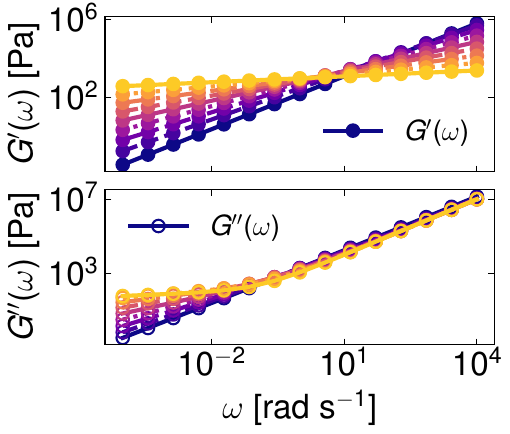} 
        \begin{picture}(0,0)
            \put(-80,120){\textbf{\Large c}} 
        \end{picture}
    \end{subfigure}
    \begin{subfigure}{1\textwidth}
        \centering
        \includegraphics[width=\linewidth]{legend_fmg.pdf} 
    \end{subfigure}
    \caption{Representative plots of the Fractional Kelvin-Voigt-D model. The plots were computed using $\mathbb{V} = 1000$~Pa~s$^{\alpha}$ and $\mathbb{G} = 1000$~Pa~s$^{\beta}$, varying $\beta$. \textbf{a)} Time-dependent creep compliance $J(t)$ under constant stress. \textbf{b)} Stress relaxation behavior depicting a decrease in the relaxation modulus $G(t)$ under constant strain over time. \textbf{c)} Oscillatory response demonstrating the storage modulus $G^{\prime}(\omega)$ and loss modulus $G^{\prime \prime}(\omega)$.}
    \label{fig:fkvd_plots}
\end{figure}

\clearpage
\subsection*{Fractional Kelvin-Voigt model (\texttt{FractionalKelvinVoigt})}

\subsubsection*{Model diagram}

\begin{figure}[H]
    \centering
    \begin{tikzpicture}
        \begin{circuitikz}
            \ctikzsetstyle{romano}
            \draw (-0.25, -0.0) -- (0, -0.0);
            \draw (0.0, -0.5) -- (0.0, 0.5);
            \draw (2.25, -0.5) -- (2.25, 0.5);
            \draw (1.25, 0.5) -- (2.25, 0.5);
            \draw (0, 0.5) -- (2.1, 0.5);
            \draw (0, -0.5) -- (2.25, -0.5);
        
            \draw (1.45, 0.5) -- (1.45, 0.9);
            \draw (1.45, .9) -- (0.85, 0.5);

            \draw (1.45, -0.5) -- (1.45, -0.1);
            \draw (1.45, -0.1) -- (0.85, -0.5);

            \draw [Latex-] (2.80, 0.0) -- (2.25, 0.0 );
        
            \draw[black, pattern=north west lines, opacity=1.0] (-0.45,0.47) rectangle ++(0.2,-1.0);
        \end{circuitikz}
    \node [] at (-1.85, 1.7) {\small $\mathbb{G},\beta$}; 
    \node [] at (-1.85, -.5) {\small $\mathbb{V},\alpha$}; 
    \end{tikzpicture}
\end{figure}

\subsubsection*{Constitutive equation}
\begin{equation}
    \begin{aligned}
    \sigma(t) &= \mathbb{V}\frac{{\rm d}^{\alpha}\gamma(t)}{{\rm d}t^{\alpha}} + \mathbb{G} \frac{{\rm d}^{\beta}\gamma(t)}{{\rm d}t^{\beta}} \\
    \tau_c &= \left(\frac{\mathbb{V}}{\mathbb{G}} \right)^{\frac{1}{\alpha-\beta}}\\
    G_c &= \mathbb{V} \tau_c^{-\alpha}
    \end{aligned}
    \label{eq:constitutive_fvk}
\end{equation}

\subsubsection*{Creep compliance}
\begin{equation}
    \begin{aligned}
    J(t) &= \frac{t^{\alpha}}{\mathbb{V}}E_{a,b}(z)\\
    a &= \alpha - \beta \\
    b &= 1 + \alpha \\
    z &= -\left(\frac{t}{\tau_c}\right)^{\alpha - \beta}
    \end{aligned}
    \label{eq:creep_fvk}
\end{equation}

\subsubsection*{Relaxation modulus}
\begin{equation}
    G(t) = \mathbb{V}\frac{t^{-\alpha}}{\Gamma(1-\alpha)} + \mathbb{G}\frac{t^{-\beta}}{\Gamma(1-\beta)}\\
    \label{eq:relaxation_fvk}
\end{equation}

\subsubsection*{Storage modulus and Loss modulus}
\begin{equation}
    \begin{aligned}
    G^{\prime}(\omega) &= \mathbb{V}\omega^{\alpha}\cos{(\frac{\pi}{2}\alpha)} + \mathbb{G}\omega^{\beta}\cos{(\frac{\pi}{2}\beta)} \\
    G^{\prime \prime}(\omega) &= \mathbb{V}\omega^{\alpha}\sin{(\frac{\pi}{2}\alpha)} + \mathbb{G}\omega^{\beta}\sin{(\frac{\pi}{2}\beta)}\\
    \end{aligned}
    \label{eq:oscillation_fvk}
\end{equation}

\subsubsection*{Representative plots}
\begin{figure}[htbp]
    \centering
    \begin{subfigure}{0.26\textwidth}
        \centering
        \includegraphics[width=\linewidth]{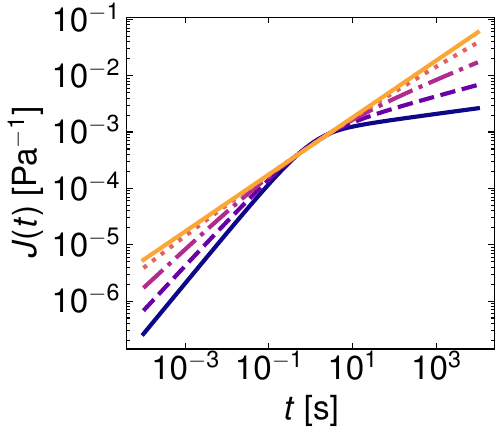} 
        \begin{picture}(0,0)
            \put(-67,120){\textbf{\Large a}} 
        \end{picture}
    \end{subfigure}
    \hfill
    \begin{subfigure}{0.26\textwidth}
        \centering
        \includegraphics[width=\linewidth]{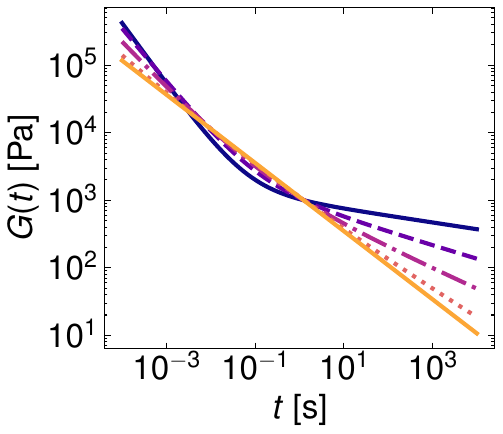} 
        \begin{picture}(0,0)
            \put(-67,120){\textbf{\Large b}} 
        \end{picture}
    \end{subfigure}
    \hfill
    \begin{subfigure}{0.26\textwidth}
        \centering
        \includegraphics[width=\linewidth]{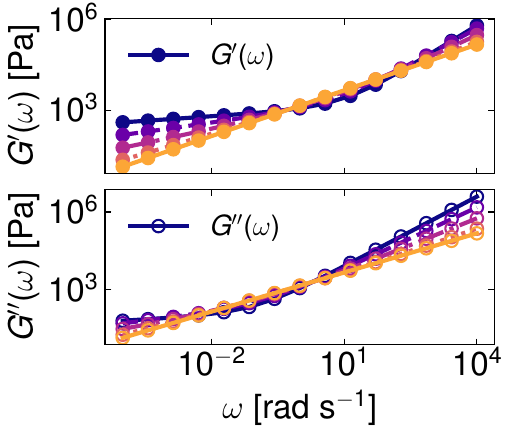} 
        \begin{picture}(0,0)
            \put(-80,120){\textbf{\Large c}} 
        \end{picture}
    \end{subfigure}
    \begin{subfigure}{1\textwidth}
        \centering
        \includegraphics[width=\linewidth]{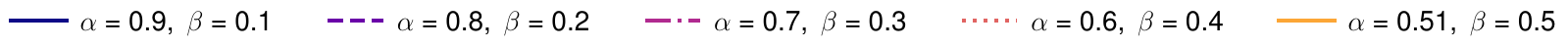} 
    \end{subfigure}
    \caption{Representative plots of the Fractional Kelvin-Voigt model. The plots were computed using $\mathbb{V} = 1000$~Pa~s$^{\alpha}$ and $\mathbb{G} = 1000$~Pa~s$^{\beta}$, varying $\alpha$ and $\beta$. \textbf{a)} Time-dependent creep compliance $J(t)$ under constant stress. \textbf{b)} Stress relaxation behavior depicting a decrease in the relaxation modulus $G(t)$ under constant strain over time. \textbf{c)} Oscillatory response demonstrating the storage modulus $G^{\prime}(\omega)$ and loss modulus $G^{\prime \prime}(\omega)$.}
    \label{fig:fkv_plots}
\end{figure}

\clearpage
\subsection*{Zener model (\texttt{Zener})}

\subsubsection*{Model diagram}

\begin{figure}[H]
    \centering
    \begin{tikzpicture}
        \begin{circuitikz}
            \ctikzsetstyle{romano}
            \draw (-0.25, -0.0) -- (0, -0.0);
            \draw (0.0, -0.5) -- (0.0, 0.5);
            \draw (2.25, -0.5) -- (2.25, 0.5);
            \draw (0, 0.5) -- (0.1, 0.5);
    
            \draw (1.1, 0.5) to[damper] ++(1.15,0);

            \draw (0.1, 0.5) to[spring] ++(1.25,0);
            \draw (0.0, -0.5) to[spring] ++(2.25,0);

            \draw [Latex-] (2.80, 0.0) -- (2.25, 0.0 );
        
            \draw[black, pattern=north west lines, opacity=1.0] (-0.45,0.47) rectangle ++(0.2,-1.0);
        \end{circuitikz}
    \node [] at (-2.05, 2) {\small $G_{1}$}; 
    \node [] at (-1.1, 2) {\small $\eta$}; 
    \node [] at (-1.6, -0.3) {\small $G_{0}$}; 
    \end{tikzpicture}
\end{figure}

\subsubsection*{Constitutive equation}
\begin{equation}
    \begin{aligned}
    \sigma(t) + \frac{\eta}{G_1} \frac{{\rm d}\sigma(t)}{{\rm d}t} &= G_0\gamma{(t)} + \eta \frac{G_0 + G_1}{G_1} \frac{{\rm d}\gamma(t)}{{\rm d}t} \\
    \tau_c &= \frac{\eta}{G_1}\\
    G_{1} &= G_{\infty} - G_0
    \end{aligned}
    \label{eq:constitutive_zener}
\end{equation}

\subsubsection*{Creep compliance}
\begin{equation}
    J(t) =  \frac{1}{G_0} - \frac{G_1}{G_0(G_0 + G_1)} \exp\left(-\frac{G_0}{G_0 + G_1}\frac{t}{\tau_c}\right)
    \label{eq:creep_zener}
\end{equation}

\subsubsection*{Relaxation modulus}
\begin{equation}
    \begin{aligned}
    G(t) &= G_0 + G_1\exp{\left(-\frac{t}{\tau_c} \right)}
    \end{aligned}
    \label{eq:relaxation_zener}
\end{equation}

\subsubsection*{Storage modulus and Loss modulus}
\begin{equation}
    \begin{aligned}
    G^{\prime}(\omega) &= G_0 + G_1\frac{(\omega\tau_{c})^{2}}{1 + (\omega\tau_{c})^2}\\
    G^{\prime \prime}(\omega) &= G_1\frac{\omega\tau_{c}}{1 + (\omega\tau_{c})^2}\\
    \end{aligned}
    \label{eq:oscillation_zener}
\end{equation}

\subsubsection*{Representative plots}
\begin{figure}[htbp]
    \centering
    \begin{subfigure}{0.3\textwidth}
        \centering
        \begin{picture}(0,0)
            \put(-67,0){\textbf{\Large a}} 
        \end{picture}
        \includegraphics[width=\linewidth]{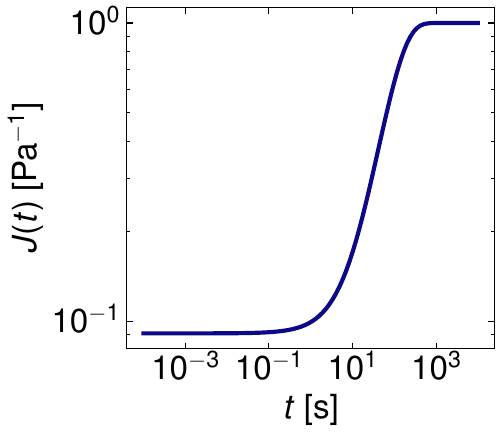} 
    \end{subfigure}
    \hfill
    \begin{subfigure}{0.3\textwidth}
        \centering
        \begin{picture}(0,0)
            \put(-67,0){\textbf{\Large b}} 
        \end{picture}
        \includegraphics[width=\linewidth]{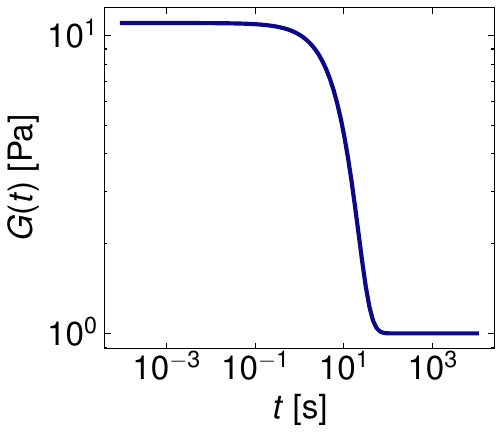} 
    \end{subfigure}
    \hfill
    \begin{subfigure}{0.3\textwidth}
        \centering
        \begin{picture}(0,0)
            \put(-67,0){\textbf{\Large c}} 
        \end{picture}
        \includegraphics[width=\linewidth]{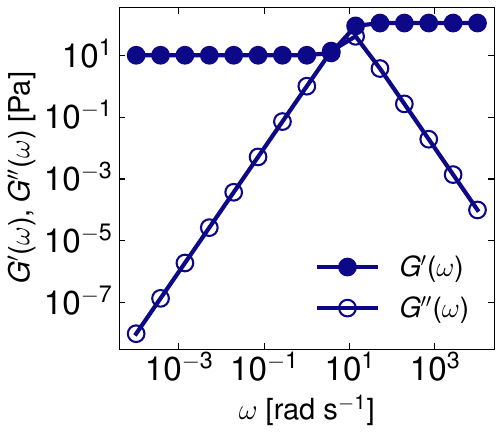} 
    \end{subfigure}
    \caption{Representative plots of the Zener model illustrating different viscoelastic behaviors under various types of stimulus. The plots were computed using fixed parameter values of $G_0 = 10$~Pa,  $G_1 = 10$~Pa, and $\eta = 100$~Pa~s. \textbf{a)} Time-dependent creep compliance $J(t)$ showing the variation of strain over time under constant stress. \textbf{b)} Stress relaxation behavior depicting a decrease in the relaxation modulus $G(t)$ under constant strain over time. \textbf{c)} Oscillatory response demonstrating the storage modulus $G^{\prime}(\omega)$ and loss modulus $G^{\prime \prime}(\omega)$ under frequency-changing load.}
    \label{fig:z_plots}
\end{figure}
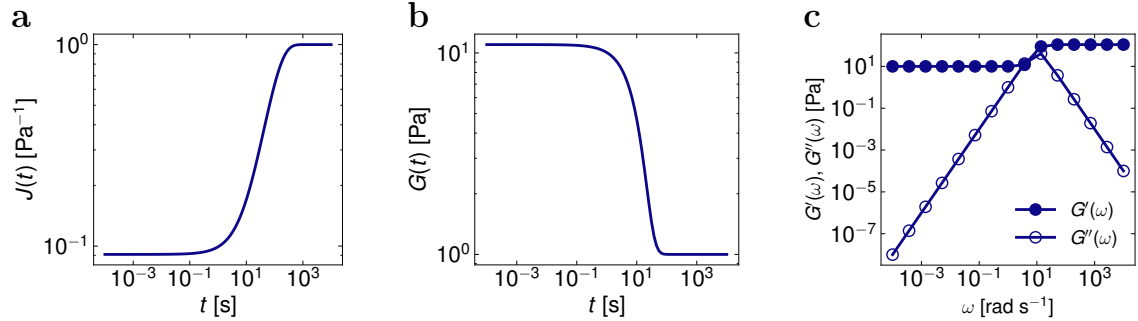

\clearpage
\subsection*{Fractional Zener Solid-S model (\texttt{FractionalZenerSolidS})}

\subsubsection*{Model diagram}

\begin{figure}[H]
    \centering
    \begin{tikzpicture}
        \begin{circuitikz}
            \ctikzsetstyle{romano}
            \draw (-0.25, -0.0) -- (0, -0.0);
            \draw (0.0, -0.5) -- (0.0, 0.5);
            \draw (2.25, -0.5) -- (2.25, 0.5);
            \draw (1.25, 0.5) -- (2.25, 0.5);
            \draw (0, 0.5) -- (0.25, 0.5);

            \draw (0.1, 0.5) to[spring] ++(1.25,0);

            \draw (2.0, 0.5) -- (2.0, 0.9);
            \draw (2.0, .9) -- (1.4, 0.5);            
            \draw (0.0, -0.5) to[spring] ++(2.25,0);

            \draw [Latex-] (2.80, 0.0) -- (2.25, 0.0 );
        
            \draw[black, pattern=north west lines, opacity=1.0] (-0.45,0.47) rectangle ++(0.2,-1.0);
        \end{circuitikz}
    \node [] at (-1.1, 2)(-2.05, 2) {\small $\mathbb{V}, \alpha$}; 
    \node [] at (-2.05, 2) {\small $G_{1}$}; 
    \node [] at (-1.6, -0.15) {\small $G_{0}$}; 
    \end{tikzpicture}
\end{figure}

\subsubsection*{Constitutive equation}
\begin{equation}
    \begin{aligned}
    \sigma(t) + \frac{\mathbb{V}}{G_1} \frac{{\rm d}^{\alpha} \sigma(t)}{{\rm d}t^{\alpha}} &= G_0 \gamma(t) +
    \mathbb{V} \frac{{\rm d}^{\alpha} \gamma(t)}{{\rm d} t^{\alpha}} +
    G_0 \frac{\mathbb{V}}{G_1} \frac{{\rm d}^{\alpha} \gamma(t)}{{\rm d} t^{\alpha}} \\
    \tau_c &= \left(\frac{\mathbb{V}}{G_1} \right)^{\frac{1}{\alpha}} \\
    G_c &= \mathbb{V} \tau_c^{-\alpha} \\
    G_{1} &= G_{\infty} - G_0.
    \end{aligned}
    \label{eq:constitutive_fzenerss}
\end{equation}

\subsubsection*{Creep compliance}
\begin{equation}
    \begin{aligned}
    J(t) = \frac{1}{G_0 + G_1} + \frac{G_1}{G_0 (G_0 + G_1)} & \left\{ 1 - E_{a,b}\left[\left(\frac{G_0}{G_0 + G_1}\right)z\right] \right\} \\
    a &= \alpha, \quad b = 1 \\
    z &= -\left(\frac{t}{\tau_c}\right)^{\alpha}
    \end{aligned}
    \label{eq:creep_fzss}
\end{equation}

\subsubsection*{Relaxation modulus}
\begin{equation}
    \begin{aligned}
    G(t) &= G_0 + G_1 E_{a,b}(z)\\
    \end{aligned}
    \label{eq:relaxation_fzenerss}
\end{equation}

\subsubsection*{Storage modulus and Loss modulus}
\begin{equation}
    \begin{aligned}
    G^{\prime}(\omega) &= G_{0} + \frac{\left( G_{1} \right)^2 \mathbb{V} \omega^\alpha \cos \left( \frac{\pi}{2} \alpha \right) + \left( \mathbb{V} \omega^\alpha \right)^2 G_{1}}{\left( \mathbb{V} \omega^\alpha \right)^2 + \left( G_{1}\right)^2 + 2 \mathbb{V} \omega^\alpha G_{1}  \cos \left( \frac{\pi}{2} \alpha\right)} \\
    G^{\prime \prime}(\omega) &= \frac{\left( G_{1} \right)^2 \mathbb{V} \omega^\alpha \sin \left(\frac{\pi}{2} \alpha \right)}{\left( \mathbb{V} \omega^\alpha \right)^2 + \left( G_{1} \right)^2 + 2 \mathbb{V} \omega^\alpha G_{1} \cos \left(\frac{\pi}{2} \alpha \right)} \\
    \end{aligned}
    \label{eq:oscillation_fzenerss}
\end{equation}

\subsubsection*{Representative plots}
\begin{figure}[htbp]
    \centering
    \begin{subfigure}{0.28\textwidth}
        \centering
        \begin{picture}(0,0)
            \put(-67,0){\textbf{\Large a}} 
        \end{picture}
        \includegraphics[width=\linewidth]{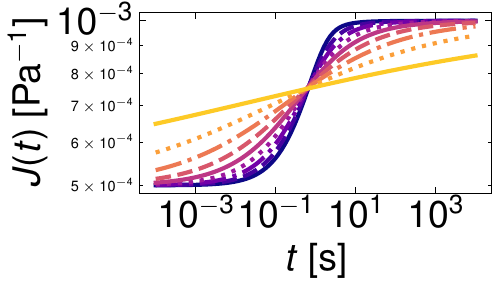} 
    \end{subfigure}
    \hfill
    \begin{subfigure}{0.26\textwidth}
        \centering
        \begin{picture}(0,0)
            \put(-67,0){\textbf{\Large b}} 
        \end{picture}
        \includegraphics[width=\linewidth]{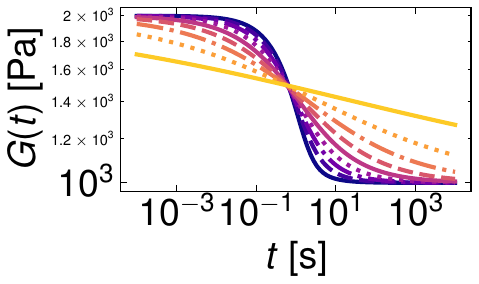} 
    \end{subfigure}
    \hfill
    \begin{subfigure}{0.40\textwidth}
        \centering
        \begin{picture}(0,0)
            \put(-100,0){\textbf{\Large c}} 
        \end{picture}
        \includegraphics[width=\linewidth]{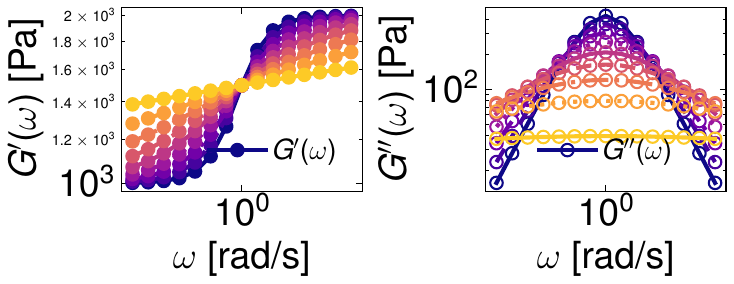} 
    \end{subfigure}
    \begin{subfigure}{1\textwidth}
        \centering
        \includegraphics[width=\linewidth]{legend_fmg.pdf} 
    \end{subfigure}
    \caption{Representative plots of the Fractional Zener Solid-S model illustrating different viscoelastic behaviors under various types of stimulus. The plots were computed using fixed parameter values of $G_0 = 1000$~Pa,  $\mathbb{V} = 1000$~Pa~s$^{\alpha}$, and $G_1 = 1000$~Pa. \textbf{a)} Time-dependent creep compliance $J(t)$ showing the variation of strain over time under constant stress. \textbf{b)} Stress relaxation behavior depicting a decrease in the relaxation modulus $G(t)$ under constant strain over time. \textbf{c)} Oscillatory response demonstrating the storage modulus $G^{\prime}(\omega)$ and loss modulus $G^{\prime \prime}(\omega)$ under frequency-changing load.}
    \label{fig:fzss_plots}
\end{figure}

\clearpage

\subsection*{Fractional Zener Liquid-S model (\texttt{FractionalZenerLiquidS})}

\subsubsection*{Model diagram}

\begin{figure}[H]
    \centering
    \begin{tikzpicture}
        \begin{circuitikz}
            \ctikzsetstyle{romano}
            \draw (-0.25, -0.0) -- (0, -0.0);
            \draw (0.0, -0.5) -- (0.0, 0.5);
            \draw (2.25, -0.5) -- (2.25, 0.5);
            \draw (0, 0.5) -- (1.25, 0.5);
    
            \draw (1.1, 0.5) to[damper] ++(1.15,0);

            \draw (1.0, 0.5) -- (1.0, 0.9);
            \draw (1.0, .9) -- (0.4, 0.5);            
            \draw (0.0, -0.5) to[spring] ++(2.25,0);

            \draw [Latex-] (2.80, 0.0) -- (2.25, 0.0 );
        
            \draw[black, pattern=north west lines, opacity=1.0] (-0.45,0.47) rectangle ++(0.2,-1.0);
        \end{circuitikz}
    \node [] at (-2.05, 2) {\small $\mathbb{G}, \beta$}; 
    \node [] at (-1.1, 2) {\small $\eta$}; 
    \node [] at (-1.6, -0.15) {\small $G_{0}$}; 
    \end{tikzpicture}
\end{figure}

\subsubsection*{Constitutive equation}
\begin{equation}
    \begin{aligned}
    \sigma(t) + \frac{\eta}{\mathbb{G}} \frac{{\rm d} ^{1 - \beta} \sigma(t)}{{\rm d}t^{1 - \beta}} &= G_0\gamma{(t)} +
    \eta \frac{{\rm d} \gamma(t)}{{\rm d} t} +
    G_0 \frac{\eta}{\mathbb{G}} \frac{{\rm d}^{1-\beta} \gamma(t)}{{\rm d} t^{1-\beta}}\\
    \tau_c &= \left(\frac{\eta}{\mathbb{G}} \right)^{\frac{1}{1-\beta}}\\
    G_c &= \eta \tau_c^{-1}
    \end{aligned}
    \label{eq:constitutive_fzenerls}
\end{equation}

\subsubsection*{Creep compliance (Laplace form)}
\begin{equation}
    \hat{J}(s) = \frac{1}{s^2} \frac{1 + \frac{\eta}{\mathbb{G}} s^{1-\beta}}{\eta + \frac{G_0}{s}+\frac{G_0 \eta}{\mathbb{G}}s^{-\beta}}
    \label{eq:creep_fzls}
\end{equation}

\subsubsection*{Relaxation modulus}
\begin{equation}
    \begin{aligned}
    G(t) &= G_0 + \mathbb{G}t^{-\beta}E_{a,b}(z)\\
    &a = 1 - \beta, \quad b = 1 - \beta \\
    z &= -\left(\frac{t}{\tau_c}\right)^{1 - \beta}
    \end{aligned}
    \label{eq:relaxation_fzenerls}
\end{equation}

\subsubsection*{Storage modulus and Loss modulus}
\begin{equation}
    \begin{aligned}
    G^{\prime}(\omega) &= G_0 + \frac{\left( \eta \omega \right)^2 \mathbb{G} \omega^\beta \cos \left(\frac{\pi}{2}\beta \right)}{\left( \eta \omega \right)^2 + \left( \mathbb{G} \omega^\beta \right)^2 + \left(2 \eta \omega \right) \mathbb{G} \omega^\beta \cos \left(\frac{\pi}{2} (1 - \beta) \right)} \\
    G^{\prime \prime}(\omega) &= \frac{\left( \mathbb{G} \omega^\beta \right)^2 \left( \eta \omega \right) + \left( \eta \omega \right)^2 \mathbb{G} \omega^\beta \sin \left(\frac{\pi}{2} \beta  \right)}{\left( \eta \omega \right)^2 + \left( \mathbb{G} \omega^\beta \right)^2 + \left(2 \eta \omega \right) \mathbb{G} \omega^\beta \cos \left(\frac{\pi}{2} (1 - \beta) \right)}\\
    \end{aligned}
    \label{eq:oscillation_fzenerls}
\end{equation}

\subsubsection*{Representative plots}
\begin{figure}[htbp]
    \centering
    \begin{subfigure}{0.26\textwidth}
        \centering
        \begin{picture}(0,0)
            \put(-67,0){\textbf{\Large a}} 
        \end{picture}
        \includegraphics[width=\linewidth]{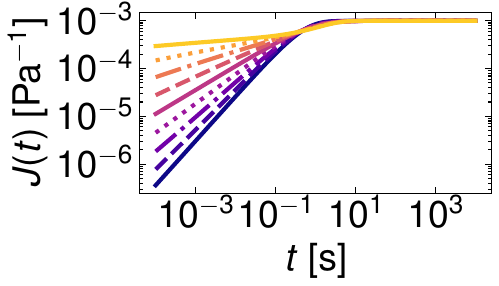} 
    \end{subfigure}
    \hfill
    \begin{subfigure}{0.24\textwidth}
        \centering
        \begin{picture}(0,0)
            \put(-67,0){\textbf{\Large b}} 
        \end{picture}
        \includegraphics[width=\linewidth]{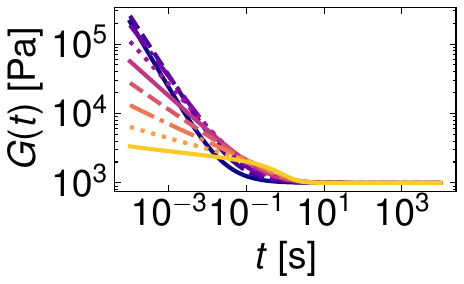} 
    \end{subfigure}
    \hfill
    \begin{subfigure}{0.38\textwidth}
        \centering
        \begin{picture}(0,0)
            \put(-110,0){\textbf{\Large c}} 
        \end{picture}
        \includegraphics[width=\linewidth]{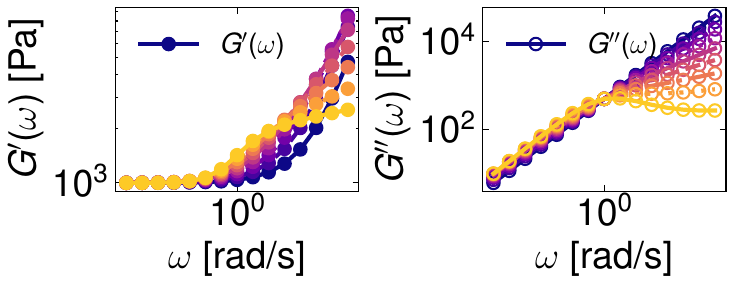} 
    \end{subfigure}
    \begin{subfigure}{1\textwidth}
        \centering
        \includegraphics[width=\linewidth]{legend_fmg.pdf} 
    \end{subfigure}
    \caption{Representative plots of the Fractional Zener Liquid-S model illustrating different viscoelastic behaviors under various types of stimulus. The plots were computed using fixed parameter values of $G_0 = 1000$~Pa,  $\mathbb{G} = 1000$~Pa~s$^{\beta}$, and $\eta = 1000$~Pa~s. \textbf{a)} Time-dependent creep compliance $J(t)$ showing the variation of strain over time under constant stress. \textbf{b)} Stress relaxation behavior depicting a decrease in the relaxation modulus $G(t)$ under constant strain over time. \textbf{c)} Oscillatory response demonstrating the storage modulus $G^{\prime}(\omega)$ and loss modulus $G^{\prime \prime}(\omega)$ under frequency-changing load.}
    \label{fig:fzls_plots}
\end{figure}

\clearpage

\subsection*{Fractional Zener Liquid-D model (\texttt{FractionalZenerLiquidD})}

\subsubsection*{Model diagram}

\begin{figure}[H]
    \centering
    \begin{tikzpicture}
        \begin{circuitikz}
            \ctikzsetstyle{romano}
            \draw (-0.25, -0.0) -- (0, -0.0);
            \draw (0.0, -0.5) -- (0.0, 0.5);
            \draw (2.25, -0.5) -- (2.25, 0.5);
            \draw (0, 0.5) -- (1.25, 0.5);
    
            \draw (1.1, 0.5) to[damper] ++(1.15,0);

            \draw (1.0, 0.5) -- (1.0, 0.9);
            \draw (1.0, .9) -- (0.4, 0.5);            
            \draw (0.0, -0.5) to[damper] ++(2.25,0);

            \draw [Latex-] (2.80, 0.0) -- (2.25, 0.0 );
        
            \draw[black, pattern=north west lines, opacity=1.0] (-0.45,0.47) rectangle ++(0.2,-1.0);
        \end{circuitikz}
    \node [] at (-2.05, 2) {\small $\mathbb{G}, \beta$}; 
    \node [] at (-1.1, 2) {\small $\eta_s$}; 
    \node [] at (-1.6, -0.3) {\small $\eta_p$}; 
    \end{tikzpicture}
\end{figure}

\subsubsection*{Constitutive equation}
\begin{equation}
    \begin{aligned}
    \sigma(t) + \frac{\eta_s}{\mathbb{G}} \frac{{\rm d}^{1-\beta}}{{\rm d}t^{1-\beta}} \sigma(t) &= (\eta_s + \eta_p) \frac{{\rm d}\gamma(t)}{{\rm d}t} + \frac{\eta_s \eta_p}{\mathbb{G}} \frac{{\rm d}^{2-\beta} \gamma(t)}{{\rm d}t^{2-\beta}}\\
    \tau_c &= \left(\frac{\eta_s}{\mathbb{G}} \right)^{\frac{1}{1-\beta}}\\
    G_c &= \eta_s \tau_c^{-1}
    \end{aligned}
    \label{eq:constitutive_fzenerld}
\end{equation}

\subsubsection*{Creep compliance (Laplace form)}
\begin{equation}
    \hat{J}(s) = \frac{1}{s}\frac{ \left( \eta_s s + \mathbb{G} s^\beta \right)}{\eta_s s \mathbb{G} s^\beta + \eta_p s \left( \eta_s s + \mathbb{G} s^\beta \right)}
    \label{eq:creep_fzld}
\end{equation}

\subsubsection*{Relaxation modulus}
\begin{equation}
    \begin{aligned}
    G(t) &= \eta_p \Delta(t) + \mathbb{G} t^{-\beta} E_{a, b} \left( z \right)\\
    &a = 1 - \beta, \quad b = 1 - \beta \\
    z &= -\left(\frac{t}{\tau_c}\right)^{1 - \beta}
    \end{aligned}
    \label{eq:relaxation_fzenerld}
\end{equation}

\subsubsection*{Storage modulus and Loss modulus}
\begin{equation}
    \begin{aligned}
    G^{\prime}(\omega) &= \frac{\eta_s^2 \mathbb{G} \omega^{\beta+2} \cos( \frac{\pi}{2} \beta)}{(\eta_s \omega)^2 + (\mathbb{G} \omega^\beta)^2 + 2 \eta_s \mathbb{G} \omega^{\beta+1} \sin( \frac{\pi}{2} \beta)}
 \\
    G^{\prime \prime}(\omega) &= \eta_p \omega + \frac{\mathbb{G}^2 \eta_s \omega^{2\beta+1} + \eta_s^2 \mathbb{G} \omega^{\beta+2} \sin( \frac{\pi}{2} \beta)}{(\eta_s \omega)^2 + (\mathbb{G} \omega^\beta)^2 + 2 \eta_s \mathbb{G} \omega^{\beta+1} \sin( \frac{\pi}{2} \beta)}\\
    \end{aligned}
    \label{eq:oscillation_fzenerld}
\end{equation}

\subsubsection*{Representative plots}
\begin{figure}[htbp]
    \centering
    \begin{subfigure}{0.26\textwidth}
        \centering
        \begin{picture}(0,0)
            \put(-67,0){\textbf{\Large a}} 
        \end{picture}
        \includegraphics[width=\linewidth]{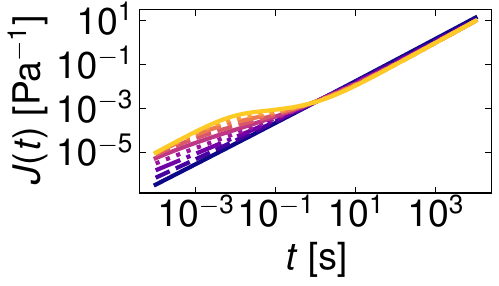} 
    \end{subfigure}
    \hfill
    \begin{subfigure}{0.24\textwidth}
        \centering
        \begin{picture}(0,0)
            \put(-67,0){\textbf{\Large b}} 
        \end{picture}
        \includegraphics[width=\linewidth]{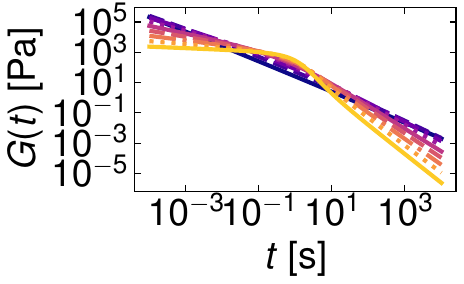} 
    \end{subfigure}
    \hfill
    \begin{subfigure}{0.38\textwidth}
        \centering
        \begin{picture}(0,0)
            \put(-110,0){\textbf{\Large c}} 
        \end{picture}
        \includegraphics[width=\linewidth]{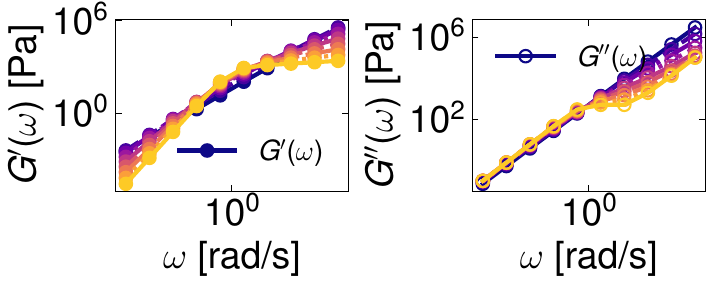} 
    \end{subfigure}
    \begin{subfigure}{1\textwidth}
        \centering
        \includegraphics[width=\linewidth]{legend_fmg.pdf} 
    \end{subfigure}
    \caption{Representative plots of the Fractional Zener Liquid-D model illustrating different viscoelastic behaviors under various types of stimulus. The plots were computed using fixed parameter values of $\eta_s = 1000$~Pa~s,  $\mathbb{G} = 1000$~Pa~s$^{\beta}$, and $\eta_p = 10$~Pa~s. \textbf{a)} Time-dependent creep compliance $J(t)$ showing the variation of strain over time under constant stress. \textbf{b)} Stress relaxation behavior depicting a decrease in the relaxation modulus $G(t)$ under constant strain over time. \textbf{c)} Oscillatory response demonstrating the storage modulus $G^{\prime}(\omega)$ and loss modulus $G^{\prime \prime}(\omega)$ under frequency-changing load.}
    \label{fig:fzld_plots}
\end{figure}

\clearpage
\subsection*{Fractional Zener-S model (\texttt{FractionalZenerS})}

\subsubsection*{Model diagram}

\begin{figure}[H]
    \centering
    \begin{tikzpicture}
        \begin{circuitikz}
            \ctikzsetstyle{romano}
            \draw (-0.25, -0.0) -- (0, -0.0);
            \draw (0.0, -0.5) -- (0.0, 0.5);
            \draw (2.25, -0.5) -- (2.25, 0.5);
            \draw (0, 0.5) -- (2.25, 0.5);
    
            \draw (1.95, 0.5) -- (1.95, 0.9);
            \draw (1.95, .9) -- (1.35, 0.5);  

            \draw (1.0, 0.5) -- (1.0, 0.9);
            \draw (1.0, .9) -- (0.4, 0.5);            
            \draw (0.0, -0.5) to[spring] ++(2.25,0);

            \draw [Latex-] (2.80, 0.0) -- (2.25, 0.0 );
        
            \draw[black, pattern=north west lines, opacity=1.0] (-0.45,0.47) rectangle ++(0.2,-1.0);
        \end{circuitikz}
    \node [] at (-2.05, 2) {\small $\mathbb{G}, \beta$}; 
    \node [] at (-1.1, 2) {\small $\mathbb{V}, \alpha$}; 
    \node [] at (-1.6, -0.3) {\small $G_0$}; 
    \end{tikzpicture}
\end{figure}

\subsubsection*{Constitutive equation}
\begin{equation}
    \begin{aligned}
    \sigma(t) + \frac{\mathbb{V}}{\mathbb{G}} \frac{{\rm d} ^{\alpha - \beta} \sigma(t)}{{\rm d}t^{\alpha - \beta}} &= G_0 \gamma(t) +
    \mathbb{V} \frac{{\rm d}^{\alpha} \gamma(t)}{{\rm d} t^{\alpha}} +
    G_0 \frac{\mathbb{V}}{\mathbb{G}} \frac{{\rm d}^{\alpha-\beta} \gamma(t)}{{\rm d} t^{\alpha-\beta}}\\
    \tau_c &= \left(\frac{\mathbb{V}}{\mathbb{G}} \right)^{\frac{1}{\alpha-\beta}}\\
    G_c &= \mathbb{V} \tau_c^{-\alpha}
    \end{aligned}
    \label{eq:constitutive_fzeners}
\end{equation}

\subsubsection*{Creep compliance (Laplace form)}
\begin{equation}
    \hat{J}(s) = \frac{1}{s}\frac{ \left( \mathbb{V}s^\alpha + \mathbb{G} s^\beta \right)}{\mathbb{V}s^\alpha \mathbb{G} s^\beta + G_0 \left( \mathbb{V}s^\alpha + \mathbb{G} s^\beta \right)}
    \label{eq:creep_fzeners}
\end{equation}

\subsubsection*{Relaxation modulus}
\begin{equation}
    \begin{aligned}
    G(t) &= G_0 + \mathbb{G} t^{-\beta} E_{a, b} \left( z \right)\\
    &a = \alpha - \beta, \quad b = 1 - \beta \\
    z &= -\left(\frac{t}{\tau_c}\right)^{\alpha - \beta}
    \end{aligned}
    \label{eq:relaxation_fzeners}
\end{equation}

\subsubsection*{Storage modulus and Loss modulus}
\begin{equation}
    \begin{aligned}
    G^{\prime}(\omega) &= G_0 + \frac{(\mathbb{G}\omega^{\beta})^{2}(\mathbb{V}\omega^{\alpha})\cos{(\frac{\pi}{2}\alpha)} + (\mathbb{V}\omega^{\alpha})^{2}(\mathbb{G}\omega^{\beta})\cos{(\frac{\pi}{2}\beta)}}{(\mathbb{V}\omega^{\alpha})^{2} + (\mathbb{G}\omega^{\beta})^{2} + 2(\mathbb{V}\omega^{\alpha})(\mathbb{G}\omega^{\beta})\cos{(\frac{\pi}{2}(\alpha-\beta)})}
 \\
    G^{\prime \prime}(\omega) &= \frac{(\mathbb{G}\omega^{\beta})^{2}(\mathbb{V}\omega^{\alpha})\sin{(\frac{\pi}{2}\alpha)} + (\mathbb{V}\omega^{\alpha})^{2}(\mathbb{G}\omega^{\beta})\sin{(\frac{\pi}{2}\beta)}}{(\mathbb{V}\omega^{\alpha})^{2} + (\mathbb{G}\omega^{\beta})^{2} + 2(\mathbb{V}\omega^{\alpha})(\mathbb{G}\omega^{\beta})\cos{(\frac{\pi}{2}(\alpha-\beta)})}\\
    \end{aligned}
    \label{eq:oscillation_fzeners}
\end{equation}

\subsubsection*{Representative plots}
\begin{figure}[htbp]
    \centering
    \begin{subfigure}{0.26\textwidth}
        \centering
        \begin{picture}(0,0)
            \put(-67,0){\textbf{\Large a}} 
        \end{picture}
        \includegraphics[width=\linewidth]{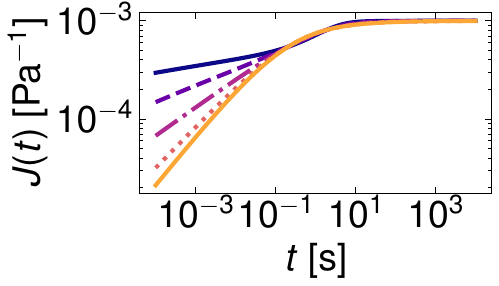} 
    \end{subfigure}
    \hfill
    \begin{subfigure}{0.25\textwidth}
        \centering
        \begin{picture}(0,0)
            \put(-67,0){\textbf{\Large b}} 
        \end{picture}
        \includegraphics[width=\linewidth]{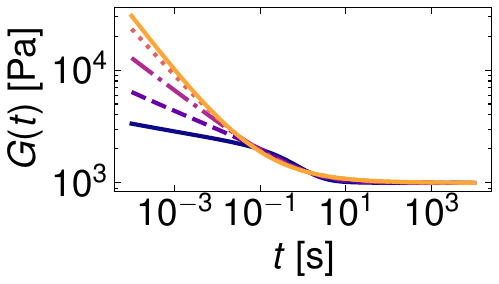} 
    \end{subfigure}
    \hfill
    \begin{subfigure}{0.38\textwidth}
        \centering
        \begin{picture}(0,0)
            \put(-110,0){\textbf{\Large c}} 
        \end{picture}
        \includegraphics[width=\linewidth]{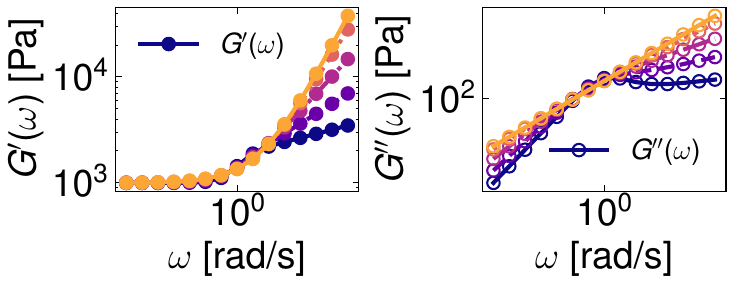} 
    \end{subfigure}
    \begin{subfigure}{1\textwidth}
        \centering
        \includegraphics[width=\linewidth]{legend_fmm.pdf} 
    \end{subfigure}
    \caption{Representative plots of the Fractional Zener-S model illustrating different viscoelastic behaviors under various types of stimulus. The plots were computed using fixed parameter values of $\mathbb{V} = 1000$~Pa~s$^{\alpha}$,  $\mathbb{G} = 1000$~Pa~s$^{\beta}$, and $G_0 = 1000$~Pa. \textbf{a)} Time-dependent creep compliance $J(t)$ showing the variation of strain over time under constant stress. \textbf{b)} Stress relaxation behavior depicting a decrease in the relaxation modulus $G(t)$ under constant strain over time. \textbf{c)} Oscillatory response demonstrating the storage modulus $G^{\prime}(\omega)$ and loss modulus $G^{\prime \prime}(\omega)$ under frequency-changing load.}
    \label{fig:fzs_plots}
\end{figure}

\clearpage

\subsection*{Viscosity models}

\subsubsection*{Herschel-Bulkley (\texttt{HerschelBulkley})}

\begin{equation}
    \label{eq:herschel_bulkley}
    \eta(\dot{\gamma}) = \frac{\sigma_0}{\dot{\gamma}} + k\dot{\gamma}^{n-1}
\end{equation}

\noindent The Herschel-Bulkley model describes a yield stress fluid where \(\eta(\dot{\gamma})\) is the apparent viscosity, \(\sigma_0\) is the yield stress, \(k\) is the consistency index, \(\dot{\gamma}\) is the shear rate, and \(n\) is the flow behavior index.

\subsubsection*{Bingham (\texttt{Bingham})}

\begin{equation}
    \label{eq:bingham}
    \eta(\dot{\gamma}) = \frac{\sigma_0}{\dot{\gamma}} + \eta_{pl}
\end{equation}

\noindent The Bingham model describes a fluid with a linear relationship between shear stress and shear rate beyond a yield stress \(\sigma_0\), where \(\eta_{pl}\) is the plastic viscosity.

\subsubsection*{Power-Law (\texttt{PowerLaw})}

\begin{equation}
    \label{eq:power_law}
    \eta(\dot{\gamma}) = k\dot{\gamma}^{n-1}
\end{equation}

\noindent The Power-Law model is used for non-Newtonian fluids, where \(\eta(\dot{\gamma})\) is the apparent viscosity, \(k\) is the consistency index, \(\dot{\gamma}\) is the shear rate, and \(n\) is the flow behavior index.

\subsubsection*{Carreau-Yasuda (\texttt{CarreauYasuda})}

\begin{equation}
    \label{eq:carreau_yasuda}
    \eta(\dot{\gamma}) = \eta_\infty + (\eta_0 - \eta_\infty) \left[ 1 + (k \dot{\gamma})^a \right]^{\frac{n-1}{a}}
\end{equation}

\noindent The Carreau-Yasuda model describes shear-thinning behavior, where \(\eta(\dot{\gamma})\) is the apparent viscosity, \(\eta_0\) is the zero-shear viscosity, \(\eta_\infty\) is the infinite-shear viscosity, \(k\) is a time constant, \(a\) is a fitting parameter, and \(n\) is the power-law index.

\subsubsection*{Cross (\texttt{Cross})}

\begin{equation}
    \label{eq:cross}
    \eta(\dot{\gamma}) = \eta_\infty + \frac{\eta_0 - \eta_\infty}{1 + (k \dot{\gamma})^n}
\end{equation}

\noindent The Cross model also describes shear-thinning behavior, where \(\eta(\dot{\gamma})\) is the apparent viscosity, \(\eta_0\) is the zero-shear viscosity, \(\eta_\infty\) is the infinite-shear viscosity, \(k\) is a time constant, and \(n\) is the power-law index.

\subsubsection*{Casson (\texttt{Casson})}

\begin{equation}
    \label{eq:casson}
    \eta(\dot{\gamma}) = \left( \frac{\sqrt{\sigma_0}}{\sqrt{\dot{\gamma}}} + \sqrt{\eta_{pl}} \right)^2
\end{equation}

\noindent The Casson model describes the behavior of certain yield stress fluids, where \(\eta(\dot{\gamma})\) is the apparent viscosity, \(\sigma_0\) is the yield stress, and \(\eta_{pl}\) is the plastic viscosity.

\clearpage

\section{Multi-Layer Perceptron classifier}

This section describes the training process of the Multi-Layer Perceptron (MLP) classifier for analyzing data related to creep, stress relaxation, oscillation, and rotation. We chose the MLP classifier because of its computational efficiency. Additionally, once trained, the MLP model file is considerably smaller than files generated by Random Forest classifiers; MLP model sizes typically range from kilobytes, while Random Forest models can reach gigabytes. The latter is important to facilitate the distribution of pyRheo.

The source code for training the MLP is available on the pyRheo GitHub page, allowing users to modify it to include additional models or adjust the number of datasets used for training. Each data type has its corresponding script, with the scripts for creep, relaxation, oscillation, and rotation being similar but tailored to fit the specific models relevant to each data type. These scripts utilize various libraries for numerical computations, data manipulation, machine learning, and data visualization, which provide the essential tools and functions needed for the tasks at hand.

Currently, the MLP is trained with the following models: Maxwell, SpringPot, FractionalMaxwellGel, FractionalMaxwellLiquid, FractionalMaxwell, and FractionalKelvinVoigt. Zener models are not included, as they overlap with the aforementioned rheological behaviors. In the future, the MLP model can be expanded to function as a multi-label classifier to incorporate Zener models.

\subsection*{Step 1: Generating synthetic data}

To train the MLP models, we started by generating synthetic data using several predefined models, such as the Maxwell Model and the Fractional Maxwell Model. Each of these models simulates a specific type of material behavior. The task of the MLP is to classify the input data the user provides into one of these predefined models. We generated 1 million synthetic datasets for each classifier, assigning random parameter values to the model functions and time (angular frequency or shear rate) intervals.

\subsection*{Step 2: Scaling synthetic data}

Before training the MLP, we scale each dataset to ensure optimal performance. First, we convert the dataset to a logarithmic scale. Then, we standardize it by subtracting the mean and scaling it to have a unit variance. This approach ensures that the user-provided data falls within the numerical limits of the MLP, enhancing the overall performance of the machine learning model.

\subsection*{Step 3: Principal Component Analysis (PCA) of the scaled synthetic data}

PCA is used to reduce the dimensionality of the data while retaining the most important information~\citep{miranda-valdez_mäkinen_coffeng_päivänsalo_jannuzzi_viitanen_koivisto_alava_2025, rickmann_lookman_kalinin_2019}. This makes it easier for machine learning models to analyze the data without the burden of using too many variables. The scaled data is transformed into a new coordinate system where the most significant features (10 principal components) are identified. This reduces the number of variables needed to describe the data while preserving important patterns.

\subsection*{Step 4: Training machine learning models}

Machine learning models are trained to classify the different types of rheological behavior based on synthetic data. The transformed data (principal components) is split into training (80\%) and testing sets (20\%). With the training set, we train the MLP classifier. The trained models predict the type of rheological behavior (model type) for the testing data.

\subsection*{Step 5: Evaluating Model Performance}

Finally, we evaluate how well the trained machine learning model can predict the type of rheological behavior on new data. The predictions made by the model using the test set are compared to the actual model types (true labels). In Supplementary Fig.\ref{fig:confusion_matrices_mlp}, we show the confusion matrices generated to visualize where the model performs well and makes errors. The MLP models have an accuracy of 70 to 80\%.

\begin{figure}[htbp]
    \centering
    \begin{subfigure}{0.47\textwidth}
        \centering
        \begin{picture}(0,0)
            \put(-90,0){\textbf{\Large a}} 
        \end{picture}
        \includegraphics[width=\linewidth]{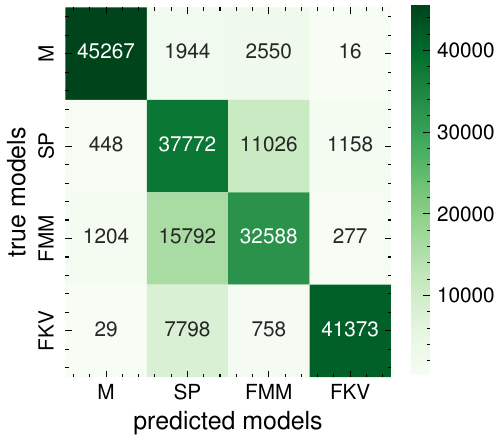} 
    \end{subfigure}
    \hfill
    \begin{subfigure}{0.47\textwidth}
        \centering
        \begin{picture}(0,0)
            \put(-90,0){\textbf{\Large b}} 
        \end{picture}
        \includegraphics[width=\linewidth]{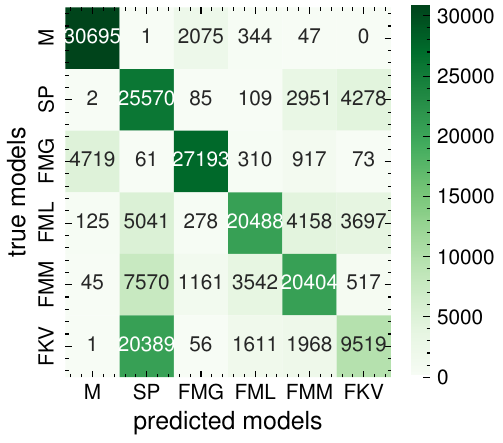} 
    \end{subfigure}
    \hfill
    \begin{subfigure}{0.47\textwidth}
        \centering
        \begin{picture}(0,0)
            \put(-105,0){\textbf{\Large c}} 
        \end{picture}
        \includegraphics[width=\linewidth]{confusiomatrix_relaxation.pdf} 
    \end{subfigure}
    \hfill
    \begin{subfigure}{0.47\textwidth}
        \centering
        \begin{picture}(0,0)
            \put(-105,0){\textbf{\Large d}} 
        \end{picture}
        \includegraphics[width=\linewidth]{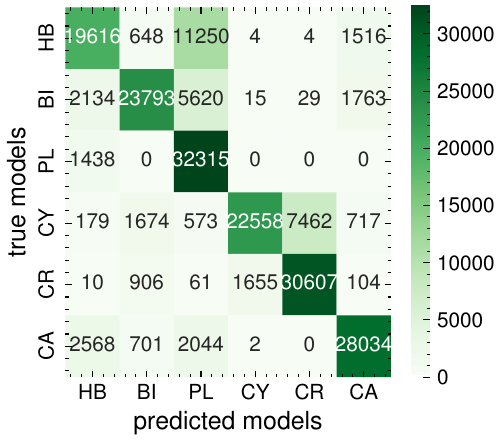} 
    \end{subfigure}
    \caption{Confusion matrices illustrating the performance of the trained MLP models in predicting rheological behavior types. Each subfigure corresponds to a different rheological data class: \textbf{a)} creep data, \textbf{b)} relaxation data, \textbf{c)} oscillation data, and \textbf{d)} rotation data. The rows represent the true labels, whereas the columns denote the predicted labels. The models achieve an accuracy between 70\% and 80\%.}

    \label{fig:confusion_matrices_mlp}
\end{figure}

\clearpage

\section{Complementary examples of predicting and fitting tasks with pyRheo}

\begin{figure}[htbp]
    \centering
    \begin{subfigure}{0.3\textwidth}
        \centering
        \begin{picture}(0,0)
            \put(-67,0){\textbf{\Large a}} 
        \end{picture}
        \includegraphics[width=\linewidth]{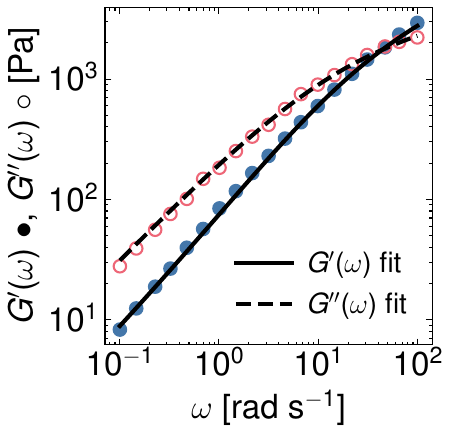} 
    \end{subfigure}
    \hfill
    \begin{subfigure}{0.3\textwidth}
        \centering
        \begin{picture}(0,0)
            \put(-67,0){\textbf{\Large b}} 
        \end{picture}
        \includegraphics[width=\linewidth]{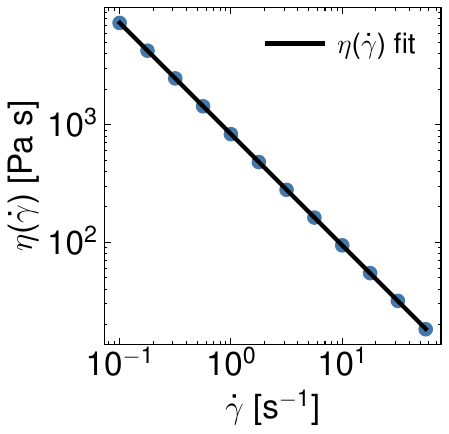} 
    \end{subfigure}
    \hfill
    \begin{subfigure}{0.3\textwidth}
        \centering
        \begin{picture}(0,0)
            \put(-67,0){\textbf{\Large c}} 
        \end{picture}
        \includegraphics[width=\linewidth]{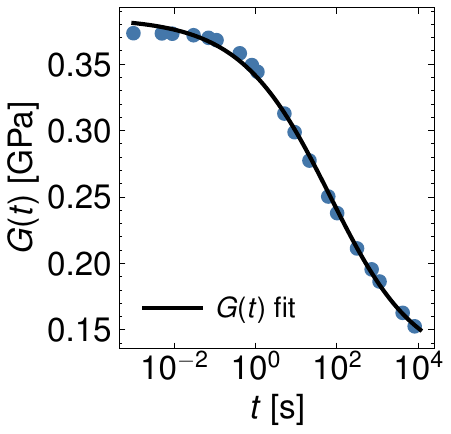} 
    \end{subfigure}
    \caption{Fitting results obtained using pyRheo. \textbf{a)} Storage and Loss modulus as a function of angular frequency $G^{\prime}(\omega)$, $G^{\prime \prime}(\omega)$ measured for a metal-coordinating polymer network fitted with \texttt{FractionalMaxwell} (data from \citet{epstein_2019_metal}). \textbf{b)}  Shear viscosity as function of shear rate $\eta(\dot{\gamma})$ measured for a cellulose nanofiber hydrogel  fitted with \texttt{PowerLaw} (data from \citet{miranda-valdez_sourroubille_mäkinen_puente-córdova_puisto_koivisto_alava_2024}). \textbf{c)} Relaxation modulus as function of time $G(t)$ measured for a polypropylene sample fitted with \texttt{FractionalZenerSolidS}. }
    \label{fig:examples_pyrheo_predict_fit_part2}
\end{figure}

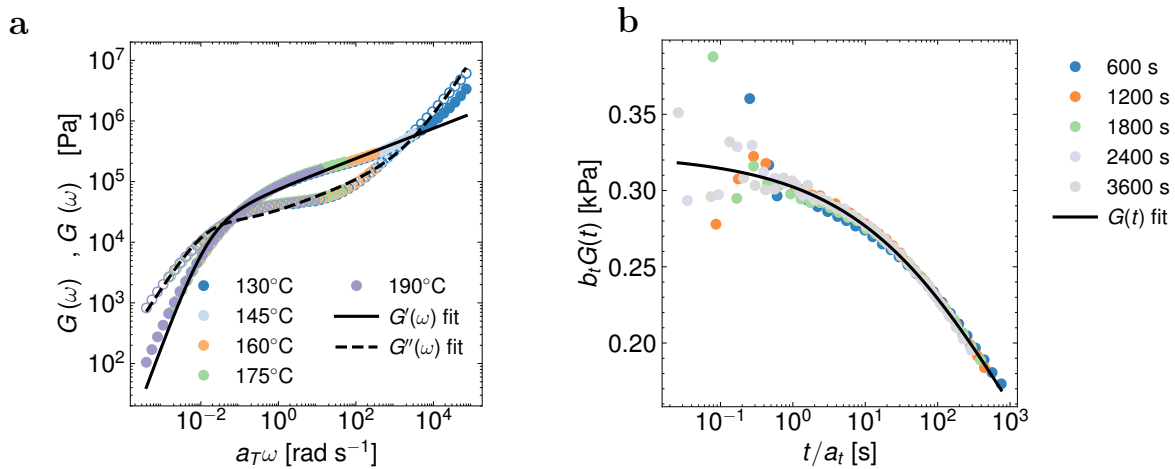
\begin{figure}[htbp]
    \centering
    \begin{subfigure}{0.385\textwidth}
        \centering
        \begin{picture}(0,0)
            \put(-105,0){\textbf{\Large a}} 
        \end{picture}
        \includegraphics[width=\linewidth]{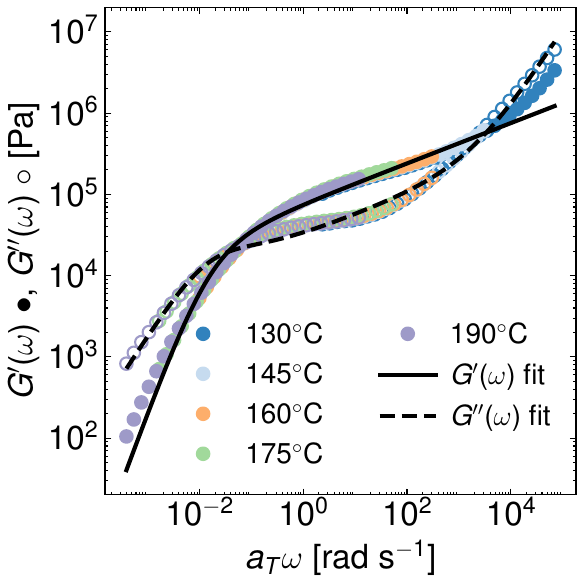} 
    \end{subfigure}
    \hfill
    \begin{subfigure}{0.54\textwidth}
        \centering
        \begin{picture}(0,0)
            \put(-105,0){\textbf{\Large b}} 
        \end{picture}
        \includegraphics[width=\linewidth]{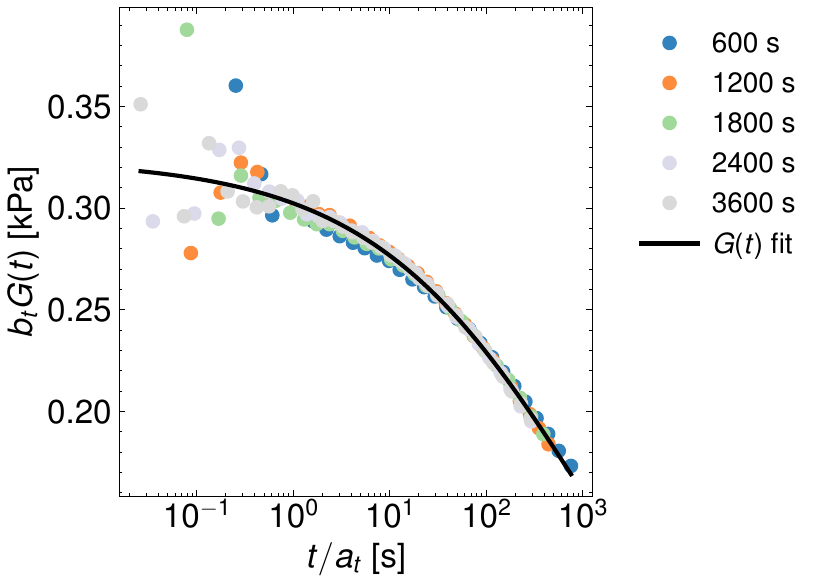} 
    \end{subfigure}
    \caption{Fitting results obtained using pyRheo. \textbf{a)} Storage and Loss modulus master curve as a function of angular frequency $G^{\prime}(\omega)$, $G^{\prime \prime}(\omega)$ measured for polystyrene ($T_{\rm{ref}}=160^{\circ}\rm{C}$) fitted with \texttt{FractionalZenerLiquidD} (data from \citet{ricarte_shanbhag_2024}). \textbf{b)} Relaxation modulus master curve as function of time $G(t)$ measured for a laponite with different aging times ($t_{\rm{ref}}=600~\rm{s}$) fitted with \texttt{FractionalMaxwellGel} (data from \citet{lennon_mcKinley_swan_2023}). }
    \label{fig:examples_pyrheo_predict_fit_part3}
\end{figure}

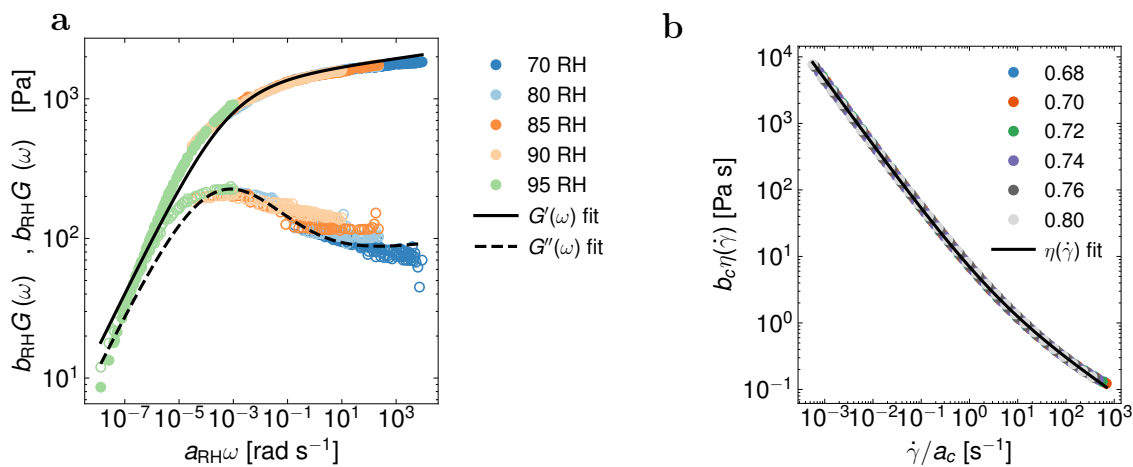
\begin{figure}[htbp]
    \centering
    \begin{subfigure}{0.54\textwidth}
        \centering
        \begin{picture}(0,0)
            \put(-105,0){\textbf{\Large a}} 
        \end{picture}
        \includegraphics[width=\linewidth]{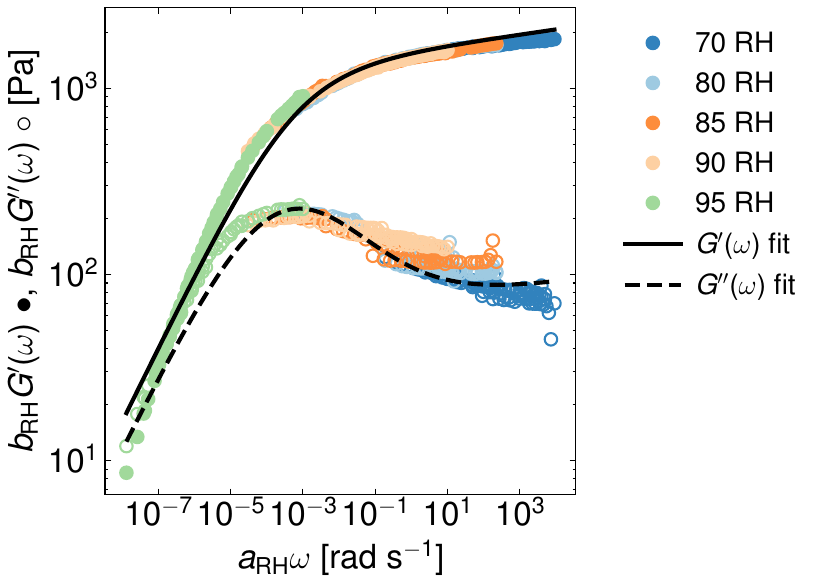} 
    \end{subfigure}
    \hfill
    \begin{subfigure}{0.38\textwidth}
        \centering
        \begin{picture}(0,0)
            \put(-105,0){\textbf{\Large b}} 
        \end{picture}
        \includegraphics[width=\linewidth]{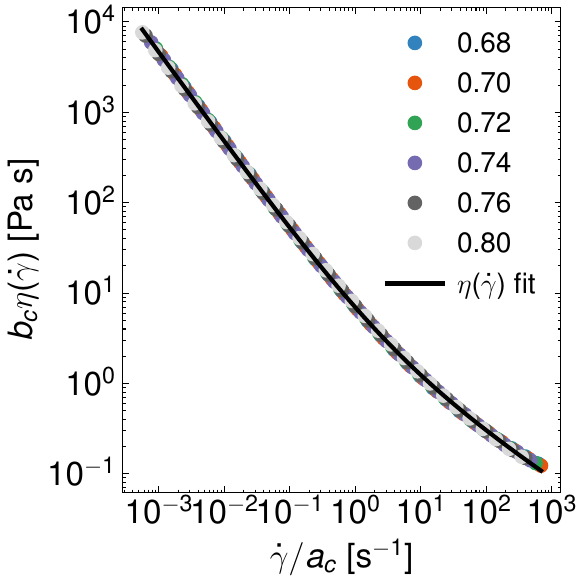} 
    \end{subfigure}
    \caption{Fitting results obtained using pyRheo. \textbf{a)} Storage and Loss modulus master curve as a function of angular frequency $G^{\prime}(\omega)$, $G^{\prime \prime}(\omega)$ measured for a viscoelastic material ($\rm{RH}_{\rm{ref}}=90$) fitted with \texttt{FractionalMaxwell} (data from \citet{lennon_mcKinley_swan_2023}). \textbf{b)} Viscosity master curve as function of shear rate $\eta(\dot{\gamma})$ measured for a castor oil emulsion with different content ($c_{\rm{ref}}=0.68$) fitted with \texttt{HerschelBulkley} (data from \citet{lennon_mcKinley_swan_2023} adapted from \citet{dekker_dinkgreve_cagny_koeze_tighe_bonn_2018}). }
    \label{fig:examples_pyrheo_predict_fit_part4}
\end{figure}

\begin{figure}[htbp]
    \centering
    \begin{subfigure}{0.7\textwidth}
        \centering
        \begin{picture}(0,0)
        \end{picture}
        \includegraphics[width=\linewidth]{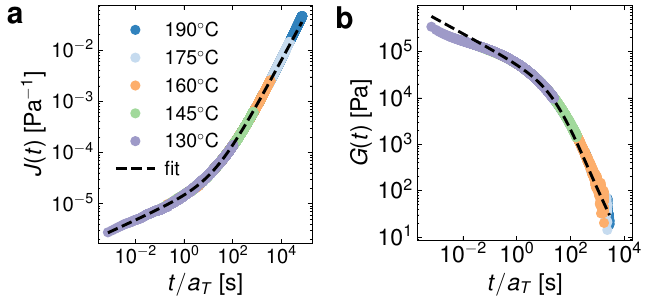} 
    \end{subfigure}
    \caption{Master curves for the linear viscoelastic behavior of polystyrene at various temperatures. \textbf{a)} Creep compliance $J(t)$ master curve ($T_{\rm ref}=160^{\circ}$C) constructed using the time-temperature superposition (TTS). The curve is fitted using the \texttt{auto} function in pyRheo, which classifies the data as a \texttt{FractionalMaxwell}. \textbf{b)} Relaxation modulus $G(t)$ master curve classified and fitted with \texttt{FractionalMaxwell}. The raw data was reproduced from \citet{ricarte_shanbhag_2024}.}
    \label{fig:examples_pyrheo_predict_fit_part5}
\end{figure}

\begin{figure}[htbp]
    \centering
    \begin{subfigure}{0.6\textwidth}
        \centering
        \begin{picture}(0,0)
        \end{picture}
        \includegraphics[width=\linewidth]{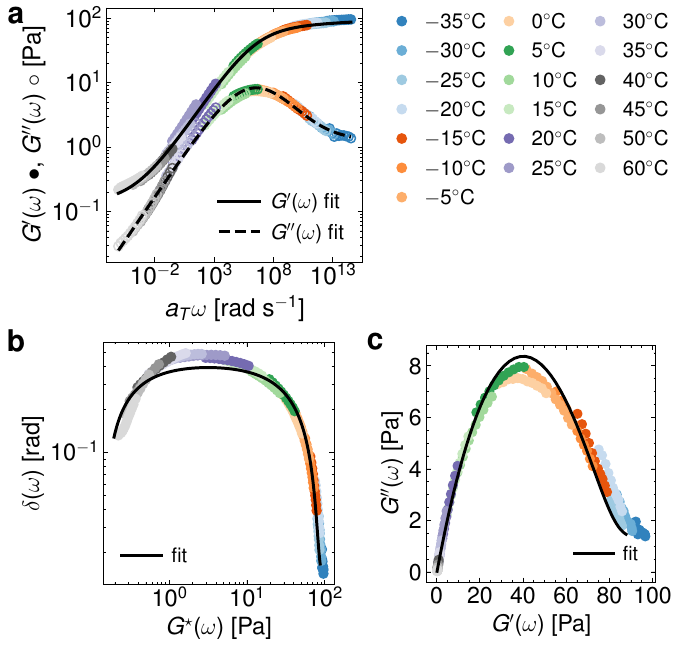} 
    \end{subfigure}
    \caption{Master curves for the linear viscoelastic behavior of a vinyl nitrile solid foam at various temperatures. \textbf{a)} Storage modulus $G^{\prime}(\omega)$ and Loss modulus $G^{\prime \prime}(\omega)$ master curves ($T_{\rm ref}=25^{\circ}$C) constructed using the time-temperature superposition (TTS) and fitted with \texttt{FractionalZenerS}, constituted by a \texttt{FractionalMaxwell} connected in parallel to a spring. \textbf{b)} Van Gurp--Palmen plot and \textbf{c)} Cole--Cole representation of the master curve. The raw data was reproduced from \citet{landauer_kafka_moser_foster_blaiszik_forster_2023}.}
    \label{fig:examples_pyrheo_predict_fit_part6}
\end{figure}

We show in Supplementary Fig.\ref{fig:examples_pyrheo_predict_fit_part1}, Supplementary Fig.\ref{fig:examples_pyrheo_predict_fit_part2}, Supplementary Fig.\ref{fig:examples_pyrheo_predict_fit_part3}, Supplementary Fig.\ref{fig:examples_pyrheo_predict_fit_part4}, Supplementary Fig.\ref{fig:examples_pyrheo_predict_fit_part5}, and Supplementary Fig.\ref{fig:examples_pyrheo_predict_fit_part6} examples of fitting with pyRheo. The parameter values of each model can be found in their corresponding Jupyter Notebook.

\begin{figure}[htbp]
    \centering
    \begin{subfigure}{0.3\textwidth}
        \centering
        \begin{picture}(0,0)
            \put(-67,0){\textbf{\Large a}} 
        \end{picture}
        \includegraphics[width=\linewidth]{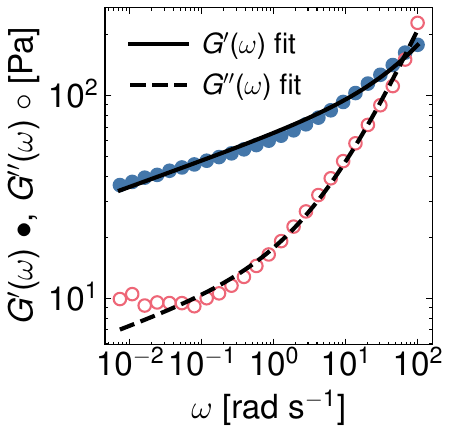} 
    \end{subfigure}
    \hfill
    \begin{subfigure}{0.3\textwidth}
        \centering
        \begin{picture}(0,0)
            \put(-67,0){\textbf{\Large b}} 
        \end{picture}
        \includegraphics[width=\linewidth]{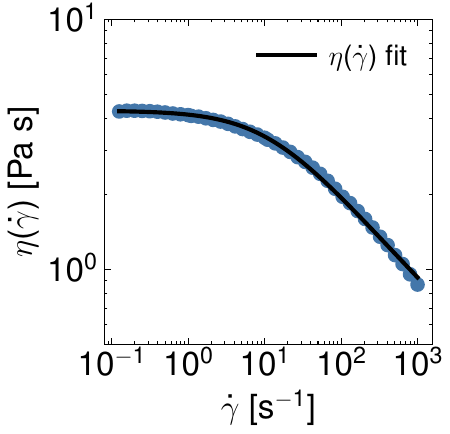} 
    \end{subfigure}
    \hfill
    \begin{subfigure}{0.3\textwidth}
        \centering
        \begin{picture}(0,0)
            \put(-67,0){\textbf{\Large c}} 
        \end{picture}
        \includegraphics[width=\linewidth]{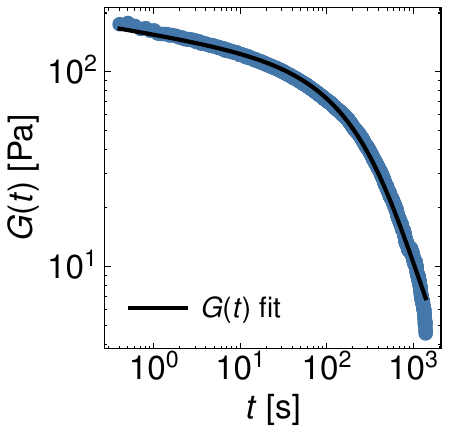} 
    \end{subfigure}
    \caption{Fitting results obtained using pyRheo. \textbf{a)} Storage and Loss modulus as a function of angular frequency $G^{\prime}(\omega)$, $G^{\prime \prime}(\omega)$ measured for a chia pudding fitted with \texttt{FractionalKelvinVoigt}. \textbf{b)} Shear viscosity as function of shear rate $\eta(\dot{\gamma})$ measured for an ethylcellulose solution in toluene fitted with \texttt{CarreauYasuda}. \textbf{c)} Relaxation modulus as function of time $G(t)$ measured for a shaving foam fitted with \texttt{FractionalMaxwell} (data from \citet{lavergne_sollich_trappe_2022}). }
    \label{fig:examples_pyrheo_predict_fit_part1}
\end{figure}

\clearpage

\section{Mittag--Leffler function in pyRheo}
\subsection*{Global Padé approximations}
In pyRheo, we have implemented the second-order global Padé approximation by \citet{zeng_chen_2015} and fourth-order global Padé approximations published by \citet{sarumi_furati_khaliq_2020}. 

\subsection*{Second-order approximation}
The second-order approximation is of the type $R^{3,2}_{a, b}(x)$. The approximation is given by
\begin{equation}
\label{eq:Zeng_gpa}
R_{a, b}^{3, 2}(-x) = \dfrac{1}{\Gamma(b - a)} \,
\dfrac{p_1 + x }{q_0 + q_1 x +  x^2 }, \qquad b > a, 
\end{equation}
with 
\begin{equation}
\begin{aligned}
p_1  & = c_{a,b} \left[
\Gamma(b) \Gamma(b + a ) - 
\frac{\Gamma(b + a)\Gamma^2(b - a)}{\Gamma(b - 2a)}
\right], 
\\ 
q_0  & = c_{a,b} \left[\
\frac{\Gamma^2(b) \Gamma(b + a )}{\Gamma(b - a)}
- \frac{\Gamma(b) \Gamma(b + a)\Gamma(b - a)}
{\Gamma(b - 2a)}
\right],
\\ 
q_1 & = c_{a,b} \left[
\Gamma(b) \Gamma(b + a ) - 
\frac{\Gamma^2(b)\Gamma(b - a)}{\Gamma(b - 2a)}
\right], 
\\
c_{a,b} &= \frac{1}{\Gamma(b + a) \Gamma(b - a) - \Gamma^2(b)},
\end{aligned}
\end{equation}
and 
\begin{equation}
R_{a, a}^{3, 2}(-x) = 
\dfrac{a}{\Gamma(1+a) + 
\frac{2\Gamma(1 - a)^2}{\Gamma(1-2a)} \, x + 
\Gamma(1 - a) \, x^2},
\qquad 0 < a < 1.
\end{equation}

\subsection*{Fourth-order approximation}

The fourth-order global Pad\'e approximation ($\nu = 4$) correspond to the types $(m,n)$ with $m+n = 9$. They include the types $(5,4)$, $(6,3)$, and $(7,2)$. pyRheo uses the approximation $R_{a, b}^{m,n}$ for $\nu = 4$, which takes the form
\begin{equation}
R_{a, b}^{m,n} (-x) = 
\begin{cases}
\label{eq:mittag_leffler_pade}
\dfrac{1}{\Gamma(b - a)} \,
\dfrac{p_1 + p_2x + p_3 x^2 + x^3}{q_0 + q_1x + q_2 x^2 + q_3 x^3 + x^4},
\quad & b > a,
\\ \\
\dfrac{-1}{\Gamma( - a)} \,
\dfrac{\hat{p}_2  + \hat{p}_3 x + x^2}{\hat{q}_0 + \hat{q}_1x + 
\hat{q}_2 x^2 + \hat{q}_3 x^3 + x^4}, & b = a.
\end{cases}
\end{equation}
The unknown coefficients are obtained by solving the following systems. For $b > a$, the coefficients satisfy the system

\subsection*{Coefficients of $R_{a, b}^{5, 4}$}

\begin{equation}
\label{sys:54ab}
\begin{bmatrix}
  1 & 0  & 0 & - \dfrac{\Gamma(b - a)}{\Gamma{(b)}}         
  & 0                                               & 0 & 0              \\
  0 & 1  & 0 & \dfrac{\Gamma(b - a)}{\Gamma{(b + a)}}  
  & -\dfrac{\Gamma(b - a)}{\Gamma{(b)}}  & 0 & 0               \\
  0 & 0  & 1 & -\dfrac{\Gamma(b - a)}{\Gamma{(b + 2a)}} 
  & \dfrac{\Gamma(b - a)}{\Gamma{(b + a)}} & -\dfrac{\Gamma(b - a)}{\Gamma{(b)}} & 0 \\
  0 & 0  & 0 & \dfrac{\Gamma(b - a)}{\Gamma{(b + 3a)}} 
  & - \dfrac{\Gamma(b - a)}{\Gamma{(b + 2a)}} & \dfrac{\Gamma(b - a)}{\Gamma{(b + a)}} 
  & - \dfrac{\Gamma{(b - a)}}{\Gamma{(b)}} \\
  1 & 0  & 0 & 0 &-1 & \dfrac{\Gamma(b - a)}{\Gamma(b - 2a)} & 
  - \dfrac{\Gamma(b - a)}{\Gamma(b - 3a)}  \\
  0 & 1  & 0 & 0 & 0 &  -1  
  & \dfrac{\Gamma(b - a)}{\Gamma(b - 2a)}  \\
  0 & 0  & 1 & 0 & 0 & 0 & -1
\end{bmatrix}
\begin{pmatrix}
  p_1                 \\
  p_2                \\
  p_3              	\\
  q_0				  \\
  q_1 \\
  q_2 \\
  q_3
\end{pmatrix} = 
\begin{pmatrix}
  0                 \\
  0                \\
  0              	\\
  -1				  \\
  -\dfrac{\Gamma(b - a)}{\Gamma(b - 4a)} \\
  \dfrac{\Gamma(b - a)}{\Gamma(b - 3a)} \\
  -\dfrac{\Gamma(b - a)}{\Gamma(b - 2a)}
\end{pmatrix}.
\end{equation}
For $b = a$ the coefficients satisfy the system
\begin{equation}
\label{sys:54aa}
\begin{bmatrix}
1  & 0 & \dfrac{\Gamma( - a)}{\Gamma{(a)}}         
& 0                                               & 0 & 0              \\
0  & 1 & -\dfrac{\Gamma( - a)}{\Gamma{(2 a)}}  
& \dfrac{\Gamma( - a)}{\Gamma{(a)}}  & 0 & 0               \\
0  & 0 & \dfrac{\Gamma( - a)}{\Gamma{(3a)}} 
& -\dfrac{\Gamma( - a)}{\Gamma{(2a)}} & \dfrac{\Gamma( - a)}{\Gamma{(a)}} & 0 \\
0  & 0 & 0 
&  -1 & -\dfrac{\Gamma( - a)}{\Gamma{(-2 a)}} 
& 0 \\
1  & 0 & 0 & 0 
& -1 &  \dfrac{\Gamma( - a)}{\Gamma(-2a)}  \\
0  & 1 & 0 & 0 &  0  
& -1 
\end{bmatrix}
\begin{pmatrix}
\hat{p}_2 \\
\hat{p}_3 \\
\hat{q}_0 \\
\hat{q}_1 \\
\hat{q}_2 \\
\hat{q}_3
\end{pmatrix} = 
\begin{pmatrix}
0                \\
0              	\\
-1				  \\
0                  \\
0                  \\
-\dfrac{\Gamma( - a)}{\Gamma( - 2a)}
\end{pmatrix}.
\end{equation}

\subsection*{Coefficients of $R_{a, b}^{6, 3}$}

For $b > a$, the coefficients satisfy the system
\begin{gather}\label{sys:63ab}
\begin{bmatrix}
1 & 0  & 0 & - \dfrac{\Gamma(b - a)}{\Gamma{(b)}}         
& 0                                               & 0 & 0              \\
0 & 1  & 0 & \dfrac{\Gamma(b - a)}{\Gamma{(b + a)}}  
& -\dfrac{\Gamma(b - a)}{\Gamma{(b)}}  & 0 & 0               \\
0 & 0  & 1 & -\dfrac{\Gamma(b - a)}{\Gamma{(b + 2a)}} 
& \dfrac{\Gamma(b - a)}{\Gamma{(b + a)}} & -\dfrac{\Gamma(b - a)}{\Gamma{(b)}} & 0 \\
0 & 0  & 0 & \dfrac{\Gamma(b - a)}{\Gamma{(b + 3a)}} 
& - \dfrac{\Gamma(b - a)}{\Gamma{(b + 2a)}} & \dfrac{\Gamma(b - a)}{\Gamma{(b + a)}} 
& - \dfrac{\Gamma{(b - a)}}{\Gamma{(b)}} \\
0 & 0  & 0 & -\dfrac{\Gamma(b - a)}{\Gamma{(b + 4a)}} & \dfrac{\Gamma(b - a)}{\Gamma{(b + 3a)}} 
& -\dfrac{\Gamma(b - a)}{\Gamma(b + 2a)} &  \dfrac{\Gamma(b - a)}{\Gamma(b + a)}  \\
0 & 1  & 0 & 0 & 0 &  -1  
& \dfrac{\Gamma(b - a)}{\Gamma(b - 2a)}  \\
0 & 0  & 1 & 0 & 0 & 0 & -1
\end{bmatrix}
\begin{pmatrix}
p_1                 \\
p_2                \\
p_3              	\\
q_0				  \\
q_1 \\
q_2 \\
q_3
\end{pmatrix} = 
\begin{pmatrix}
0                 \\
0                \\
0              	\\
-1				  \\
\dfrac{\Gamma(b - a)}{\Gamma(b)} \\
\dfrac{\Gamma(b - a)}{\Gamma(b - 3a)} \\
-\dfrac{\Gamma(b - a)}{\Gamma(b - 2a)}
\end{pmatrix},
\end{gather}
For $b = a$, the coefficients satisfy the system
\begin{gather}\label{sys:63aa}
\begin{bmatrix}
		1  & 0 & \dfrac{\Gamma( - a)}{\Gamma{(a)}}         
		& 0                                               & 0 & 0              \\
		0  & 1 & -\dfrac{\Gamma( - a)}{\Gamma{(2 a)}}  
		& \dfrac{\Gamma( - a)}{\Gamma{(a)}}  & 0 & 0               \\
		0  & 0 & \dfrac{\Gamma( - a)}{\Gamma{(3a)}} 
		& -\dfrac{\Gamma( - a)}{\Gamma{(2a)}} & \dfrac{\Gamma( - a)}{\Gamma{(a)}} & 0 \\
		0  & 0 & -\dfrac{\Gamma( - a)}{\Gamma{(4a)}} 
		&  \dfrac{\Gamma( - a)}{\Gamma{(3a)}} & -\dfrac{\Gamma( - a)}{\Gamma{(2 a)}} 
		& \dfrac{\Gamma{( - a)}}{\Gamma{(a)}} \\
		1  & 0 & 0 & 0 
		& -1 &  \dfrac{\Gamma( - a)}{\Gamma(-2a)}  \\
		0  & 1 & 0 & 0 &  0  
		& -1 
\end{bmatrix}
\begin{pmatrix}
	\hat{p}_2 \\
	\hat{p}_3 \\
	\hat{q}_0 \\
	\hat{q}_1 \\
	\hat{q}_2 \\
	\hat{q}_3
\end{pmatrix} = 
\begin{pmatrix}
		0                \\
		0              	\\
		-1				  \\
		0                  \\
		\dfrac{\Gamma( - a)}{\Gamma( - 3a)} \\
		-\dfrac{\Gamma( - a)}{\Gamma( - 2a)}
\end{pmatrix}.
\end{gather}

\subsection*{Coefficients of $R_{a, b}^{7, 2}$}
For $b > a$, the coefficients satisfy the system
\begin{gather}\label{sys:72ab}
\begin{bmatrix}
1 & 0  & 0 & - \dfrac{\Gamma(b - a)}{\Gamma{(b)}}         
& 0                                               & 0 & 0              \\
0 & 1  & 0 & \dfrac{\Gamma(b - a)}{\Gamma{(b + a)}}  
& -\dfrac{\Gamma(b - a)}{\Gamma{(b)}}  & 0 & 0               \\
0 & 0  & 1 & -\dfrac{\Gamma(b - a)}{\Gamma{(b + 2a)}} 
& \dfrac{\Gamma(b - a)}{\Gamma{(b + a)}} & -\dfrac{\Gamma(b - a)}{\Gamma{(b)}} & 0 \\
0 & 0  & 0 & \dfrac{\Gamma(b - a)}{\Gamma{(b + 3a)}} 
& - \dfrac{\Gamma(b - a)}{\Gamma{(b + 2a)}} & \dfrac{\Gamma(b - a)}{\Gamma{(b + a)}} 
& - \dfrac{\Gamma{(b - a)}}{\Gamma{(b)}} \\
0 & 0  & 0 & -\dfrac{\Gamma(b - a)}{\Gamma{(b + 4a)}} & \dfrac{\Gamma(b - a)}{\Gamma{(b + 3a)}} 
& -\dfrac{\Gamma(b - a)}{\Gamma(b + 2a)} &  \dfrac{\Gamma(b - a)}{\Gamma(b + a)}  \\
0 & 0  & 0 & \dfrac{\Gamma(b - a)}{\Gamma{(b + 5a)}} & -\dfrac{\Gamma(b - a)}{\Gamma{(b + 4a)}} 
& \dfrac{\Gamma(b - a)}{\Gamma(b + 3a)} &  -\dfrac{\Gamma(b - a)}{\Gamma(b + 2a)}  \\
0 & 0  & 1 & 0 & 0 & 0 & -1
\end{bmatrix}
\begin{pmatrix}
p_1  \\
p_2   \\
p_3 \\
q_0	\\
q_1 \\
q_2 \\
q_3
\end{pmatrix} = 
\begin{pmatrix}
0  \\
0   \\
0   \\
-1	\\
\dfrac{\Gamma(b - a)}{\Gamma(b)} \\
-\dfrac{\Gamma(b - a)}{\Gamma(b + a)} \\
-\dfrac{\Gamma(b - a)}{\Gamma(b - 2a)}
\end{pmatrix}.
\end{gather}

For $b = a$, the coefficients satisfy the system
\begin{gather}\label{sys:72aa}
\begin{bmatrix}
1  & 0 & \dfrac{\Gamma( - a)}{\Gamma{(a)}}         
& 0                                               & 0 & 0              \\
0  & 1 & -\dfrac{\Gamma( - a)}{\Gamma{(2 a)}}  
& \dfrac{\Gamma( - a)}{\Gamma{(a)}}  & 0 & 0               \\
0  & 0 & \dfrac{\Gamma( - a)}{\Gamma{(3a)}} 
& -\dfrac{\Gamma( - a)}{\Gamma{(2a)}} & \dfrac{\Gamma( - a)}{\Gamma{(a)}} & 0 \\
0  & 0 & -\dfrac{\Gamma( - a)}{\Gamma{(4a)}} 
&  \dfrac{\Gamma( - a)}{\Gamma{(3a)}} & -\dfrac{\Gamma( - a)}{\Gamma{(2 a)}} 
& - \dfrac{\Gamma{( - a)}}{\Gamma{(a)}} \\
0  & 0 & \dfrac{\Gamma( - a)}{\Gamma{(5a)}} 
&  -\dfrac{\Gamma( - a)}{\Gamma{(4a)}} & \dfrac{\Gamma( - a)}{\Gamma{(3 a)}} 
& - \dfrac{\Gamma{( - a)}}{\Gamma{(2a)}} \\
0  & 1 & 0 & 0 &  0  
& -1 
\end{bmatrix}
\begin{pmatrix}
	\hat{p}_2 \\
	\hat{p}_3 \\
	\hat{q}_0 \\
	\hat{q}_1 \\
	\hat{q}_2 \\
	\hat{q}_3
\end{pmatrix} = 
\begin{pmatrix}
0  \\
0  	\\
-1  \\
0  \\
\dfrac{\Gamma( - a)}{\Gamma( - 3a)} \\
-\dfrac{\Gamma( - a)}{\Gamma( - 2a)}
\end{pmatrix}.
\end{gather}

\begin{figure}[htbp]
    \centering
    \begin{subfigure}{0.47\textwidth}
        \centering
        \begin{picture}(0,0)
            \put(-90,0){\textbf{\Large a}} 
        \end{picture}
        \includegraphics[width=\linewidth]{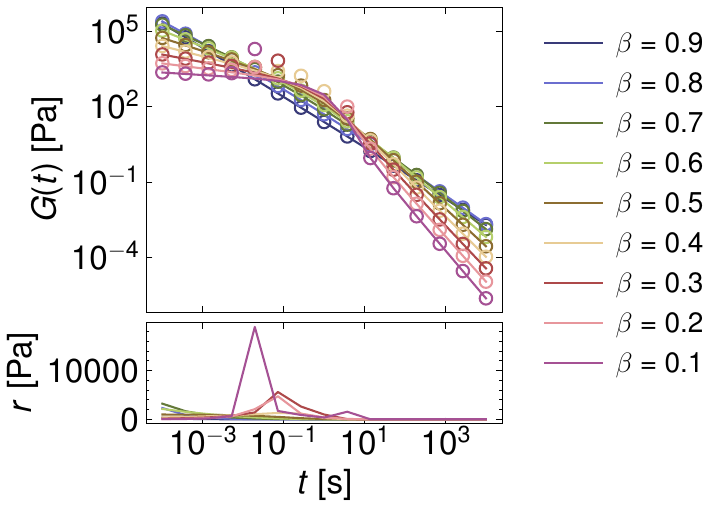} 
    \end{subfigure}
    \hfill
    \begin{subfigure}{0.47\textwidth}
        \centering
        \begin{picture}(0,0)
            \put(-90,0){\textbf{\Large b}} 
        \end{picture}
        \includegraphics[width=\linewidth]{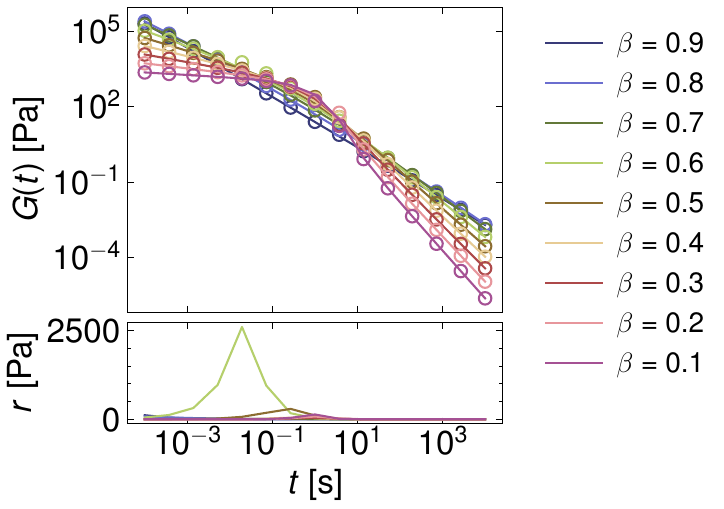} 
    \end{subfigure}
    \hfill
    \begin{subfigure}{0.47\textwidth}
        \centering
        \begin{picture}(0,0)
            \put(-105,0){\textbf{\Large c}} 
        \end{picture}
        \includegraphics[width=\linewidth]{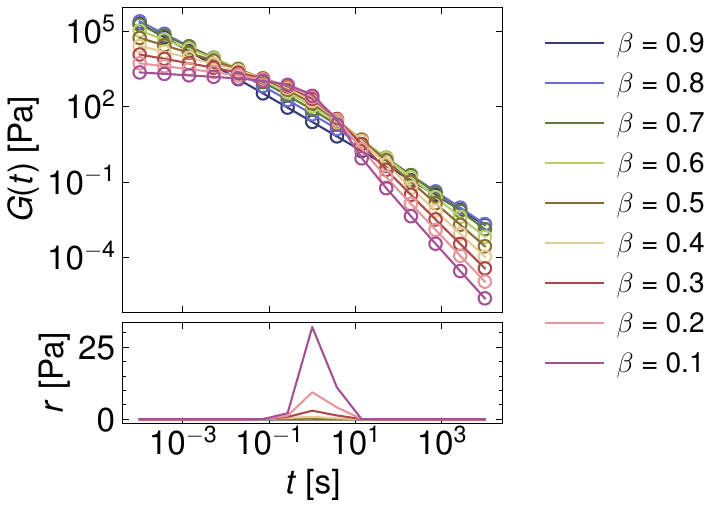} 
    \end{subfigure}
    \hfill
    \begin{subfigure}{0.47\textwidth}
        \centering
        \begin{picture}(0,0)
            \put(-105,0){\textbf{\Large d}} 
        \end{picture}
        \includegraphics[width=\linewidth]{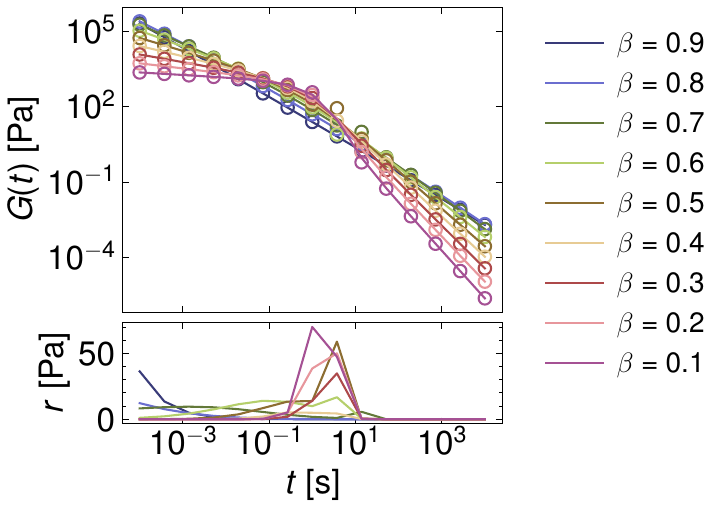} 
    \end{subfigure}
    \caption{Stress relaxation functions  $G(t)$  computed for different fractional orders $\beta$ using the \texttt{FractionalMaxwellLiquid}. The solid lines represent the functions computed using Garrappa's algorithm~\citep{garrappa_numerical_2015}, while the empty markers represent those computed using the global Padé approximation~\citep{zeng_chen_2015, sarumi_furati_khaliq_2020} \textbf{a)} $R_{a, b}^{3, 2}$, \textbf{b)} $R_{a, b}^{5, 4}$, \textbf{c)} $R_{a, b}^{6, 3}$, \textbf{d)} $R_{a, b}^{7, 2}$. Each figure also includes a residual $r$ plot, highlighting the main differences between both approaches. All the functions were computed by assigning a value of 1000 to $\mathbb{G}$ and $\eta$.}
    \label{fig:comparison_pade_garrappa_fml}
\end{figure}

\begin{figure}[htbp]
    \centering
    \begin{subfigure}{0.47\textwidth}
        \centering
        \begin{picture}(0,0)
            \put(-90,0){\textbf{\Large a}} 
        \end{picture}
        \includegraphics[width=\linewidth]{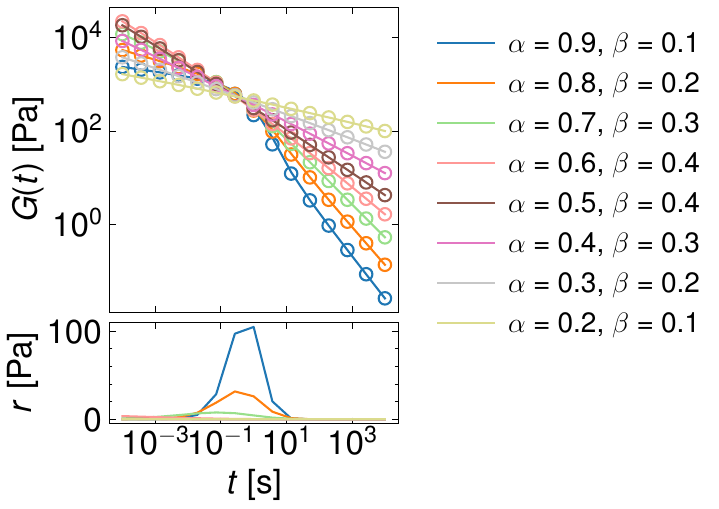} 
    \end{subfigure}
    \hfill
    \begin{subfigure}{0.47\textwidth}
        \centering
        \begin{picture}(0,0)
            \put(-90,0){\textbf{\Large b}} 
        \end{picture}
        \includegraphics[width=\linewidth]{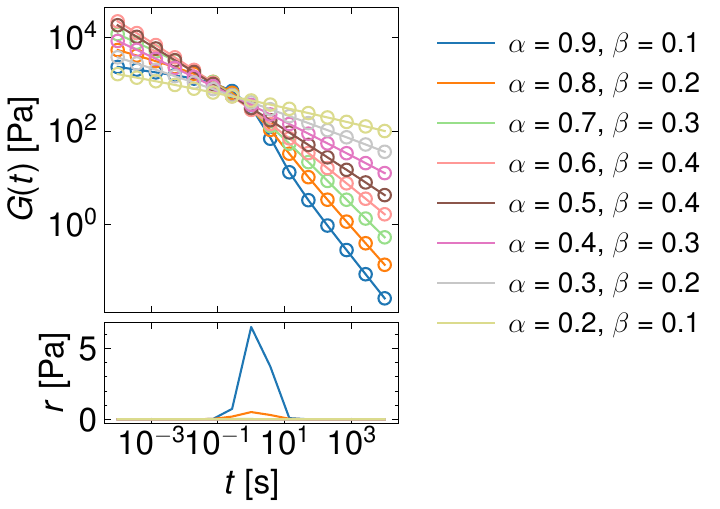} 
    \end{subfigure}
    \hfill
    \begin{subfigure}{0.47\textwidth}
        \centering
        \begin{picture}(0,0)
            \put(-105,0){\textbf{\Large c}} 
        \end{picture}
        \includegraphics[width=\linewidth]{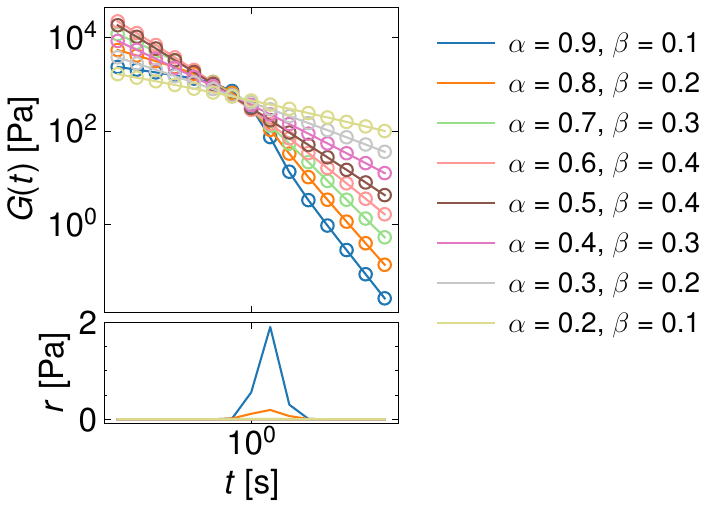} 
    \end{subfigure}
    \hfill
    \begin{subfigure}{0.47\textwidth}
        \centering
        \begin{picture}(0,0)
            \put(-105,0){\textbf{\Large d}} 
        \end{picture}
        \includegraphics[width=\linewidth]{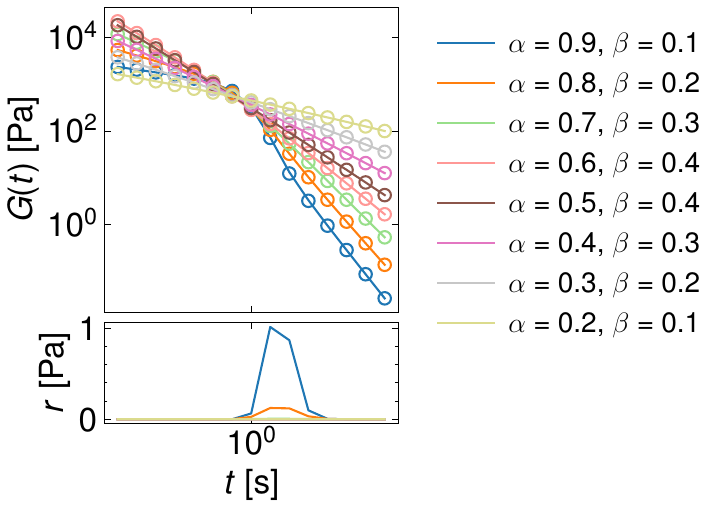} 
    \end{subfigure}
    \caption{Stress relaxation functions  $G(t)$  computed for different fractional orders $\alpha$ and $\beta$ using the \texttt{FractionalMaxwell}. The solid lines represent the functions computed using Garrappa's algorithm~\citep{garrappa_numerical_2015}, while the empty markers represent those computed using the global Padé approximation~\citep{zeng_chen_2015, sarumi_furati_khaliq_2020} \textbf{a)} $R_{a, b}^{3, 2}$, \textbf{b)} $R_{a, b}^{5, 4}$, \textbf{c)} $R_{a, b}^{6, 3}$, \textbf{d)} $R_{a, b}^{7, 2}$. Each figure also includes a residual $r$ plot, highlighting the main differences between both approaches. All the functions were computed by assigning a value of 1000 to $\mathbb{V}$ and $\mathbb{G}$.}
    \label{fig:comparison_pade_garrappa_fmm}
\end{figure}

\clearpage
\section{Comparison between global Padé approximations and Garrappa's algorithm }

pyRheo also provides users with the option to use Garrappa's algorithm~\citep{garrappa_numerical_2015} for computing the Mittag--Leffler function. The algorithm calculates the Mittag--Leffler function based on the numerical inversion of its Laplace transform (LT). An optimal parabolic contour is chosen based on the distance to and strength of the singularities of the Laplace Transform, which aims to minimize computational effort and reduce error propagation. Next, we show how the Padé approximations perform compared to Garrappa's algorithm. In Supplementary Fig.~\ref{fig:comparison_pade_garrappa_fmg}, Supplementary Fig.~\ref{fig:comparison_pade_garrappa_fml}, and Supplementary Fig.~\ref{fig:comparison_pade_garrappa_fmm}, we simply compute Eq.~\ref{eq:relaxation_fmg}, Eq.~\ref{eq:relaxation_fml}, and Eq.~\ref{eq:relaxation_fmm} respectively and show the difference in the stress relaxation $G(t)$ function when using Padé approximations and Garrappa's algorithm.

As seen in Supplementary Fig.\ref{fig:comparison_pade_garrappa_fmg}, Supplementary Fig.~\ref{fig:comparison_pade_garrappa_fml}, and Supplementary Fig.~\ref{fig:comparison_pade_garrappa_fmm}, the Padé approximation shows a residual $r$ peak when the Fractional Maxwell models approach traditional Maxwellian behaviors. The residual, at the same time, reduces its intensity when using more comprehensive Padé approximations. The user might consider using Garrappa's implementation of the Mittag--Leffler function when access to computational resources is not limited, or the task requires outstanding accuracy.

\begin{figure}[htbp]
    \centering
    \begin{subfigure}{0.47\textwidth}
        \centering
        \begin{picture}(0,0)
            \put(-90,0){\textbf{\Large a}} 
        \end{picture}
        \includegraphics[width=\linewidth]{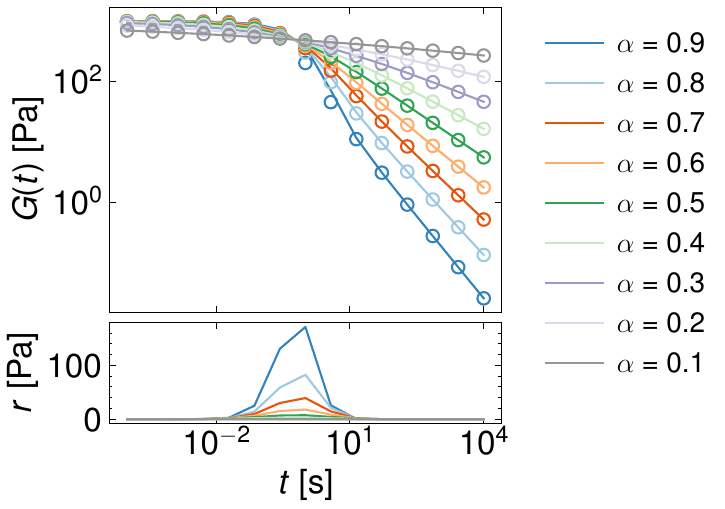} 
    \end{subfigure}
    \hfill
    \begin{subfigure}{0.47\textwidth}
        \centering
        \begin{picture}(0,0)
            \put(-90,0){\textbf{\Large b}} 
        \end{picture}
        \includegraphics[width=\linewidth]{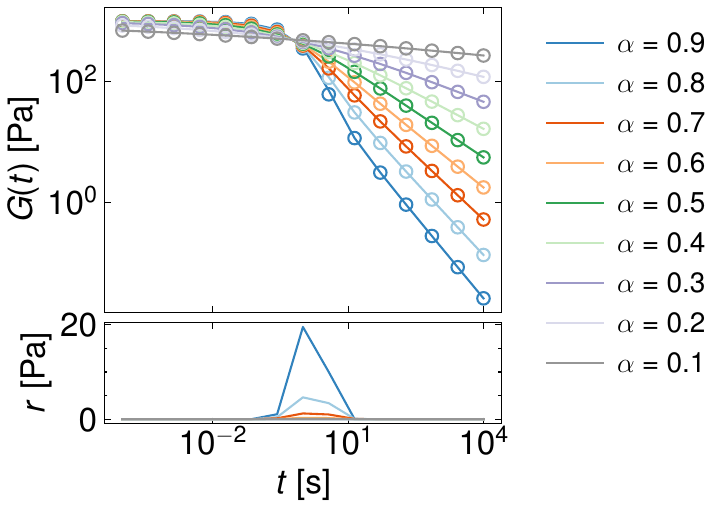} 
    \end{subfigure}
    \hfill
    \begin{subfigure}{0.47\textwidth}
        \centering
        \begin{picture}(0,0)
            \put(-105,0){\textbf{\Large c}} 
        \end{picture}
        \includegraphics[width=\linewidth]{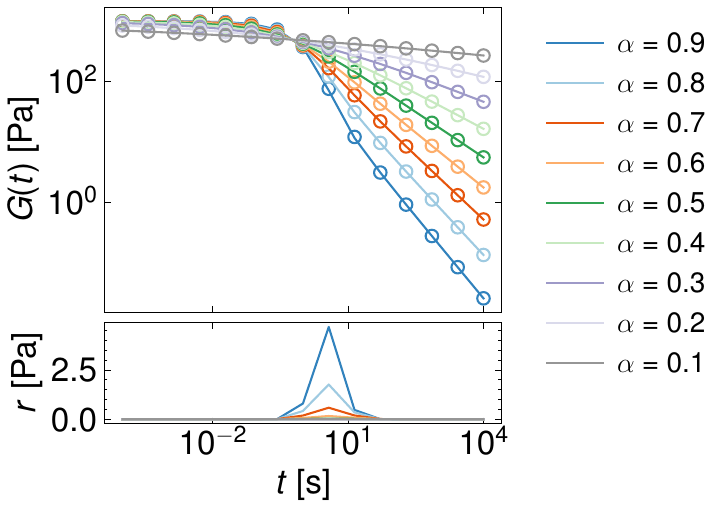} 
    \end{subfigure}
    \hfill
    \begin{subfigure}{0.47\textwidth}
        \centering
        \begin{picture}(0,0)
            \put(-105,0){\textbf{\Large d}} 
        \end{picture}
        \includegraphics[width=\linewidth]{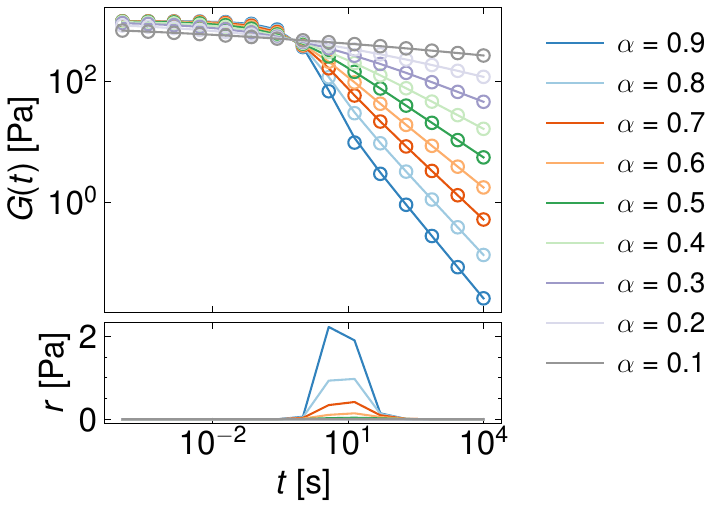} 
    \end{subfigure}
    \caption{Stress relaxation functions  $G(t)$  computed for different fractional orders $\alpha$ using the \texttt{FractionalMaxwellGel}. The solid lines represent the functions computed using Garrappa's algorithm~\citep{garrappa_numerical_2015}, while the empty markers represent those computed using the global Padé approximation~\citep{zeng_chen_2015, sarumi_furati_khaliq_2020} \textbf{a)} $R_{a, b}^{3, 2}$, \textbf{b)} $R_{a, b}^{5, 4}$, \textbf{c)} $R_{a, b}^{6, 3}$, \textbf{d)} $R_{a, b}^{7, 2}$. Each figure also includes a residual $r$ plot, highlighting the main differences between both approaches. All the functions were computed by assigning a value of 1000 to $\mathbb{V}$ and $G$.}
    \label{fig:comparison_pade_garrappa_fmg}
\end{figure}

\clearpage

\section{Initial guesses with Bayesian optimization}
Due to the sensitivity of fitting routines to initial parameter guesses, pyRheo employs three methodologies to help prevent convergence to local minima during parameter optimization. The first method involves using \texttt{manual} initial guesses, which require user expertise. The second method is to restart the fitting process using \texttt{random} initial guesses for a fixed number of iterations. In each iteration, pyRheo attempts to minimize the weighted residual sum of squares (${\rm RSS}_{w_{i}}$) and ultimately selects the best iteration, based on the lowest ${\rm RSS}_{w_{i}}$. The \texttt{random} method is robust. For example, the fittings in the main manuscript were generated with a maximum of 10 iterations.

A method that pyRheo offers for defining initial guesses is Bayesian Optimization (BO)~\citep{head2018scikit, miranda-valdez_viitanen_macintyre_puisto_koivisto_alava_2022, valtteri2024improving}. In this approach, pyRheo creates a mapping from the parameter space $\mathcal{P}$ to the error space $\mathcal{E}$ using Gaussian Process Regression (GPR), represented as  $g: \mathcal{P} \to \mathcal{E}$, where $g$ is the Gaussian Process. The surrogate model $\epsilon = g(p)$ (with $\epsilon \in \mathcal{E}$ and $p \in \mathcal{P}$) is developed by computing the constitutive equation of the target model with fixed parameter values and then recording the difference (residuals) between this computation and the data being analyzed. The goal of BO is to minimize $\epsilon$ by exploring various combinations of parameter values, guided by an acquisition function known as Expected Improvement, which balances exploration and exploitation of the parameter space. Afterward, pyRheo uses the BO solution as the initial guess for the minimization algorithm.

In Supplementary Fig.~\ref{fig:bo_mucus}, we show how the BO works when used as a method to define the initial guesses. In the case of Supplementary Fig.~\ref{fig:bo_mucus}, we fit the creep data of a mucus gel~\citep{liegeois_braunreuther_charbit_raymond_tang_woodruff_christenson_castro_erzurum_israel_etal_2024} using a \texttt{FractionalKelvinVoigtD} model. The Bayesian optimization learns from every iteration $n_i$ and, in the end, finds the best combination of parameter values, which pyRheo uses to initialize the final minimization algorithm.

\begin{figure}[htbp]
    \centering
    \begin{subfigure}{0.47\textwidth}
        \centering
        \includegraphics[width=\linewidth]{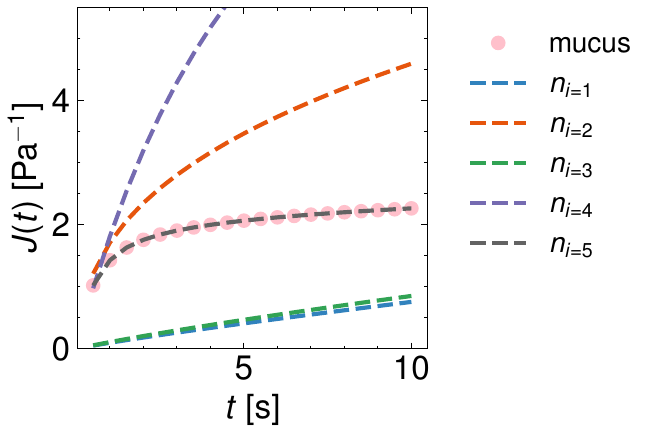} 
    \end{subfigure}
    \hfill
    \caption{Iterative fitting with Bayesian Optimization (BO) as a method to define initial guesses in the \texttt{FractionalKelvinVoigtD} model when fitting the creep data of a mucus gel. Every iteration $n_i$ in the BO process refines the understanding of the parameter space, effectively guiding the selection of the optimal parameter combination. The approach uses the Expected Improvement acquisition function to explore and minimize the error space $\epsilon$. The final best combination of parameters is then used as the initial guess for subsequent minimization algorithms. The BO method shows significant effectiveness in preventing convergence to local minima, improving the robustness of parameter optimization.}
    \label{fig:bo_mucus}
\end{figure}

\clearpage

\section{Graphical user interface (GUI) tutorial}

This tutorial provides step-by-step instructions on how to run a Python file using Anaconda on both Windows and Linux. The GUI for pyRheo is launched by executing either \texttt{pyRheo\_creep\_GUI.py}, \texttt{pyRheo\_relaxation\_GUI.py}, \texttt{pyRheo\_oscillation\_GUI.py}, or \texttt{pyRheo\_rotation\_GUI.py}. We note that for simplicity, the GUI works only using \texttt{auto} bounds and \texttt{random} initial guesses.

\subsection*{Prerequisites}
Before getting started, ensure that you meet all the necessary requirements for using pyRheo. A recommended approach is to use Anaconda for managing dependencies and environments. Download Anaconda and follow the installation instructions suitable for your operating system.

\subsection*{Step 1: Clone the GitHub repository}
Download pyRheo package from \href{GitHub}{https://github.com/mirandi1/pyRheo}.

\subsubsection*{Windows and Linux}
\begin{enumerate}
    \item Open a terminal (Windows: Anaconda Prompt or Command Prompt; Linux: Terminal).
    \item Change the directory to where you want to clone the repository:
    \begin{verbatim}
    cd /path/to/your/directory
    \end{verbatim}
    \item Clone the GitHub repository using \texttt{git} (assuming Git is installed):
    \begin{verbatim}
    git clone https://github.com/mirandi1/pyRheo
    \end{verbatim}
    \item Navigate into the cloned repository and to the GUI folder:
    \begin{verbatim}
    cd pyRheo/gui
    \end{verbatim}
\end{enumerate}

\subsection*{Step 2: Run the Python file}
Execute the specific Python file. For example, in this tutorial, we execute \texttt{pyRheo\_creep\_GUI.py}.

\subsubsection*{Windows and Linux}
\begin{enumerate}
    \item Ensure you are in the correct directory:
    \begin{verbatim}
    cd pyRheo/gui
    \end{verbatim}
    \item Run the Python file:
    \begin{verbatim}
    python pyRheo_creep_GUI.py
    \end{verbatim}
    \item The GUI should prompt as a new window, see Supplementary Fig.\ref{fig:gui_initial}
\end{enumerate}

\noindent Upon running the command, the GUI window titled "pyRheo Creep Model Fitting" will appear.

\begin{figure}[!ht]
    \centering
    \includegraphics[width=\textwidth]{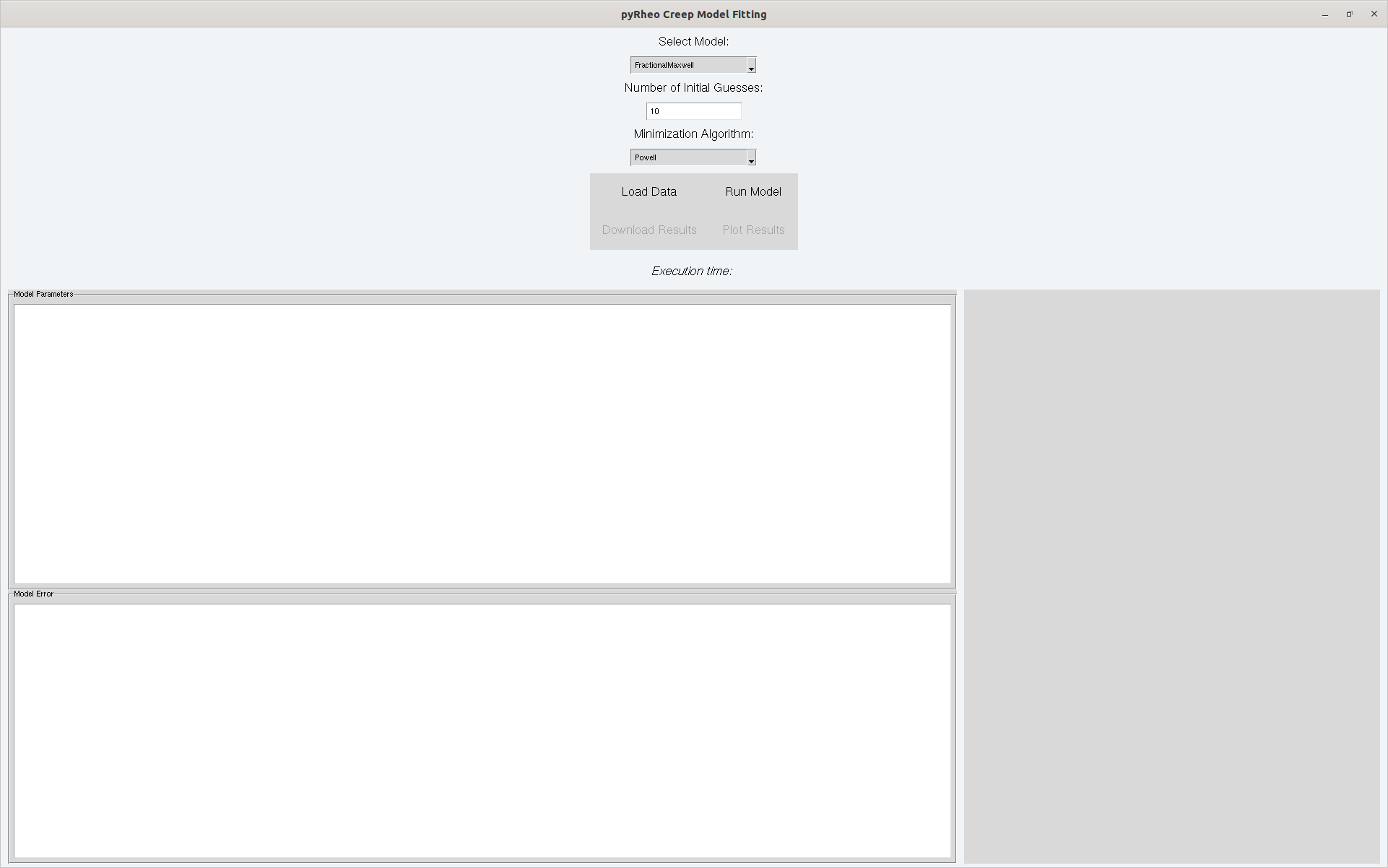}
    \caption{Screenshot of the pyRheo GUI - Initial screen.}
    \label{fig:gui_initial}
\end{figure}

\subsection*{Step 3: Using the GUI}

\subsubsection*{Loading data}
\begin{enumerate}
    \item Click on the \textbf{"Load Data"} button to select the data file to use. The data file should be a CSV file with columns with header "Time" and "Creep Compliance", followed by their corresponding data. The user can look at the file \texttt{creep\_ps190\_data.csv} to understand the requirements in the file structure before loading. The delimiter between columns should be comma "," and decimal ".". Column headers in case of relaxation: "Time" and "Relaxation Modulus". Column headers in case of oscillation: "Angular Frequency" and "Storage Modulus", and "Loss Modulus". Column headers in case of rotation: "Shear Rate" and "Viscosity".
    \item After successfully loading the data, a message box will notify you.
\end{enumerate}

\vspace{1em}
\begin{figure}[h!]
    \centering
    \includegraphics[width=0.5\textwidth]{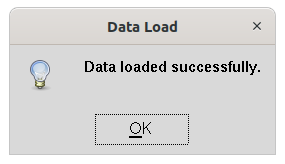}
    \caption{Successful data load prompt.}
\end{figure}

\subsubsection*{Selecting the model and algorithm}
\begin{enumerate}
    \item Use the drop-down menu labeled \textbf{"Select Model:"} to choose the desired model.
    \item Enter the number of initial guesses in the field labeled \textbf{"Number of Initial Guesses:"}. The default value is set to 10.
    \item Use the drop-down menu labeled \textbf{"Minimization Algorithm:"} to select the optimization algorithm.
\end{enumerate}

\subsubsection*{Running the model}
\begin{enumerate}
    \item Click the \textbf{"Run Model"} button to start the model fitting process.
    \item The execution time will be displayed in the bottom part of the GUI.
    \item After the model has run, the model parameters and fitting error will be displayed in the respective sections.
\end{enumerate}

\subsubsection*{Plotting results}
\begin{enumerate}
    \item After running the model, click the \textbf{"Plot Results"} button to visualize the fitting results. The plot will display the original data points and the fitted model as in Supplementary Fig.\ref{fig:gui_plot}.
\end{enumerate}

\begin{figure}
    \centering
    \includegraphics[width=\textwidth]{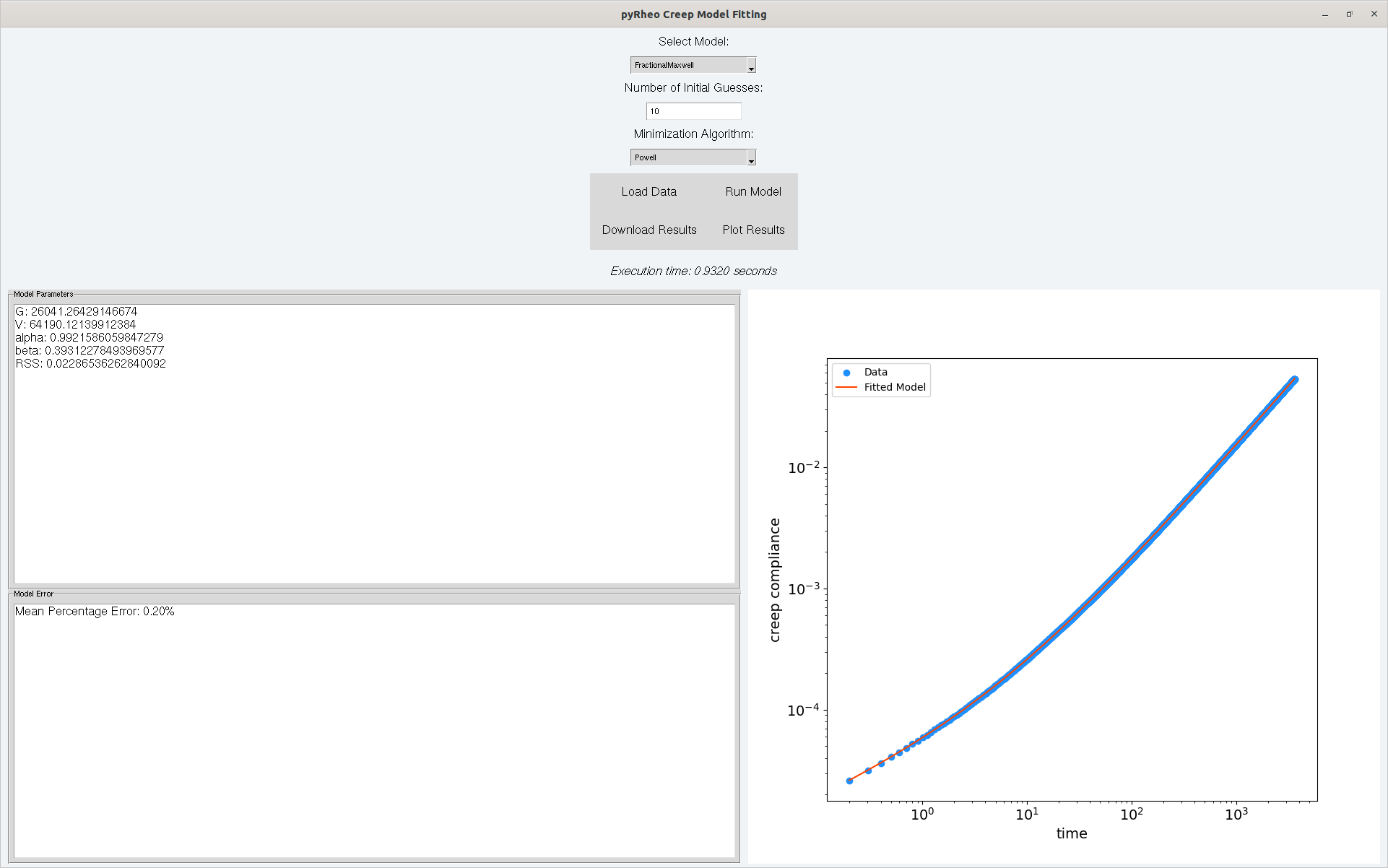}
    \caption{Screenshot of pyRheo GUI - After loading and modeling data.}
    \label{fig:gui_plot}
\end{figure}

\subsubsection*{Downloading Results}
\begin{enumerate}
    \item Click the \textbf{"Download Results"} button to save the fitting results.
    \item You will be prompted to choose a location and filename for the results. The results will be saved as a CSV file.
\end{enumerate}

\clearpage

\bibliographystyle{references} 
\bibliography{supplementary}